%% file: paper.tex
\documentclass[sigconf, nonacm]{acmart}

\newcommand{\final}{1}
\newcommand{\supp}{1}
\newcommand{\resultsupp}{1}

\input{package}
\input{macros}
\input{symbols}

\input{graphicspath}
\input{keywords}

\input{uist2021}

\hypersetup{draft} %
\begin{document}

\input{title_author}

\input{format}

\input{abstract}

\input{ccs}

\input{teaser_fig}

\maketitle

\input{introduction}
\input{prior}

\input{interface}

\input{method}

\input{evaluation}

\input{conclusion}
\input{limitation}

\bibliographystyle{acmart}
{
\bibliography{paper,misc}
}

\balance{}

\ifdefined\supp
\input{supp}

\fi

\ifthenelse{\equal{\final}{0}}
{
\normalsize
\clearpage
\input{plan}
\input{video}
\input{note}
\input{cemetery}
}
{}

\end{document}

%% file: package.tex
\usepackage{color}
\usepackage{ifthen}
\usepackage{float}
\usepackage{alltt}
\usepackage{amsmath}
\usepackage{newlfont} %
\usepackage{floatflt}
\usepackage{wrapfig}
\usepackage{subfig} %
\usepackage{multirow}
\usepackage{booktabs}
\usepackage{cleveref}
\usepackage{algorithmic}
\usepackage[export]{adjustbox}
\usepackage{mathtools, cuted}
\usepackage{CJKutf8} %
\usepackage{balance}

\usepackage[T1]{fontenc}   %

\usepackage[ruled]{algorithm2e} %

\usepackage{booktabs}
\usepackage{textcomp}

\usepackage{microtype}        %
\usepackage{ccicons}          %

\usepackage{enumitem}
\usepackage{capt-of}
\usepackage{tabularx}

%% file: macros.tex
\def\textrightarrow{$\rightarrow$}

\newcommand{\Caption}[2]{\caption[#1]{{\em #1} #2}}

\definecolor{SithColor}{rgb}{0.7,0,0} %
\newcommand{\liyi}[1]{{\color{SithColor} Li-Yi: #1 $\qed$}}
\definecolor{ConsularColor}{rgb}{0,0.4,0} %
\definecolor{GuardianColor}{rgb}{0,0,0.8} %
\definecolor{WinduColor}{rgb}{0.56,0.34,0.62} %

\newcommand{\yilan}[1]{{\color{GuardianColor} Yilan: #1 $\qed$}}
\newcommand{\yl}[1]{{\color{GuardianColor}#1}} %
\newcommand{\kc}[1]{{\color{WinduColor} KC: #1 $\qed$}}
\newcommand{\hongbo}[1]{{\color{ConsularColor} Hongbo: #1 $\qed$}}
\newcommand{\hbc}[1]{}

\newcommand{\krwood}[1]{{\color{WinduColor} Kristin: #1 $\qed$}} %
\newcommand{\rubaiat}[1]{{\color{WinduColor} Rubaiat: #1 $\qed$}} 
\newcommand{\saquib}[1]{{\color{WinduColor} Saquib: #1 $\qed$}} 

\newcommand{\warning}[1]{{\it\color{red} #1}}
\definecolor{NoteColor}{rgb}{0.8,0.5,0.2}
\newcommand{\note}[1]{{\it\color{NoteColor} #1}}
\newcommand{\nothing}[1]{}

\definecolor{AudioColor}{rgb}{0.56,0.34,0.62}

\definecolor{VideoColor}{rgb}{0.44,0.66,0.38}

\definecolor{TodoColor}{rgb}{0.9,0.4,0} %
\newcommand{\todo}[1]{{\bf\color{TodoColor} ETA: #1}}
\definecolor{DoneColor}{rgb}{0.7,0.7,0.7} %

\definecolor{OldColor}{rgb}{0.5,0.5,0.5} %
\newcommand{\old}[1]{{\color{OldColor}#1}}

\definecolor{figred}{rgb}{1,0,0}
\definecolor{figgreen}{rgb}{0,0.6,0}
\definecolor{figblue}{rgb}{0,0,1}
\definecolor{figpink}{rgb}{1,0.63,0.63}

\ifthenelse{\equal{\final}{1}}
{
\renewcommand{\liyi}[1]{}
\renewcommand{\yilan}[1]{}
\renewcommand{\hongbo}[1]{}

\renewcommand{\kc}[1]{}
\renewcommand{\rubaiat}[1]{}
\renewcommand{\saquib}[1]{}
\renewcommand{\krwood}[1]{}
\renewcommand{\warning}[1]{}
\renewcommand{\note}[1]{}
\renewcommand{\todo}[1]{}
\renewcommand{\old}[1]{}
\renewcommand{\yl}[1]{#1}
}
{}

\ifdefined\uist %
\newcommand{\forwardref}[2]{{#2}}
\else
\newcommand{\forwardref}[2]{{#1}}
\fi

\newcommand{\filename}[1]{\url{#1}}
\newcommand{\foldername}[1]{\url{#1}}

\ifdefined\email
\else
\newcommand{\email}[1]{\url{#1}}
\fi

\if@twocolumn
\newcommand{\twocol}{1}
\fi

%% file: symbols.tex
\newcommand{\cityu}{City University of Hong Kong}
\newcommand{\adobe}{Adobe Research}

\newcommand{\energy}{\phi}

\newcommand{\dist}{d}

\newcommand{\refImage}{I}
\newcommand{\dstRegion}{M}
\newcommand{\orientMap}{O}
\newcommand{\radiusMap}{R} %

\newcommand{\inputElemSet}{E}
\newcommand{\outputElemSet}{X}
\newcommand{\stroke}{s}
\newcommand{\elem}{\stroke}
\newcommand{\nbrElem}{\elem^{\prime}}
\newcommand{\outputElem}{\elem_{o}}
\newcommand{\nbrOutputElem}{\outputElem^{\prime}}
\newcommand{\inputElem}{\elem_{i}}
\newcommand{\nbrInputElem}{\inputElem^{\prime}}

\newcommand{\numElemsSelected}{k}

\newcommand{\neighborhood}{\mathbf{N}}
\newcommand{\diff}{\hat{u}}
\newcommand{\elemCentroid}{p}
\newcommand{\elemDir}{v}
\newcommand{\cellCentroid}{\bar{\elemCentroid}}

\newcommand{\weight}{\mu}

\newcommand{\distNeigh}{\dist_{neigh}}
\newcommand{\distRefNeigh}{\dist_{\refImage}}

\newcommand{\lightnessVal}{l}
\newcommand{\gradientMag}{g}
\newcommand{\modelCoeff}{\mathbf{t}} %
\newcommand{\radius}{r}

\newcommand{\autostrokes}[1]{{\color{red}#1}}
\newcommand{\manualstrokes}[1]{{\color{black}#1}}

%% file: graphicspath.tex
\graphicspath{
{figs/}
{figs/userDrawings/}
}%

%% file: keywords.tex
\def\plainkeywords{Interactive System; Autocompletion; Digital Drawing; Prediction; Texture Synthesis}

%% file: uist2021.tex
\setcopyright{acmlicensed}
\copyrightyear{2021}
\acmYear{2021}
\acmDOI{10.1145/1122445.1122456}

\acmConference[UIST 2021]{UIST 2021}{Oct 10--13, 2021}{Virtual}
\acmBooktitle{UIST '21: ACM Symposium on User Interface Software and Technology, Oct 10--13, 2021, Virtual}
\acmPrice{15.00}
\acmISBN{978-1-4503-XXXX-X/18/06}

\acmSubmissionID{6912}

%% file: title_author.tex
\newcommand{\plaintitle}{Autocomplete Image-Guided Sketching} %
\renewcommand{\plaintitle}{Autocomplete Repetitive Stroking with Image Guidance} %

\title{\plaintitle}

\newcommand{\emptyauthor}{}

\author{Yilan Chen}
\affiliation{\institution{\cityu{}}}

\author{Kin Chung Kwan}
\affiliation{\institution{University of Konstanz}}

\author{Li-Yi Wei}
\affiliation{\institution{\adobe{}}}

\author{Hongbo Fu}
\affiliation{\institution{\cityu{}}}

%% file: format.tex
\makeatletter
\def\url@leostyle{%
  \@ifundefined{selectfont}{
    \def\UrlFont{\sf}
  }{
    \def\UrlFont{\small\bf\ttfamily}
  }}
\makeatother
\urlstyle{leo}

\def\pprw{8.5in}
\def\pprh{11in}

\setlength{\paperwidth}{\pprw}
\setlength{\paperheight}{\pprh}
\setlength{\pdfpagewidth}{\pprw}
\setlength{\pdfpageheight}{\pprh}

\definecolor{linkColor}{RGB}{6,125,233}
\hypersetup{%
  pdftitle={\plaintitle},
  pdfauthor={\emptyauthor},
  pdfkeywords={\plainkeywords},
  pdfdisplaydoctitle=true, %
  bookmarksnumbered,
  pdfstartview={FitH},
  colorlinks,
  citecolor=black,
  filecolor=black,
  linkcolor=black,
  urlcolor=linkColor,
  breaklinks=true,
  hypertexnames=false
}

\pagenumbering{arabic}

%% file: abstract.tex
\begin{abstract}

Image-guided drawing can compensate for the lack of skills but often requires a significant number of repetitive strokes to create textures.
Existing automatic stroke synthesis methods are usually limited to predefined styles or require indirect manipulation that may break the spontaneous flow of drawing.
We present a method to autocomplete repetitive short strokes during users' normal drawing process.
Users can draw over a reference image as usual.
At the same time, our system silently analyzes the input strokes and the reference to infer strokes that follow users' input style when certain repetition is detected.
Users can accept, modify, or ignore the system predictions and continue drawing, thus maintaining the fluid control of drawing.
Our key idea is to jointly analyze image regions and operation history for detecting and predicting repetitions.
The proposed system can effectively reduce users' workload in drawing repetitive short strokes and facilitates users in creating results with rich patterns.

\end{abstract}

%% file: ccs.tex
\ccsdesc[500]{Computing methodologies~Computer graphics}
\ccsdesc[500]{Human-centered computing~User interface design}

\keywords{\plainkeywords}

%% file: teaser_fig.tex
\begin{teaserfigure}
\centering
\subfloat[user input]{
\label{fig:teaser:input}
\includegraphics[width=0.19\textwidth]{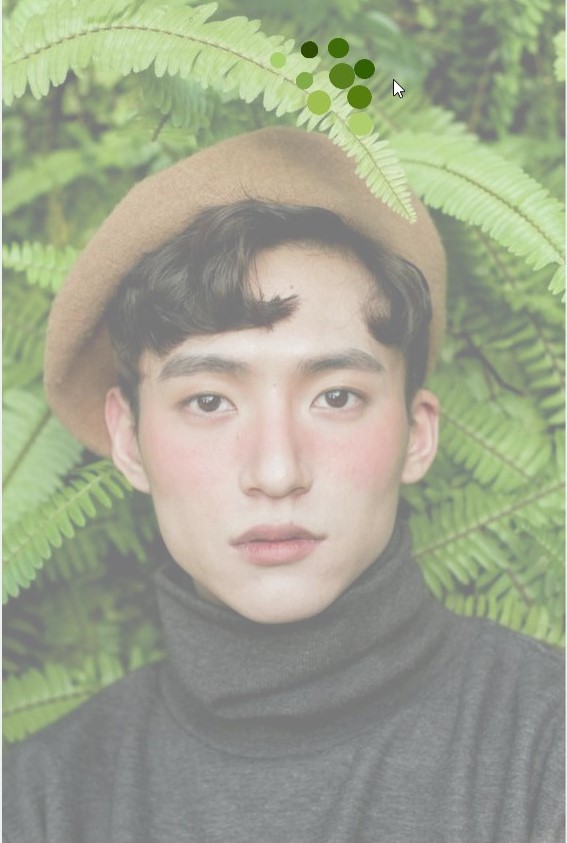}
}%
\subfloat[suggestion]{
\label{fig:teaser:suggest}
\includegraphics[width=0.19\textwidth]{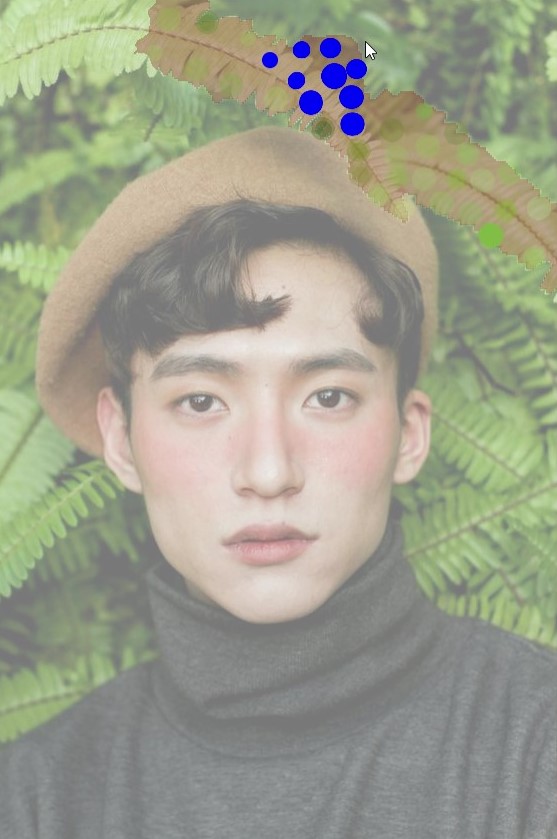}
}%
\subfloat[accept]{
\label{fig:teaser:accept}
\includegraphics[width=0.19\textwidth]{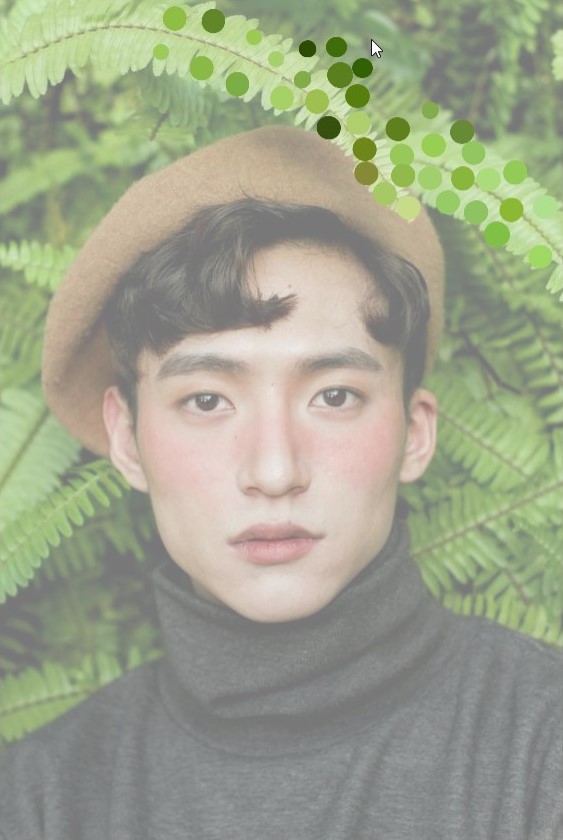}
}%
\subfloat[type visualization]{
\label{fig:teaser:vis}
\includegraphics[width=0.19\textwidth]{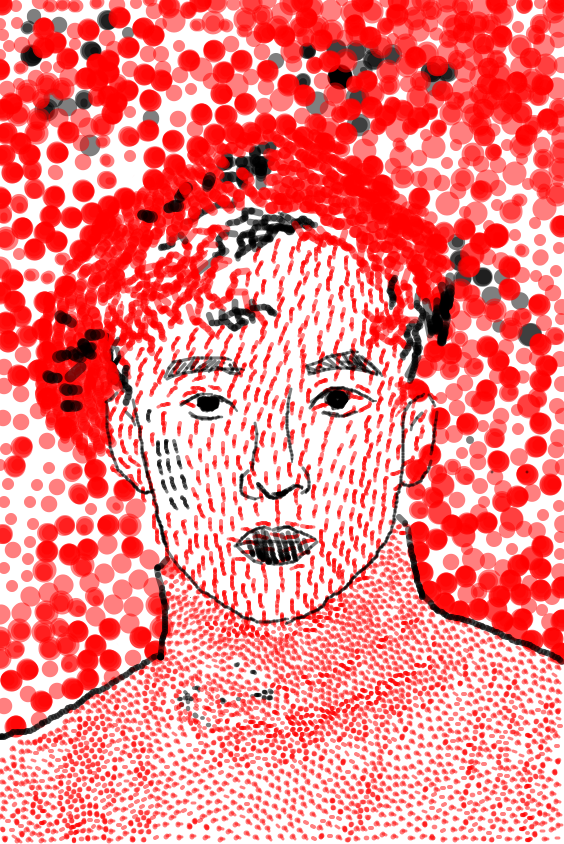}
}%
\subfloat[result]{
\label{fig:teaser:output}
\includegraphics[width=0.19\textwidth]{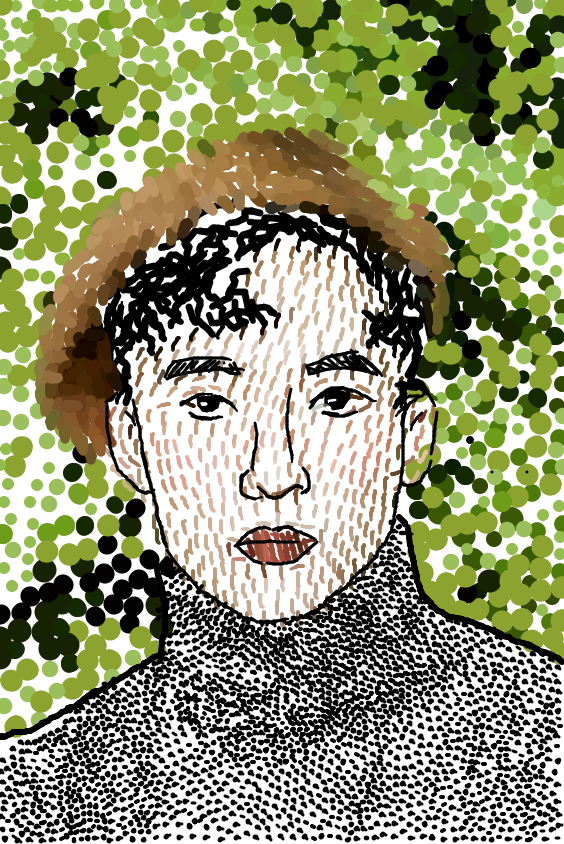}
}%
\Caption{Example of our system workflow.}
{%
A user stipples over a leaf region of a reference image
\subref{fig:teaser:input} while our system predicts what she might draw next \subref{fig:teaser:suggest} (blue strokes: inferred exemplars; pale red region: inferred target region; semi-transparent strokes: system suggestions), which is then accepted by the user \subref{fig:teaser:accept} (green strokes: user inputs or accepted suggestions in this scene).
\subref{fig:teaser:vis} visualizes all the manually drawn content in black (261 strokes) and autocompleted content in red (3510 strokes).
\subref{fig:teaser:output} shows the final result with different repetitive stroke patterns over different regions.
Our autocomplete system can reduce tedious repetitive inputs, while being fully under user control.
}%
\label{fig:teaser}
\end{teaserfigure}

%% file: introduction.tex
\section{Introduction}
\label{sec:intro}

Drawing is a common form of artistic expression.
By varying the stroke, texture, and shading, artists can create drawings with various styles \cite{Dunn:2015:PID}.
Yet, it remains a largely manual process that may require significant artistic expertise and repetitive manual labor.

\input{style_comp_fig}

Various methods have been proposed to synthesize user-initiated repetitive strokes \cite{Kazi:2012:VIT,Xing:2014:APR} to reduce the manual labor.
However, such methods still require sufficient artistic expertise or experience for high-level picture composition.
One common way to overcome this skill barrier is to use a reference photo as a scaffold for drawing, i.e., tracing a reference photo physically via transparent papers or digitally via layers in digital drawing applications.
With a given reference, many methods exist to automate the synthesis of details, such as contours, textures, or strokes \cite{Winkenbach:1994:CPI,Salisbury:1994:IPI,Hertzmann:2001:IA,Alves:2010:SBS,Lu:2012:CST,Benedetti:2014:PBA,Fiser:2016:SIE,Li:2017:CS,Tsai:2017:ULA,Li:2019:ICP}, with the effects tunable via input parameters or exemplars.
However, since these algorithms largely predefine the behaviors, their results may look canned (\Cref{fig:style_comp}) and cannot give users a sense of ownership.
Furthermore, tweaking parameters or providing exemplars can break the spontaneous flow of direct drawing manipulation, which is important to creative decision making \cite{Jacobs:2017:SEP} and essential to a user's enjoyment and exploration \cite{Shneiderman:1987:DMS}.

Manual drawing provides sufficient freedom for individual expressing even when scaffolded with a reference image \cite{Xie:2014:PFS}, and its typical interface (e.g., brush, eraser) is familiar to general users.
Thus, we aim to enhance the manual drawing process and the typical UI design, by automating tedious repetitions.
Our idea is to bridge the two extremes: {\em manual drawing}, which allows full control but can be tedious; and {\em image-based algorithmic synthesis}, which saves efforts but provides limited user control and interactivity. 
As the first attempt towards this goal, our approach focuses on autocompleting repetitive {\em short} strokes, which are very common in pen-and-ink drawing (\Cref{fig:artwork_eg}), under the guidance of a reference image. 
Like typical digital drawing applications, users can draw freely on a reference image with our system.
Meanwhile, our system analyzes the relationships between user inputs and the reference image, detects potential repetitions, and suggests what users might want to draw next.
Users can accept, reject, or ignore the suggestions and continue drawing, thus maintaining the fluid control of drawing. 
See \Cref{fig:teaser} for an example scenario.

The challenge of autocompletion is to predict suggestions that respect both users' inputs and the reference image.
Our method is inspired by image analogy \cite{Hertzmann:2001:IA} and operation history analysis and synthesis \cite{Xing:2014:APR} while leveraging two key insights.
First, since the act of drawing repetitive strokes usually indicates specific intentions (e.g., filling an object or hatching a shading region), we use the common image features among the coherent repetitive strokes to infer the intended regions.
Second, the drawing usually relates to the underlying reference image (e.g., the density of strokes with respect to the image lightness).
Therefore, we analyze the properties of both the drawing and the reference image to infer possible relationships as contextual constraints for stroke prediction.

We implemented a prototype and conducted a pilot study with participants in different backgrounds to evaluate its utility and usability.
The quantitative analysis and qualitative feedback, as well as various drawing results created by users, 
suggest that our system effectively reduces users' workload in drawing repetitive short strokes
and facilitates users in creating results with rich patterns.

%% file: style_comp_fig.tex
\begin{figure}[t]
	\begin{minipage}[t]{0.48\textwidth}
	\centering
	\subfloat[\copyright Alphonso Dunn]{
		\includegraphics[height=0.15\textheight]{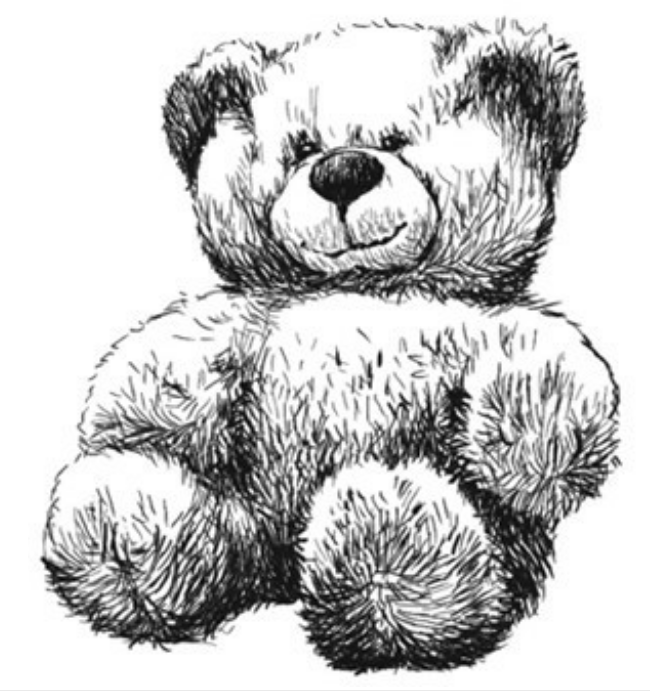}
	}
	\subfloat[\copyright Vincent van Gogh]{
		\includegraphics[height=0.15\textheight]{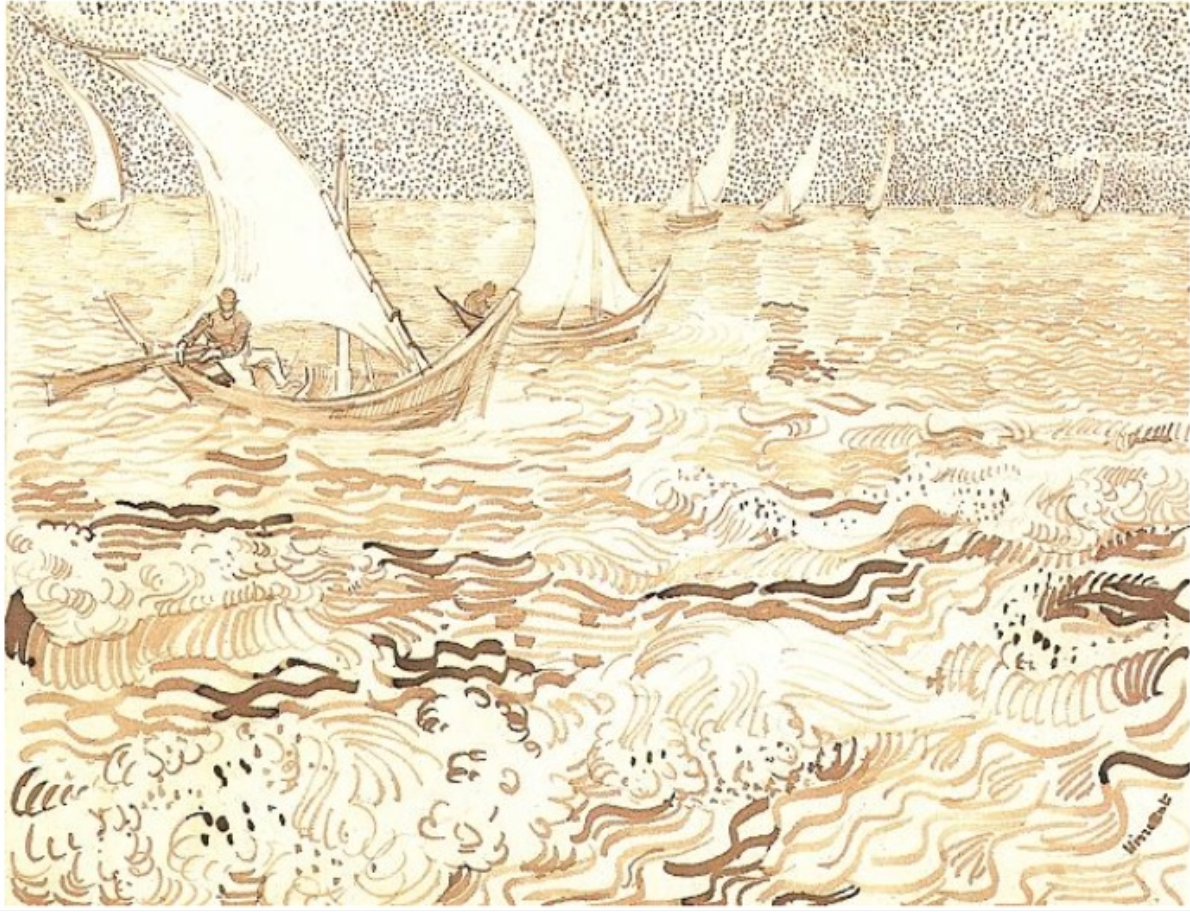}
	}
	\Caption{Inspiring manual drawings by artists.}
	{%
	}
	\label{fig:artwork_eg}
\end{minipage}
\ifdefined\twocol
\end{figure}
\begin{figure}[t]
\vspace{-2mm}
\else
\hfill
\fi
\begin{minipage}[t]{0.48\textwidth}
	\centering
	\subfloat[our result]{
		\label{fig:style_comp:ours}
		\includegraphics[height=0.15\textheight]{man_result_drawing.png}
	}
	\subfloat[produced with \protect\cite{Adobe:2017:PSS}]{\makebox[0.34\linewidth][c]{
		\label{fig:style_comp:paint}
		\includegraphics[height=0.15\textheight]{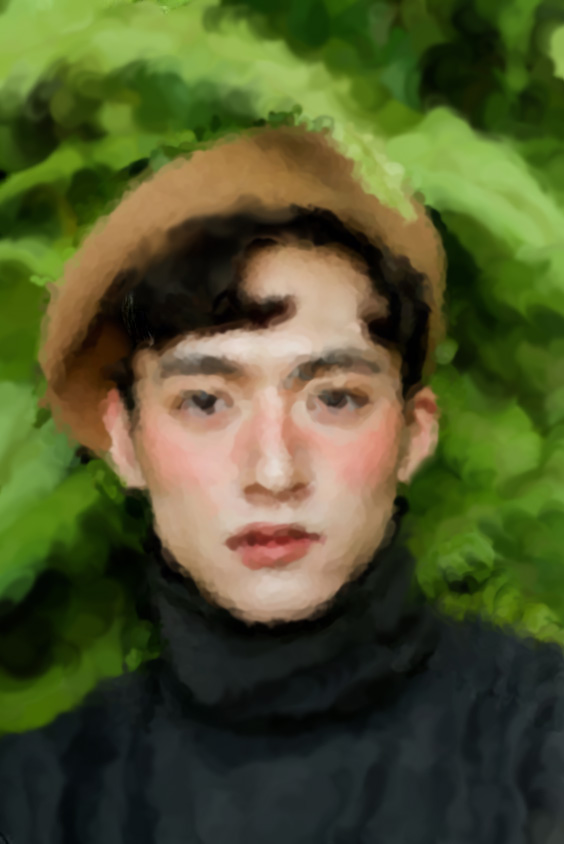}
	}}
	\subfloat[produced with \protect\cite{Martin:2017:SDS}]{\makebox[0.34\linewidth][c]{
		\label{fig:style_comp:stipple}
		\includegraphics[height=0.15\textheight]{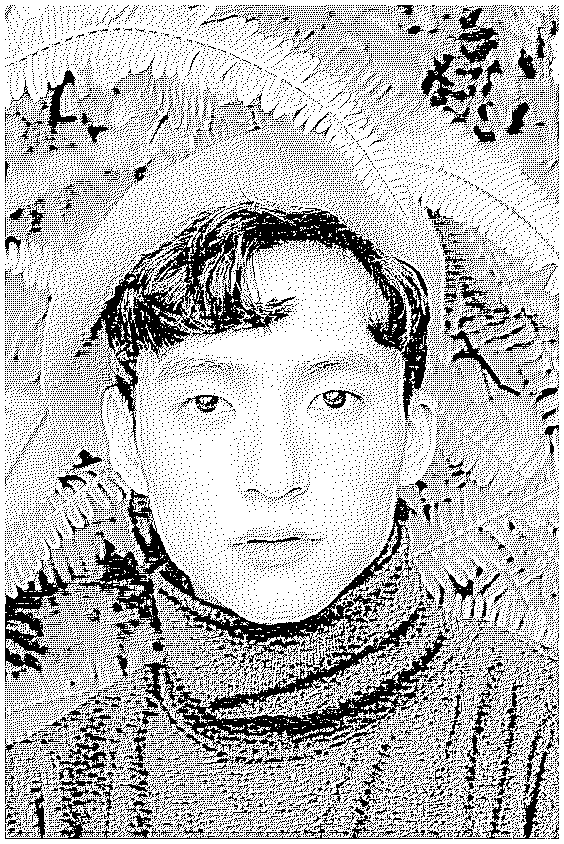}
	}}
	\Caption{}
	{%
		Our work is designed to reduce the workload of completing repetitive patterns during the manual drawing process. 
		The full control of the drawing process leads to more dynamic results
		 than \subref{fig:style_comp:paint} Photoshop's Art History Brush Tool \cite{Adobe:2017:PSS} and
		 \subref{fig:style_comp:stipple} StippleShop \cite{Martin:2017:SDS}.
	}
	\label{fig:style_comp}
\end{minipage}
\end{figure}

%% file: prior.tex
\section{Related Work}
\label{sec:prior}

\subsection{Image-assisted Drawing}%

Many drawing support tools adopt reference images and provide intelligent assistance to novices, e.g.,
beautifying users' sketches with extracted image features \cite{Kang:2005:ISG,Su:2014:ETO,Xie:2014:PFS,Li:2017:CS}, 
or providing educational guidance to novice users \cite{Iarussi:2013:DAA,Matsui:2017:DFD,Williford:2019:DAN}.
We share a similar goal to \cite{Haeberli:1990:PNA,Benedetti:2014:PBA,Tsai:2017:ULA} so as to reduce the user workload.
However, these works use predefined algorithms to generate strokes along cursor movement and only take users’ input as an indicator of where to render, thus greatly limiting users' artistic freedom.
In contrast, we aim to provide more flexibility between automatic synthesis and manual artistic control by autocompleting tedious repetitions during users' normal drawing processes.

\subsection{Image-based Artistic Rendering}

Our work is related to image-based artistic rendering (IB-AR) \cite{Kyprianidis:2013:SAA}, especially stroke-based methods and example-based methods.

\emph{Stroke-based methods} create artistic results from images by strategically generating brushstrokes whose properties (e.g., position, density, orientation, color, size) are related to the image properties (e.g., gradient, edge, color, salience) \cite{Hegde:2013:PRT}.
Among those methods, the closest to ours are the early image-based pen-and-ink rendering methods \cite{Salisbury:1997:OTI,Hiller:2003:BSM}, which allow users to input sample elements for distribution.
However, users have to prepare the sample elements separately (usually as a standalone file) and then tweak parameters to view the rendered output.
In contrast, our system lets users directly specify exemplars on a reference image while silently inferring the distribution properties.

\emph{Example-based methods} aim to model the visual features of example images for transferring.
There are two major modeling approaches: the parametric approach \cite{Kalogerakis:2012:LHP,Gerl:2013:TSI,Gatys:2017:CPF} that is based on the summary statistics of stroke characteristics and thus preserves the global textures better, and the non-parametric approach \cite{Hertzmann:2001:IA,Kaspar:2015:STT,Fiser:2016:SIE} that is based on patch-wise mapping and thus captures the local structures better.
We combine both methods for generating strokes:
the parametric approach to infer statistical relationships between stroke properties and image features, and the patch-wise matching method to preserve the local arrangements of strokes.
Stylit \cite{Fiser:2016:SIE} allows users to stylize a rendered ball and simultaneously propagates the style to arbitrary 3D shapes.
Our method shares a similar idea in interactive style propagation but with two main differences.
First, instead of propagating a style globally, we propagate a style to its perceptually similar local areas so that users can conveniently define different styles in different areas.
Second, we represent drawings as discrete stroke operations instead of raster textures for better preserving their structures and enabling procedural editing \cite{Schwarz:2007:MRP}, such as changing the color or size of the drawn strokes.%

\subsection{Operation History-assisted Authoring}%

Operation histories~\cite{Nancel:2014:CCM} have been leveraged in different authoring tasks, such as sketching \cite{Xing:2014:APR}, animation \cite{Xing:2015:AHA,Peng:2020:AKS}, modeling \cite{Peng:2018:A3S,Suzuki:2018:TED}, beautification of freehand drawings \cite{Fiser:2015:SDB}, and handwritings \cite{Zitnick:2013:HBU}.
Our work is most closely related to that by 
Xing et al.'s \shortcite{Xing:2014:APR}, which autocompletes repetitive sketching by analyzing the dynamic operations recorded during authoring.
Our method extends their work to consider additional information from a reference image and thus enables the propagation of strokes to regions with similar image attributes such as color or semantic meaning.

In our use scenario, an operation is an input stroke, so our work is also related to stroke pattern analysis and synthesis \cite{Barla:2006:IHS,Ijiri:2008:EBP,Alves:2010:SBS,Kazi:2012:VIT,Hsu:2020:AEF}.
These works disregard the temporal relationship among past strokes and do not use image guidances and thus are different from ours.

To sum up, we list our major differences from the discussed closely related works in \Cref{table:prior:diff}.

\input{method_comparison_tab}

%% file: method_comparison_tab.tex
\begin{table}[tbh!p]
\footnotesize
\Caption{The differences between our tool and closely related works.}
{%
	``batch'' means the generation is performed in a batch, based on predefined attributes;
	``dynamic'' means the generation is performed based on dynamic operation history.
	``direct'' means users can specify a style by directly operating on the output.
	``Y'' and ``N'' represent yes and no, respectively, for using image references.
}
\label{table:prior:diff}
\begin{tabularx}{\linewidth}{|X|XXXXXX|}
	\hline
	Method & \cite{Hiller:2003:BSM}  & \cite{Hertzmann:2001:IA} & \cite{Gerl:2013:TSI} & \cite{Kazi:2012:VIT} & \cite{Xing:2014:APR} & Ours \\
	\hline
	Reference          & Y  & Y & Y & N  & N  & Y  \\
	Process  & batch  & batch  & batch & batch  & dynamic  & dynamic  \\
	Format & stroke & pixel & stroke & stroke & stroke & stroke \\
	Operate & indirect & indirect & indirect & direct & direct & direct \\
	\hline
\end{tabularx}
\end{table}

%% file: interface.tex
\section{User Interface}
\label{sec:interface}

\input{ui_fig}

Our prototype follows a standard digital drawing interface, with the added autocomplete feature, as shown in \Cref{fig:ui}.
A user draws on top of a reference image displayed semi-transparently on the main canvas, 
while our system analyzes the input strokes and the reference image in the background.

\subsection{Autocomplete}
\input{autocomplete_fig}

In the autocomplete mode, our system automatically analyzes whenever the user finishes a new stroke.
When a potential repetition is detected, our system highlights the currently repetitive strokes and an inferred propagation region, updates the inferred parameters in the filling property panel, and generates autocompletion suggestions.
Users can accept or reject all the suggestions via hotkeys, accept part of them via lasso selection, or ignore them and continue to draw (\Cref{fig:autocomplete}).
The suggestions will keep updating according to user inputs.

\subsection{Interactive Editing}
\input{interactive_edit_fig}

Our system provides a set of tools to refine the autocompleted results.

\begin{description}
	
	\item[Propagation region editing.]
	Users can create/add/subtract a new region using the intelligent scissors tool \cite{Mortensen:1995:ISI} or expand an existing region by a fixed width (\Cref{fig:ui}e) for stroke autocompletion.
	\Cref{fig:region_edit} shows an example of creating a new region for stroke regeneration.
	
	\item[Density editing.]
	Users can tweak three parameters to adjust the density of the generated strokes: the average {\em spacing}, the {\em lightness} coefficient and the {\em gradient} coefficient.
	The latter two define the relationships between density and image lightness/gradient, respectively.
	Our system automatically updates these parameters upon prediction, and the updated parameters provide
	a starting point for users to manipulate.
	\Cref{fig:density_edit} shows an example.%
	
	\item[Orientation editing.]
	Our system automatically predicts whether the input exemplar correlates with the image flow, which can also be tweaked by users manually.
	Users can also modify the image flow field via the gesture brush, and the touched strokes will be rotated to align with the gesture direction.
	See \Cref{fig:orient_edit} for an example.

\end{description}

\subsection{Auxiliary Functions}
Our prototype also includes the auxiliary functions below.
These are not unique to our system but can facilitate the usual drawing processes.
\begin{description}
	\item[Post-edit stroke properties.]
	Users can select the existing strokes and edit their properties, such as size and color.
    \item[Auto-color.]
    This function, when toggled on, can automatically colorize strokes with color from the reference image.
    \item[Switch view.]
    Users can press the space key to switch between the canvas view, reference view, and pure drawing view.
\end{description}

%% file: ui_fig.tex
\begin{figure}[th]
	\centering
	\includegraphics[width=0.98\linewidth]{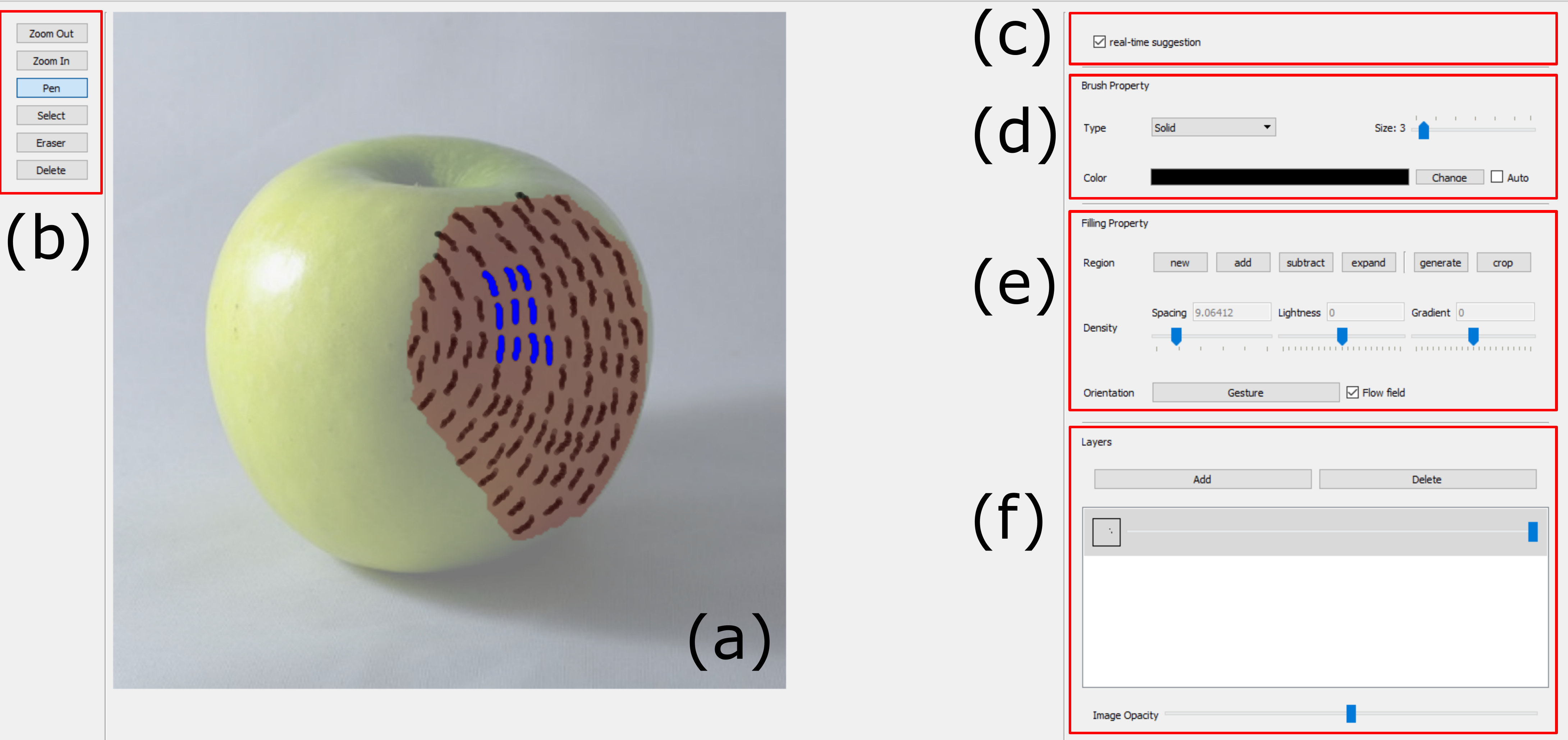}
	\Caption{User interface,}
	{%
		consisting of a central drawing canvas (a), a toolbar for drawing and selection (b), a toggle-switch of the autocomplete mode (c), a brush property toolbar (d), a filling property toolbar (e), and a layers panel (f).
	}
	\label{fig:ui}
\end{figure}

%% file: autocomplete_fig.tex
\begin{figure}[t]
  \centering
  \subfloat[selection]{
			\label{fig:autocomplete:select}
			\includegraphics[width=0.23\linewidth]{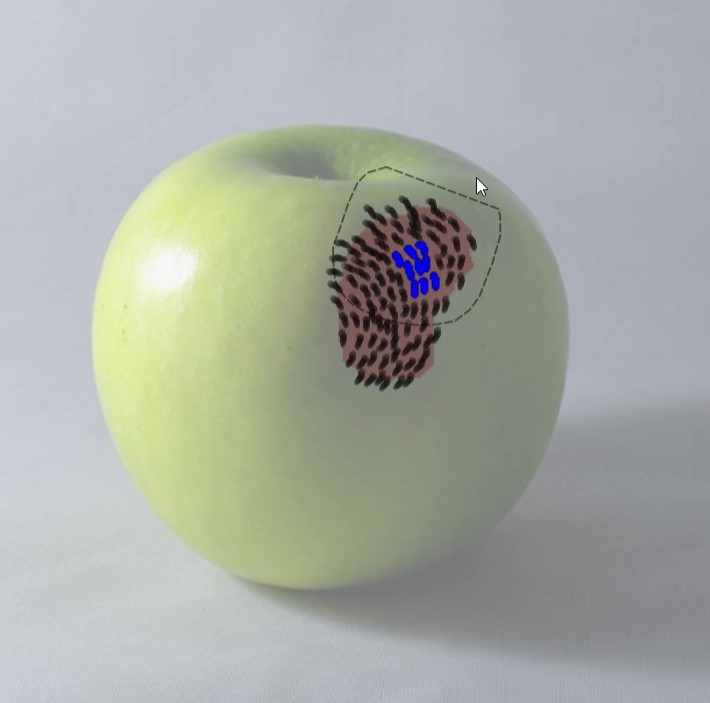}
		}
		\subfloat[result]{
			\label{fig:autocomplete:result}
			\includegraphics[width=0.23\linewidth]{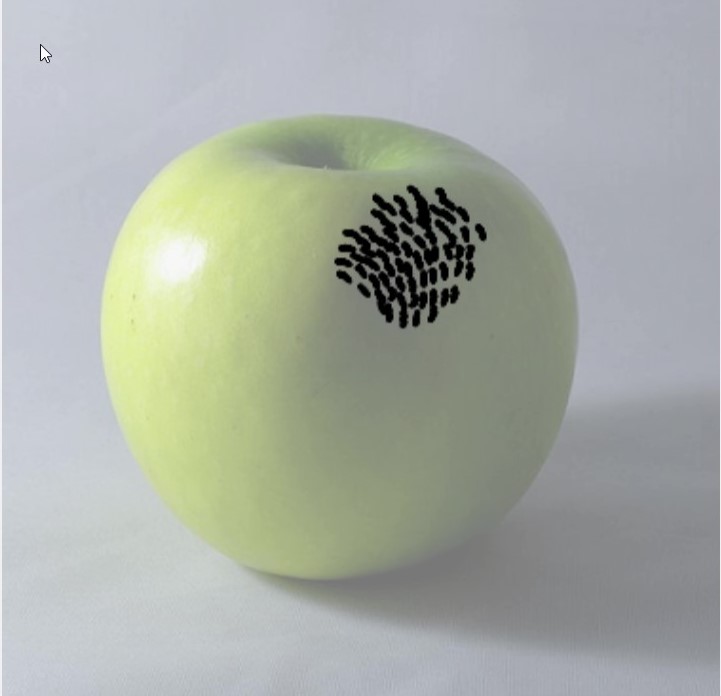}
		}
		\subfloat[updated suggestions]{
			\label{fig:autocomplete:continue}
			\includegraphics[width=0.23\linewidth]{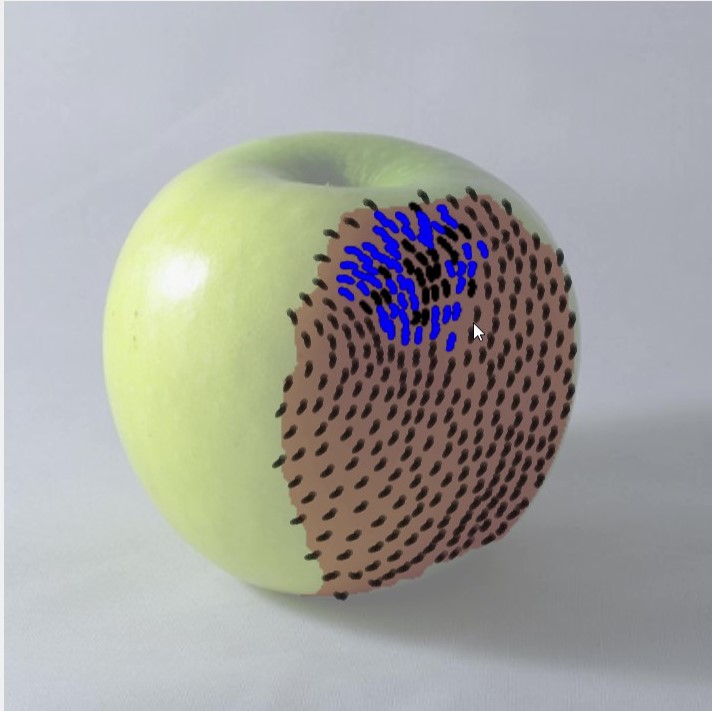}
		}
		\subfloat[final]{
			\label{fig:autocomplete:accept}
			\includegraphics[width=0.23\linewidth]{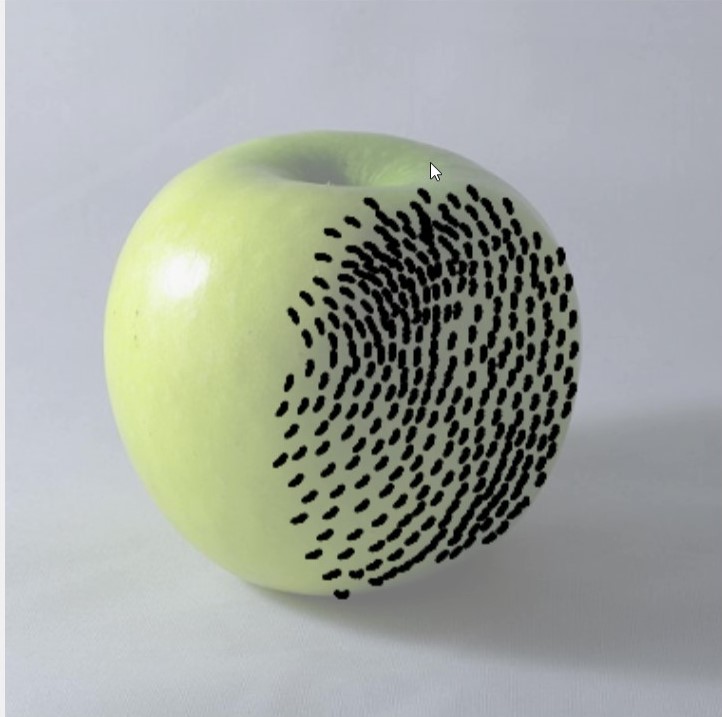}
		}

		\Caption{An example of autocompletion.}
		{%
			The user selects part of the suggestions via the lasso selection tool \subref{fig:autocomplete:select} with the result in \subref{fig:autocomplete:result}, continues to draw leading to the updated suggestions \subref{fig:autocomplete:continue}, and accepts all the suggestions via a hotkey \subref{fig:autocomplete:accept}.
The blue strokes in \subref{fig:autocomplete:select} and \subref{fig:autocomplete:continue} indicate inferred exemplars from user-input strokes.
		}
		\label{fig:autocomplete}
		\label{fig:autocomplete:suggestion} %
\end{figure}

%% file: interactive_edit_fig.tex
\input{region_edit_fig}
\input{density_edit_fig}
\input{orient_edit_fig}

%% file: region_edit_fig.tex
\begin{figure}[tbh]
\vspace{-2mm}
\centering
\subfloat[initial]{
\label{fig:region_edit:init}
\includegraphics[width=0.24\linewidth]{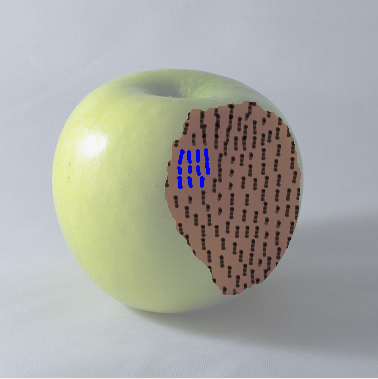}%
}
\subfloat[new region]{
\label{fig:region_edit:edit}
\includegraphics[width=0.24\linewidth]{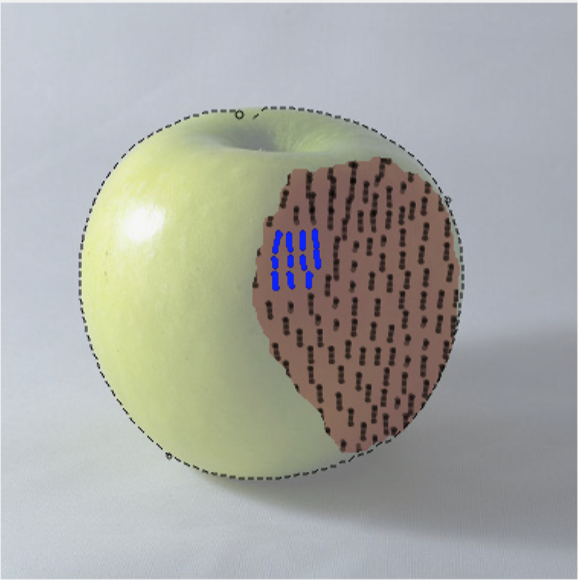}%
}
\subfloat[result]{
\label{fig:region_edit:result}
\includegraphics[width=0.24\linewidth]{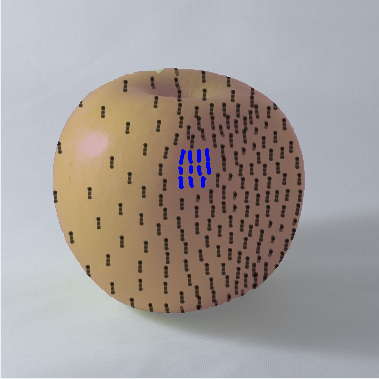}%
}
\Caption{Region editing example.}
{%
The initial prediction \subref{fig:region_edit:init} contains only the brown region.
The user-specified region \subref{fig:region_edit:edit} contains the entire apple, with the corresponding synthesis result in \subref{fig:region_edit:result}.
}
\label{fig:region_edit}
\end{figure}

%% file: density_edit_fig.tex
\begin{figure}[tbh]
\vspace{-2mm}
\centering
\subfloat[(8, 0, 0)]{
	\label{fig:density_edit:init}
	\includegraphics[width=0.23\linewidth]{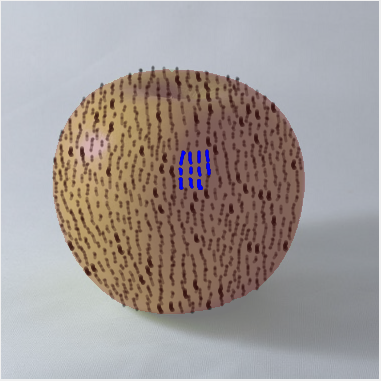}
}
\subfloat[(15, 0, 0)]{
	\label{fig:density_edit:slope0}
	\includegraphics[width=0.23\linewidth]{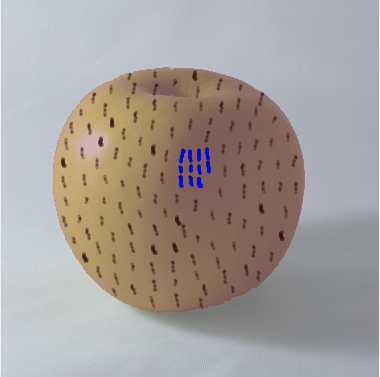}
}
\subfloat[(8, 0.2, 0)]{
	\label{fig:density_edit:slope50}
	\includegraphics[width=0.23\linewidth]{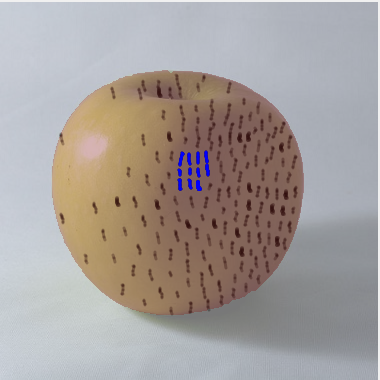}
}
\subfloat[(8, 0, 0.6)]{
	\label{fig:density_edit:both}
	\includegraphics[width=0.23\linewidth]{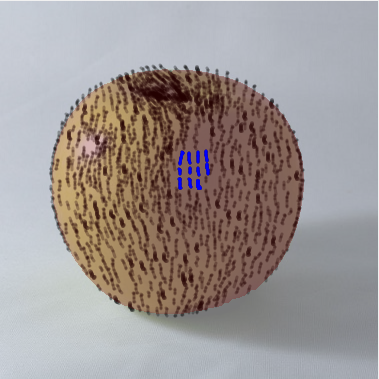}
}

\Caption{Density editing example with different values of {\em spacing}, {\em lightness} and {\em gradient} parameters.}
{%
Larger spacing parameters lead to sparser strokes, while larger lightness and gradient parameters lead to larger stroke density variations.
}
\label{fig:density_edit}
\end{figure}

%% file: orient_edit_fig.tex
\begin{figure}[tbh]
\vspace{-2mm}
\centering
\subfloat[]{
	\label{fig:orient_edit:gesture}
	\includegraphics[width=0.23\linewidth]{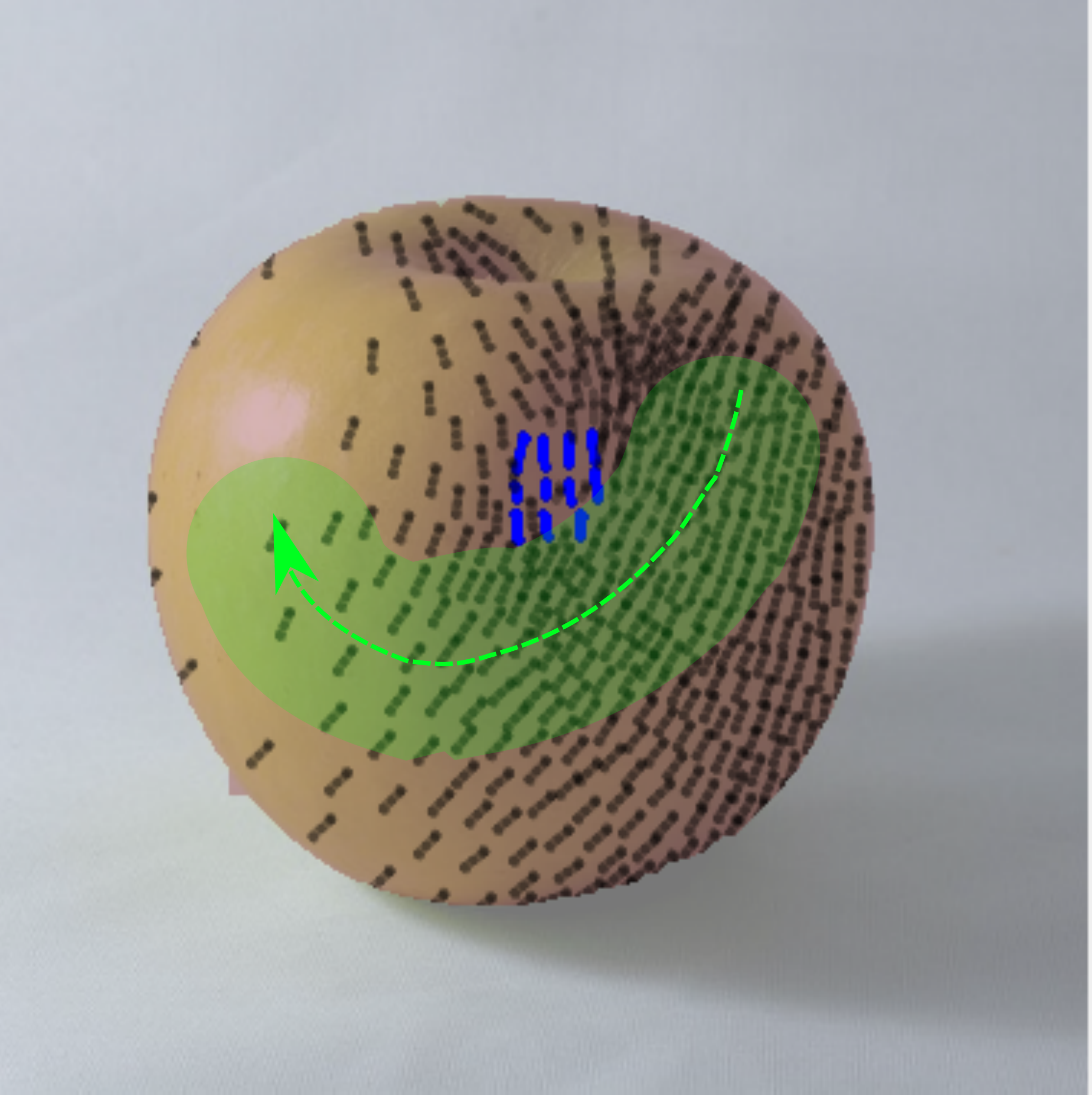}
}
\subfloat[]{
	\label{fig:orient_edit:newfield}
	\includegraphics[width=0.23\linewidth]{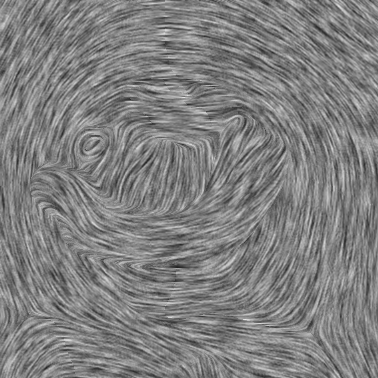}
}
\subfloat[]{
	\label{fig:orient_edit:fieldresult}
	\includegraphics[width=0.23\linewidth]{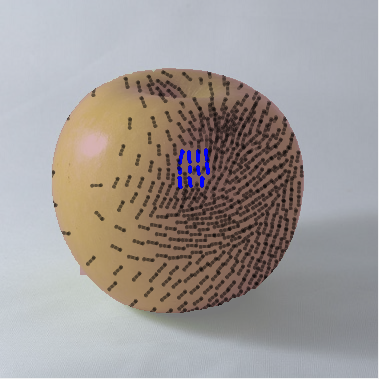}
}
\subfloat[]{
	\label{fig:orient_edit:global}
	\includegraphics[width=0.23\linewidth]{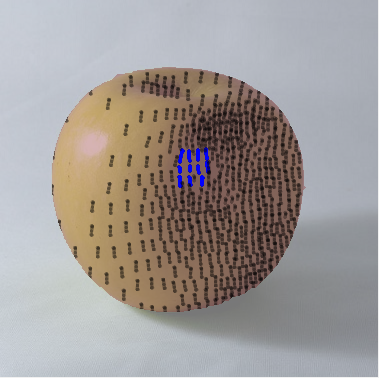}
}
\Caption{Orientation editing example.}
{
	\subref{fig:orient_edit:gesture} User gesture.
	\subref{fig:orient_edit:newfield} Orientation field updated based on the user gesture and the original image flow field.
	\subref{fig:orient_edit:fieldresult} Updated result.
	\subref{fig:orient_edit:global} A result without any orientation field.
}
\label{fig:orient_edit}
\end{figure}

%% file: method.tex
\section{Our Approach} %
\label{sec:method}

\input{method_overview}
\input{method_synthesis}

\input{method_analysis}

%% file: method_overview.tex
To support the autocomplete functionalities described in \Cref{sec:interface},
our system involves two key algorithm steps:
(1) inferring the input exemplar, the output region, and the contextual constraints from the stroke history and the reference image;
(2) synthesizing suggestive strokes accordingly.
This section first describes how to {\em synthesize} \forwardref{~(\Cref{sec:method:synthesis})}{} strokes, assuming all the information is available, and then explains how to {\em infer} \forwardref{~(\Cref{sec:method:analysis})}{} the necessary information for synthesis.

%% file: method_synthesis.tex
\subsection{Stroke Synthesis}
\label{sec:method:synthesis}

\paragraph{Problem statement}
The inputs to our stroke synthesis method include an exemplar $\inputElemSet$ consisting of repetitive strokes, the reference image $\refImage$, a target region mask $\dstRegion$, an orientation map $\orientMap$, and a radius map $\radiusMap$.
Pixel values of $\radiusMap$ denote the extents of stroke spacing: a smaller value leads to a denser distribution.
Our goal is to compute an output set of strokes $\outputElemSet$ over the output region $\dstRegion$, such that $\outputElemSet$ is similar to $\inputElemSet$ with respect to $\refImage$.
We describe how to infer $\inputElemSet$, $\dstRegion$, $\orientMap$, and $\radiusMap$ from user interactions with $\refImage$ in \forwardref{\Cref{sec:method:analysis}}{the {\em prediction} section}.

\paragraph{Key idea}
We extend the discrete element texture synthesis method \cite{Ma:2011:DET,Xing:2014:APR}, which represents strokes as point samples and iteratively improves the sample distribution by minimizing the neighborhood difference between the exemplar and the output, with an additional reference image.
First, we combine sample neighborhoods \cite{Ma:2011:DET} with image features \cite{Hertzmann:2001:IA} for measuring neighborhood difference. %
Second, the range and orientation of sample neighborhoods are determined by the radius and orientation maps inferred from the reference image. 
\Cref{fig:method_illustration} shows our key idea.

\input{method_representation}

\paragraph{Initialization}
We pre-process the target region mask $\dstRegion$ by removing the area occupied by existing strokes in the same layer to avoid cluttering,
and then initialize the output $\outputElemSet$ by generating sample positions with Poisson-disk sampling based on the radius map $\radiusMap$.
For each sampled position, we copy the input stroke with the smallest image feature distance $\distRefNeigh$, which will be explained in \Cref{eqn:dist:neigh}.
We then optimize the output for a few objectives, as detailed below.

\paragraph{Neighborhood term}

We define the neighborhood of a stroke $\elem$ as both its neighboring strokes as well as an $\radiusMap(\elem) \times \radiusMap(\elem)$ image patch around its centroid, where $\radiusMap(\elem)$ is the radius value at $\elem$.
Prior methods (e.g. \cite{Ma:2011:DET}) determine the neighboring strokes by spatial distances.
Thus, the neighborhood radius should be large enough in order to capture an underlying pattern.
However, this might include redundant strokes and thus decrease the performance.
Therefore, we adopt Zhao et al.’s method \shortcite{Zhao:2011:CPR} to automatically find a minimum representative neighborhood, considering not only the spatial distance between strokes but also their locations.
As depicted in \Cref{fig:synthesis:neighborhood}, we set the neighborhood radius of the center stroke $\elem$ to $2\radiusMap(\elem)$.
We then divide all the strokes within the neighborhood radius into four quadrants with respect to the local frame defined by the orientation at $\orientMap(\elem)$,
and collect the $n$ nearest strokes from each quadrant as the representative neighborhood, denoted as $\neighborhood(\elem)$.
In our implementation, we set $n=4$ for the input exemplar and $n=1$ for the output strokes to ensure that each output neighborhood can be maximally matched.

For a stroke $\elem$ and a neighboring stroke $\nbrElem\in\neighborhood(\elem)$, we compute their difference in position and direction as:
\begin{align}
\diff(\nbrElem, \elem)
=\left(\frac{1}{\radiusMap(\elem)} \orientMap(\elem)^{-1} \left(\elemCentroid(\nbrElem) - \elemCentroid(\elem) \right),
\orientMap(\elem)^{-1} \left(\elemDir(\nbrElem) - \elemDir(\elem) \right)\right),
\end{align}
which is computed in the local frame defined by the radius map $\radiusMap$ and orientation map $\orientMap$.
Therefore, the neighborhood distance between an output stroke $\outputElem$ and an input stroke $\inputElem$ is:
\begin{equation}
\distNeigh(\outputElem, \inputElem) =
\sum_{\nbrOutputElem\in\neighborhood(\outputElem)}
\left| \diff(\nbrOutputElem, \outputElem) - \diff(\nbrInputElem, \inputElem)\right| ^{2}
+ \weight 
\underbrace{\left| \refImage(\outputElem) - \refImage(\inputElem)\right| ^{2}}_{\distRefNeigh},
\label{eqn:dist:neigh}
\end{equation}
where
$\nbrInputElem$ is the matched input sample for $\nbrOutputElem$ via the Hungarian algorithm \cite{Ma:2011:DET,Ma:2013:DET},
the second term measures the image feature distance $\distRefNeigh$, and $\weight$ ($=0.1$ in our implementation) controls the relative weighting.
We use the mean $Lab*$ color of an $\radius \times \radius$ patch at the stroke centroid as the image feature vector.
The overall neighborhood term to minimize is:
\begin{equation} \label{eq:energy:nbh}
	\energy_{neigh}(\outputElemSet, \inputElemSet) = 
	\sum_{\outputElem\in\outputElemSet} \min_{\inputElem\in\inputElemSet} \distNeigh(\outputElem, \inputElem).
\end{equation}

\paragraph{Correction term}
Since the neighborhood term is a one-way matching from the output neighborhoods to the input neighborhoods, \yl{sometimes the optimization would tend to leave out some void regions}.
Besides, the neighborhood term does not preserve strokes' alignment to the image (e.g., \Cref{fig:iteration:nocomp}).
To address these issues, we apply a correction term.
We compute a weighted centroidal Voronoi diagram from all the strokes' center points, using $\frac{1}{\radiusMap}$ as weight, and denote the computed region centroids as $\{\cellCentroid\}$.
Thus we can minimize the distance between each output stroke centroid and the region centroid, defined as follows:
\begin{equation}\label{eq:energy:corr}
	\energy_{corr}(\outputElemSet) = \sum_{\outputElem\in\outputElemSet} 
	\left| \elemCentroid(\outputElem) - \cellCentroid(\outputElem)\right| ^{2}.
\end{equation}

\paragraph{Solver}

\input{iteration_fig}

The energy function we aim to minimize is defined as:
\begin{equation}
\energy(\outputElemSet, \inputElemSet) = (1-w) \energy_{neigh} + w \energy_{corr}.
\end{equation}

We iteratively minimize the energy function following the EM methodology in \cite{Ma:2011:DET}.
In each iteration, for each output stroke $\outputElem$, we search for the most matched input stroke $\inputElem$ to minimize $\energy_{neigh}$, compute the Voronoi diagram centroid $\cellCentroid$ to minimize $\energy_{corr}$, and solve a least-squares system combining both terms.
Let $m$ be the total number of iterations. For the $i-$th iteration, we set $w = (i / m)^{2}$, which means that more weight is given to $\energy_{neigh}$ in the beginning of iterations,
so that we can optimize the neighborhood distribution first before doing corrections, which leads to better results.

\Cref{fig:iteration:init,fig:iteration:intermediate,fig:iteration:final} 
show the iterative optimization process of both the objectives.
In comparison, \Cref{fig:iteration:nocomp} shows the result without the correction term and \Cref{fig:iteration:noimg} shows the result without using the image neighborhood in both initialization and optimization.

%% file: method_representation.tex
\paragraph{Stroke representation}
\label{sec:method:representation}

\input{method_illustration_fig}

As shown in \Cref{fig:synthesis:element}, a stroke $\elem$ is an ordered list of sample points, each with a timestamp and appearance attributes such as thickness and color. 
Here we focus on autocompleting short strokes, so we represent each stroke by its centroid $\elemCentroid$ and the average direction $\elemDir$ for efficiency
during synthesis, without considering any other information of the original stroke.
To take the drawing order into consideration, we obtain the dominant direction by averaging the vectors from the start point to each subsequent point.
After synthesis, we reconstruct all the sample points according to the updated centroid and direction.

%% file: method_illustration_fig.tex
\begin{figure}[t]
	\begin{minipage}{0.48\textwidth}
		\centering
		
		\subfloat{
			\includegraphics[height=0.18\textheight]{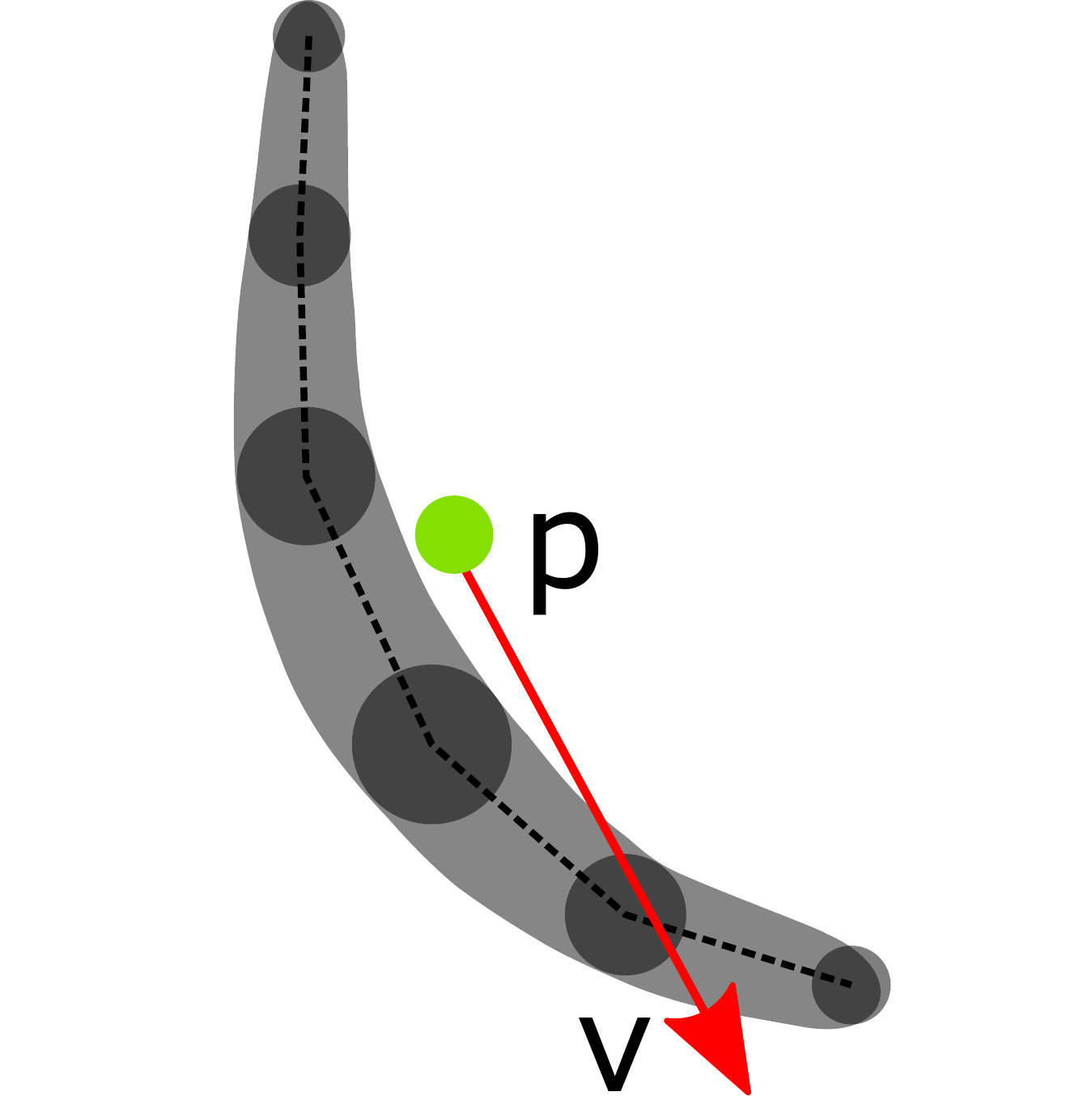}
			\label{fig:synthesis:element}
		}%
		\subfloat{
			\includegraphics[height=0.18\textheight]{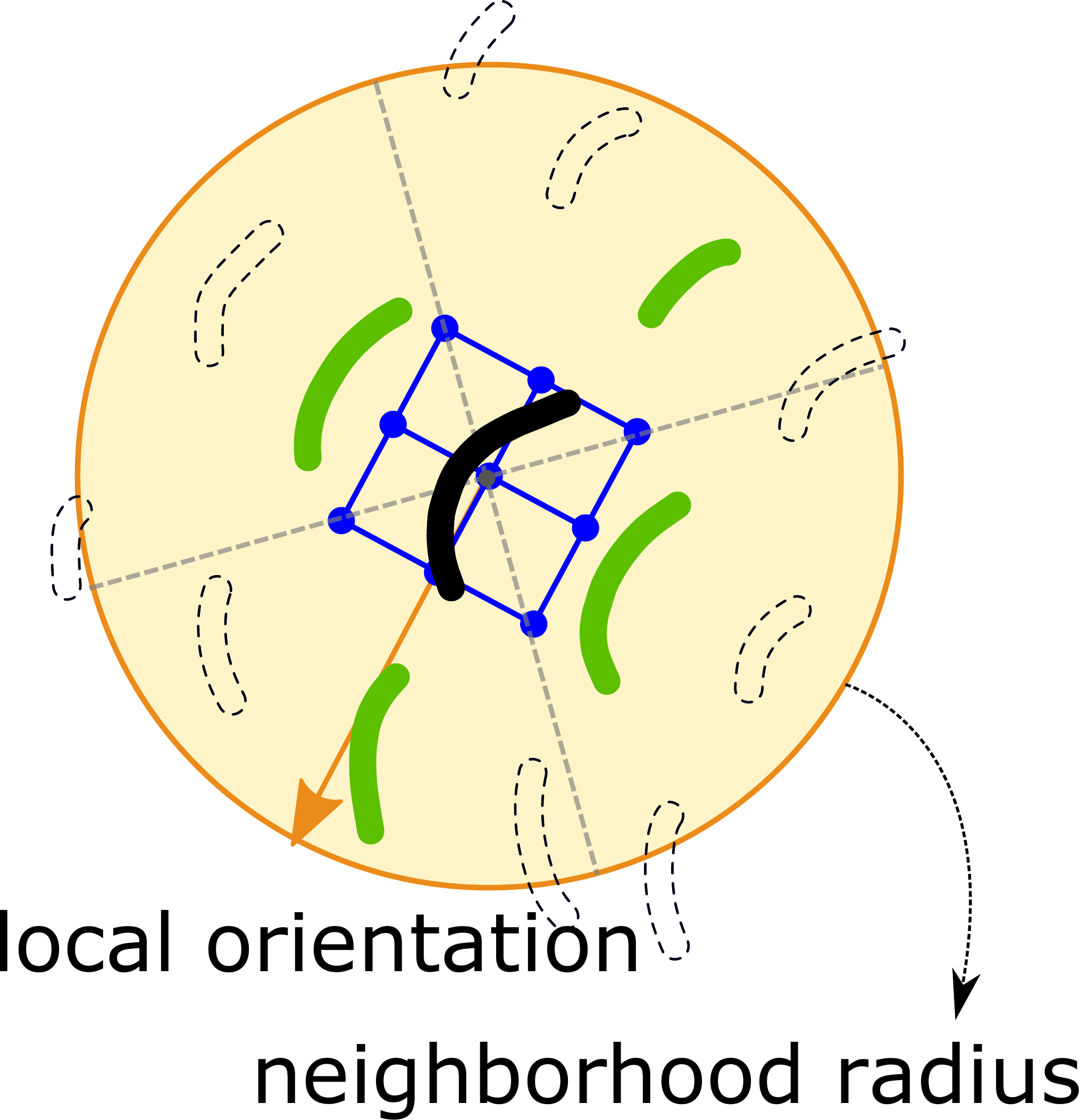}
			\label{fig:synthesis:neighborhood}
		}%
		\Caption{}
		{%
			\subref{fig:synthesis:element} A stroke, with centroid $\elemCentroid$ and dominant direction $\elemDir$.
			\subref{fig:synthesis:neighborhood} The neighborhood of the black stroke includes
			the $n$ ($n=1$ in this example) closest strokes (in green) from each quadrant
			and the middle image patch (blue pixel grid).
		}
		\label{fig:representation}
	\end{minipage}
\ifdefined\twocol
\end{figure}
\begin{figure}[t]
\else
\hfill
\fi
	\begin{minipage}{0.48\textwidth}
	  \centering
	  \includegraphics[width=0.88\linewidth]{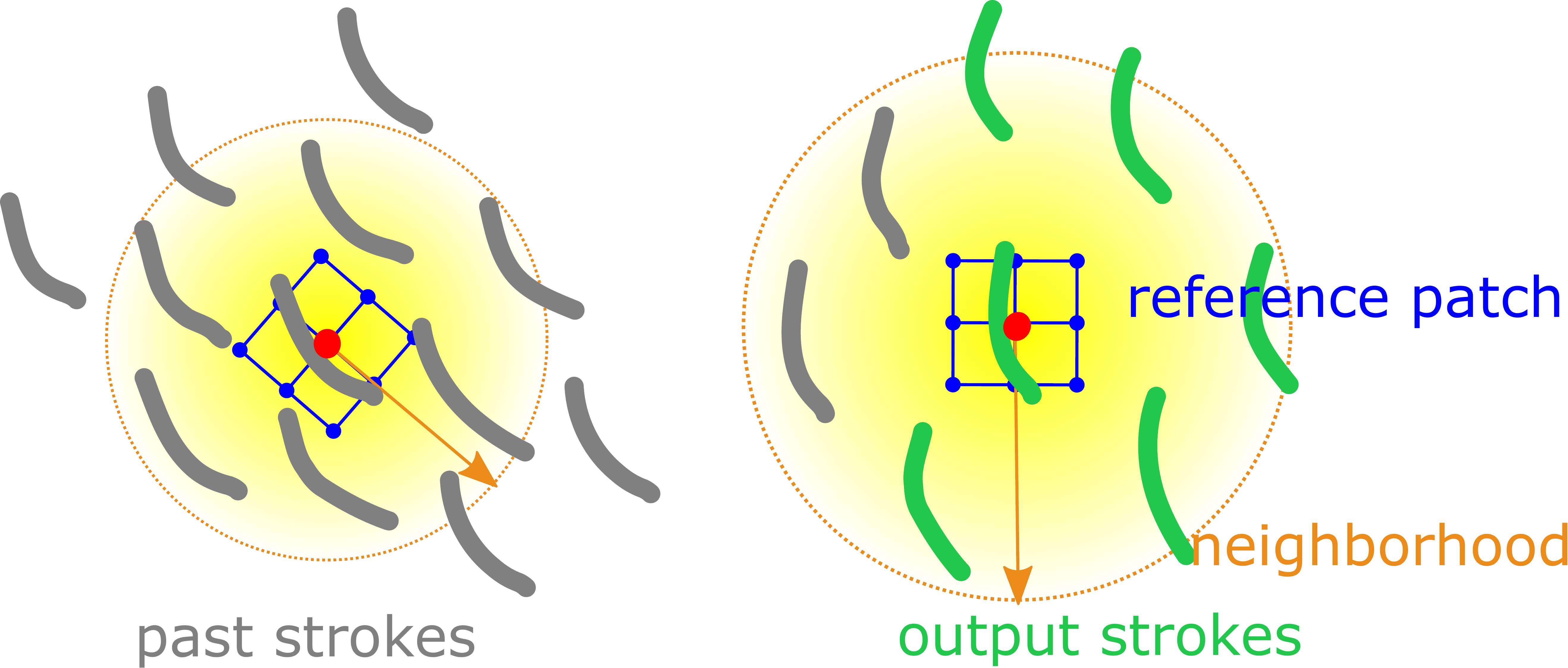}
	  \Caption{Illustration of our synthesis algorithm. %
	  }
	  {%
	We synthesize the predicted strokes (in green) from previously drawn strokes (in gray) by matching their neighborhoods.
	  }
	  \label{fig:method_illustration}
	\end{minipage}
\end{figure}

%% file: iteration_fig.tex
\begin{figure*}[t]
	\centering
	\subfloat[input]{%
		\label{fig:iteration:input}
		\includegraphics[width=0.15\textwidth]{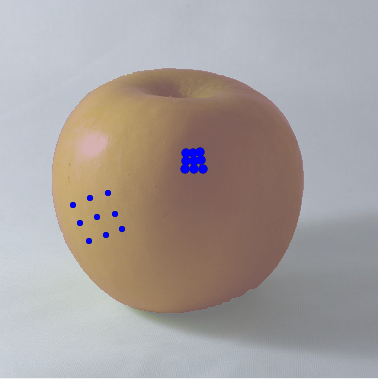}
	}%
	\subfloat[initialization]{%
		\label{fig:iteration:init}
		\includegraphics[width=0.15\textwidth]{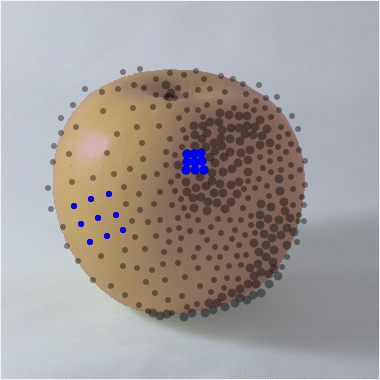}
	}%
	\subfloat[iteration 5]{%
		\label{fig:iteration:intermediate}
		\includegraphics[width=0.15\textwidth]{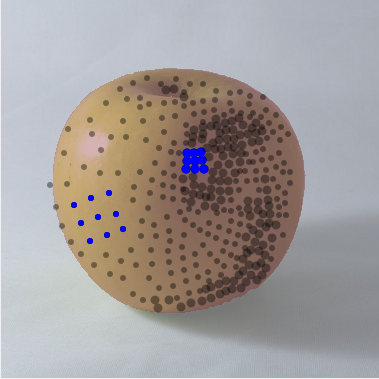}
	}%
	\subfloat[iteration 15]{%
		\label{fig:iteration:final}
		\includegraphics[width=0.15\textwidth]{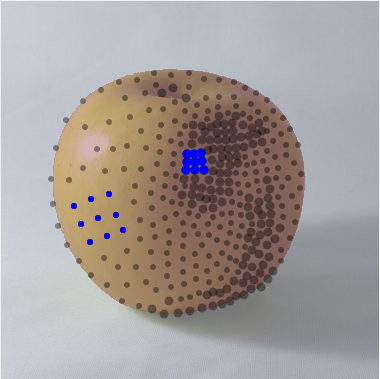}
	}%
	\subfloat[w/o $\energy_{corr}$]{%
		\label{fig:iteration:nocomp}
		\includegraphics[width=0.15\textwidth]{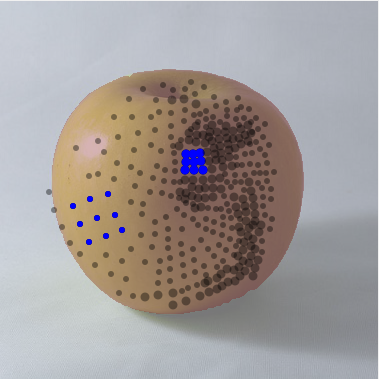}
	}%
	\subfloat[w/o $\distRefNeigh$]{%
		\label{fig:iteration:noimg}
		\includegraphics[width=0.15\textwidth]{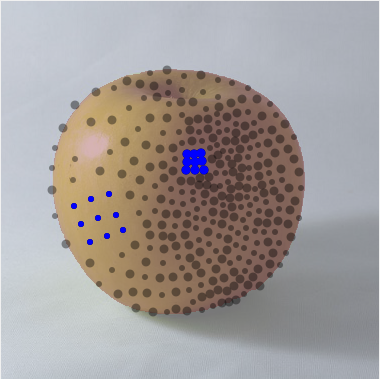}%
	}
	\Caption{Iteration process in \protect\subref{fig:iteration:init} to \protect\subref{fig:iteration:final} and ablation studies in \protect\subref{fig:iteration:nocomp} and \protect\subref{fig:iteration:noimg}.}
        {%
Without the correction term $\energy_{corr}$ the predicted strokes tend to clutter together as in \subref{fig:iteration:nocomp}.
Without the image term $\distRefNeigh$ the predicted strokes might not follow the reference sufficiently as in \subref{fig:iteration:noimg}.
}
\label{fig:iteration}
\end{figure*}

%% file: method_analysis.tex
\subsection{Inference}
\label{sec:method:analysis}

In this section, we describe how to infer $\inputElemSet$, $\dstRegion$, $\orientMap$, and $\radiusMap$ used for our synthesis method in \forwardref{\Cref{sec:method:synthesis}}{the previous section} from user interactions with $\refImage$.

\input{method_grouping}

\input{method_prediction}

\input{method_relation}

%% file: method_grouping.tex
\subsubsection{Input exemplar $\inputElemSet$}
\label{sec:method:grouping}

\input{grouping_fig}

This step aims to detect whether stroke repetitions exist and obtain the repetitive group as an exemplar for the synthesis process.
Since people usually draw strokes in a coherent manner \cite{Xing:2014:APR} and they usually have specific intentions when drawing repetitive strokes,
we assume the example strokes to be temporally consecutive and have certain similar properties.

We start from the last stroke input by the user 
and search backward in the stroke sequence to incrementally find strokes that have similar shape and image features to the last stroke.
Specifically, the stroke shape similarity is measured with the Fr\'echet distance, and the image features include $Lab*$ color (weighted by 0.12, 0.44, and 0.44 to suppress the impact of lightness) and precomputed semantic segmentation \cite{Zhao:2017:PSP} at a stroke's center.
We compare the standard deviation of a feature in the traversed $\numElemsSelected$ strokes against a threshold (15/255 for the color feature, 1 for the segmentation feature) for similarity measurement.
The back-traversal stops when the next stroke does not contain any similar feature or $\numElemsSelected > 50$.
These $\numElemsSelected$ strokes serve as the input exemplar for the synthesis process.
See \Cref{fig:grouping} for an example of the incremental searching process.

%% file: grouping_fig.tex
\begin{figure}[tb]
	\centering
	\includegraphics[width=\linewidth]{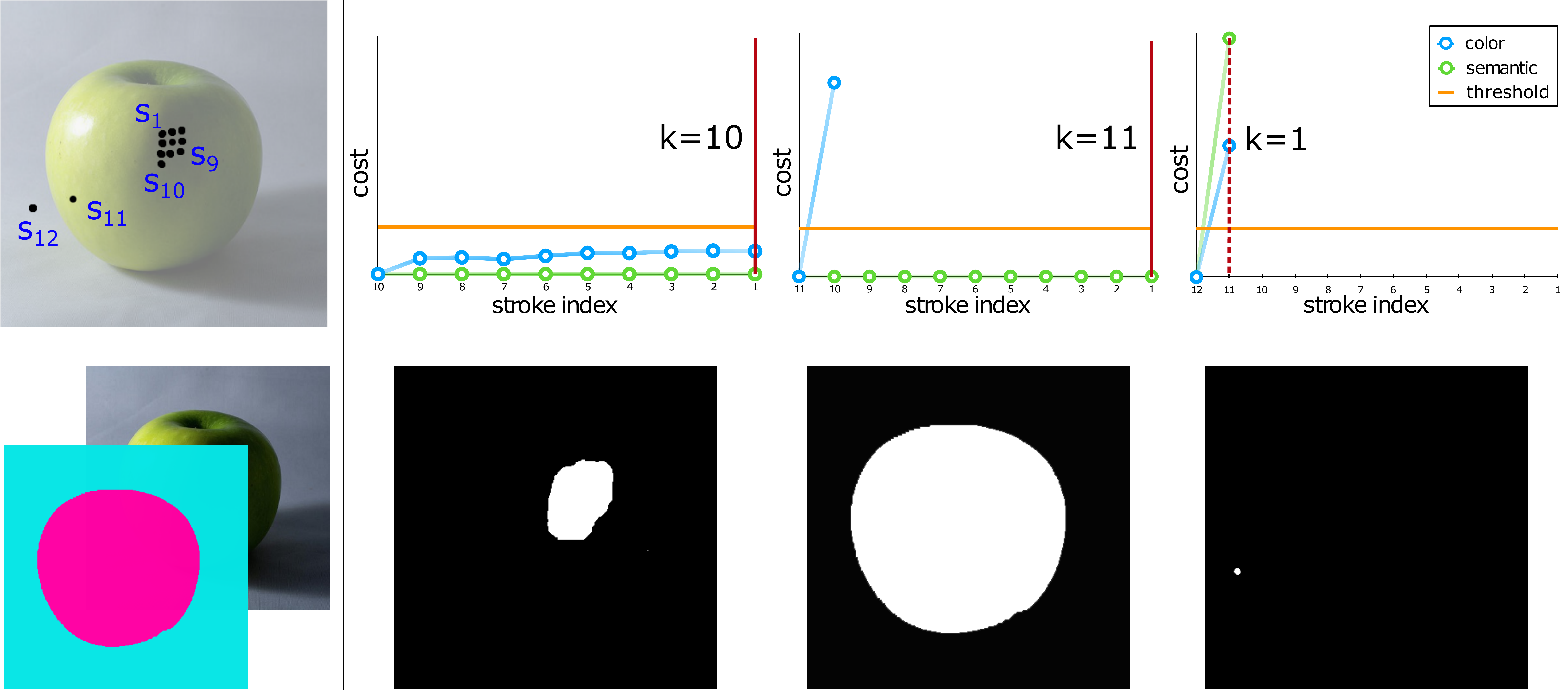}
	\Caption{An example of predicting the input exemplar and output region.}
	{%
The left column shows the input stroke sequence visualized in black dots (only a few indices are shown for clarity) on the reference image (top) and the image features (bottom).
The right columns show the threshold lines and the image feature cost curves for $\elem_{10}$, $\elem_{11}$, $\elem_{12}$ respectively (top), and the corresponding predicted output regions (bottom).
The cumulative number $\numElemsSelected$ is determined when both cost curves exceed the threshold.
Note that the third region prediction result is only for demonstration: since the exemplar only contains one stroke (i.e., $\numElemsSelected = 1$), it is not considered a valid exemplar and will not be used for synthesis.
}
\label{fig:grouping}
\end{figure}

%% file: method_prediction.tex
\subsubsection{Output region $\dstRegion$}
\label{sec:method:Prediction}

The shared features of the obtained stroke exemplar also indicate the intended region.
For instance, if all of the exemplar strokes are inside the same object segmentation region, it is very likely that the user intends to fill that region.
Therefore, we use the shared features obtained in the exemplar grouping process to find a similar region for output.

Since there are only two features in our implementation, we simply obtain the region by GrabCut \cite{Rother:2004:GIF} if the $Lab*$ color feature is shared among the exemplar strokes, directly take the corresponding segmentation if the semantic feature is shared, and take the intersection if both features are shared.
See \Cref{fig:grouping} for an example.
When there are multiple disconnected regions, we retain the nearest region to the user's last stroke and discard the rest, because it is less natural to propagate to distant regions.

%% file: method_relation.tex
\subsubsection{Contextual constraints}
\label{sec:method:property}

Since the drawing usually relates to the underlying reference image, we analyze the properties of both the drawn strokes and the reference image to infer possible relationships that control the global distribution of strokes.

\paragraph{Orientation $\orientMap$}
Artists usually adjust the stroke directions to convey curvatures, but they may sometimes randomize or fix the stroke orientation regardless of the depicted objects to create different visual effects.
Therefore, the problem is to decide which case the input exemplar implies.
We first compute the edge tangent field (ETF) \cite{Kyprianidis:2011:CEF} for the reference image and then calculate the angles between the exemplar strokes and the ETF directions at their centroids.
If the standard deviation of the angles is small (less than 15 degrees), we consider the stroke orientations to be related to the ETF and take the ETF as the orientation field; 
otherwise, we set a default global coordinate frame to each point of the orientation field.

\paragraph{Radius $\radiusMap$}
Since density is inversely proportional to the spacing between strokes, we reframe the problem as predicting a radius map that controls the extent of stroke neighborhoods.
First, we compute the distance from each exemplar stroke to its nearest neighbor.
We assume a linear relationship between these minimum distances $\radius$ and the image features, including image lightness $\lightnessVal$ and gradient strength $\gradientMag$ at a stroke's centroid, represented as:
\begin{equation}
	\radius = \begin{pmatrix}
	\lightnessVal & \gradientMag & 1
	\end{pmatrix}
	\cdot \modelCoeff,
\end{equation}
where $\modelCoeff$ denotes the coefficients to solve.
With the fitted linear model, if the squared correlation value is lower than 0.5 (the closer to 1, the better explanation), we use the model to compute a radius map.
Otherwise, we consider the density as uniform and create a constant radius map with the average spatial distance of the exemplar.
We then update the UI with the computed coefficients.

%% file: evaluation.tex
\section{Evaluation}
\label{sec:eval}

We conducted a pilot study to evaluate the utility and usability of our approach.
We compared three modes through quantitative analysis and qualitative feedback.
\begin{description}%
	\item[Autocomplete]
	Users have full access to our prototype, including autocomplete and interactive editing.
	\item[Interactive batch filling]
	(aka {\em batch mode})
	Users are required to create a texture example first and then manually specify the properties for batch filling.
	It simulates the sequential procedure in many IB-AR methods (e.g., \cite{Salisbury:1997:OTI}), although they rarely allow users to directly define examples on target images.
	This mode is performed on our system with the autocomplete function off.
	\item[Fully manual drawing]
	(aka {\em manual mode})
	Users have to manually draw each stroke without any automatic synthesis.
\end{description}
We also tested the expressiveness of our system through an open creation session and obtained comments for future improvements.

\input{new_study}

%% file: new_study.tex
\subsection{Target Session}
\input{target_session_fig}

The goal of this session is to compare the three interaction modes in utility and usability.
Since we aim to facilitate image-scaffolded drawing, we hope to include general users from different background while focusing more on less skillful users, who are more likely to use reference images.
We thus recruited 12 participants, including nine novices with little drawing experiences, two amateurs with some experiences (P3, P4), and a student majored in illustration (P5).
Most of the studies were conducted on a Lenovo Miix 520 tablet with stylus in a lab environment, except two studies conducted remotely with mouse due to the pandemic.

The study procedure consisted of the following parts and took each participant about two hours in total.

\textbf{Tutorial.}
Each participant was first given a brief introduction to our system and then asked to fill the apple in \Cref{fig:ui} with short hatches as a warm-up task.
They were encouraged to vary the density and orientation of input strokes and get familiar with the features of our system.

\textbf{Target tasks.}
We used a within-subjects design, where each participant was asked to reproduce two target drawings (\Cref{fig:target_tasks}) in all the three modes: autocomplete, interactive batch filling, and fully manual drawing.
The target drawings include an object and a landscape, which are common illustration topics (e.g., \Cref{fig:artwork_eg}).
The assigned order of modes was counter-balanced among all the participants.
Since we focus on region filling, we asked the participants to draw the outlines of both images in advance, so that they could focus on drawing the textures during the study.
We encouraged the participants to finish each drawing as soon as possible, preferably in a dozen of minutes, but without any hard time limit.
After completing the two drawings in each mode, each participant filled in a NASA-TLX questionnaire \cite{Hart:1988:DNT}.
At the end, we asked the participants about their preferred mode, usage experience and other comments.

\subsection{Open session}
The goal of this session is to observe users' interaction with our system and learn about users' subjective experience.
We invited seven participants (one professional artist, two amateurs and four novices) for this session.
They were asked to create a drawing freely from the same reference image (\Cref{fig:open_result:reference}) with our system. 
The reference image was a portrait photo, which is also common in illustrations.
The only requirement was that the drawings should contain some repetitive content. 
We again gave a tutorial in the beginning and conducted the task on a Lenovo Miix 520 tablet with stylus.
The participants were encouraged to think aloud and describe their thought process and interactions during this session.
After this task, participants could optionally create more drawings with any images they want.
Since our prototype does not contain all common functions in commercial drawing tools, we allow the participants to retouch the result drawings without adding more strokes in Photoshop.

\subsection{Results and Observations}%
\input{target_session_data}

\paragraph{Workload}
\Cref{fig:compare_study:nasa} shows the perceived workload scores from the target session.
Generally, the autocomplete mode received the lowest (i.e., best) scores for almost all the factors.
One-way ANOVA showed the three modes have significant difference in physical demand (F=10.69, p < 0.001) while no significant difference in other factors.
Regarding the physical demand, post-hoc pairwise tests showed that the autocomplete mode and batch mode were both rated significantly lower than manual mode, while had no significant difference from each other.
This matches our expectation, since automatic synthesis should only reduce physical load and not cause extra pressure than manual work. %

\paragraph{Efficiency}
We calculate the average completion time (\Cref{fig:target_data:time}) and stroke count (\Cref{fig:target_data:stroke}) in each mode and each task.
Generally, the system synthesized 
about $82\%$ strokes in the autocomplete mode and about $92\%$ strokes in the batch mode.
Although the manual mode took the shortest time for the participants to complete, it also resulted in the fewest total number of strokes.
We thus calculated the strokes per minute for each mode: autocomplete (111.03, SD=38.76), batch (101.98, SD=45.13), manual (115.95, SD=46.73).
It turns out automatic generation did not improve the efficiency, probably because the users spent extra time adjusting and experimenting with the generated effects instead of just drawing strokes.
It should be noted that such directed tasks omit the time for exploring alternative patterns, which, however, might be high in a fully manual case.

\paragraph{Quality}
We asked 30 external volunteers to evaluate the quality of participants' drawings, as shown in
\ifdefined\supp
\Cref{fig:user_drawings}.
\else
the supplementary material.
\fi
We randomized all the drawings created by the participants, showed each output drawing alongside the target drawing, and asked volunteers to rate the resemblance of the output drawing to the target drawing, on a scale from 1 (very dissimilar) to 5 (very similar).
The volunteers were instructed to focus more on the overall stroke distributions and flows instead of individual stroke thickness and detailed shapes.
\yl{We calculated the average scores for each mode: autocomplete (3.10, SD=1.24), batch (3.09, SD=1.21), manual (2.98, SD=1.20).
The quality of the drawings created with automatic synthesis is slightly better than the fully manual drawings, but without significant difference.
From the participants' perspective, three novices commented the automated strokes were better than their manual strokes, because they tend to lose patience when manually drawing all strokes, which results in worse quality.}

\paragraph{Preferred Mode}
Seven participants preferred the autocomplete mode while the rest five participants preferred the batch mode.
Generally, the autocomplete mode is considered more convenient, yet less precise; the batch mode is considered more precise, but requires too many interactions.
P12 commented, ``\textit{the autocomplete mode is more straightforward, because you can see the filled effects instantly without doing a lot of manipulation beforehand; while in the batch mode, you have to remember the meaning of parameters and tweak them in order to create strokes.}''
P10 also said, ``\textit{Compared with batch filling, the autocomplete mode provides a quick guess of filled regions and allows me to get the results more quickly with 
less work.}''
However, the autocomplete mode is ``\textit{less accurate at some vague and detailed regions, such as the shadows of the boat, where it tends to include some unwanted regions, so I have to manually subtract those regions, which is a bit tedious}'', according to P3.
The professional, P5, also preferred the batch mode for being able to precisely select the regions.
Therefore, we consider the autocomplete function and the interactive editing function are complementary in usability.

\input{open_session_fig}
\input{results_fig}
\input{results_supp_fig}

\paragraph{Creation Results and Experience}
\Cref{fig:open_result} shows the outcomes from the open session.
Although from the same reference image and widely using repetitive short strokes, the study participants were able to create different results by varying the stroke shapes and arrangement.
\Cref{fig:results,fig:results:supp} demonstrate some sample results.
Regarding the creation experience,
one user said ``\textit{it is playful, the final result is also good}'';
two users described it as ``\textit{encouraging}'', because the system allows beginners to quickly create stylistic drawings;
one user commented that she ``\textit{felt creative when drawing with this system}'', because she could test out patterns over image regions conveniently and she was more comfortable with drawing from a reference image than from scratch.
The professional suggested that the tool itself was somewhat limited to pointillism and hatching styles, but can be helpful in adding interesting textures into color paintings (e.g., \Cref{fig:results:ladyhat}).
Two users commented that the reduction of workload is useful, but they also complained about some inaccurate inference of autocompletion.
We will discuss about this problem in \Cref{sec:limit}.

%% file: target_session_fig.tex
\begin{figure}[t]
\centering
\subfloat[bear]{
	\label{fig:target_tasks:bear}
	\includegraphics[width=0.23\linewidth]{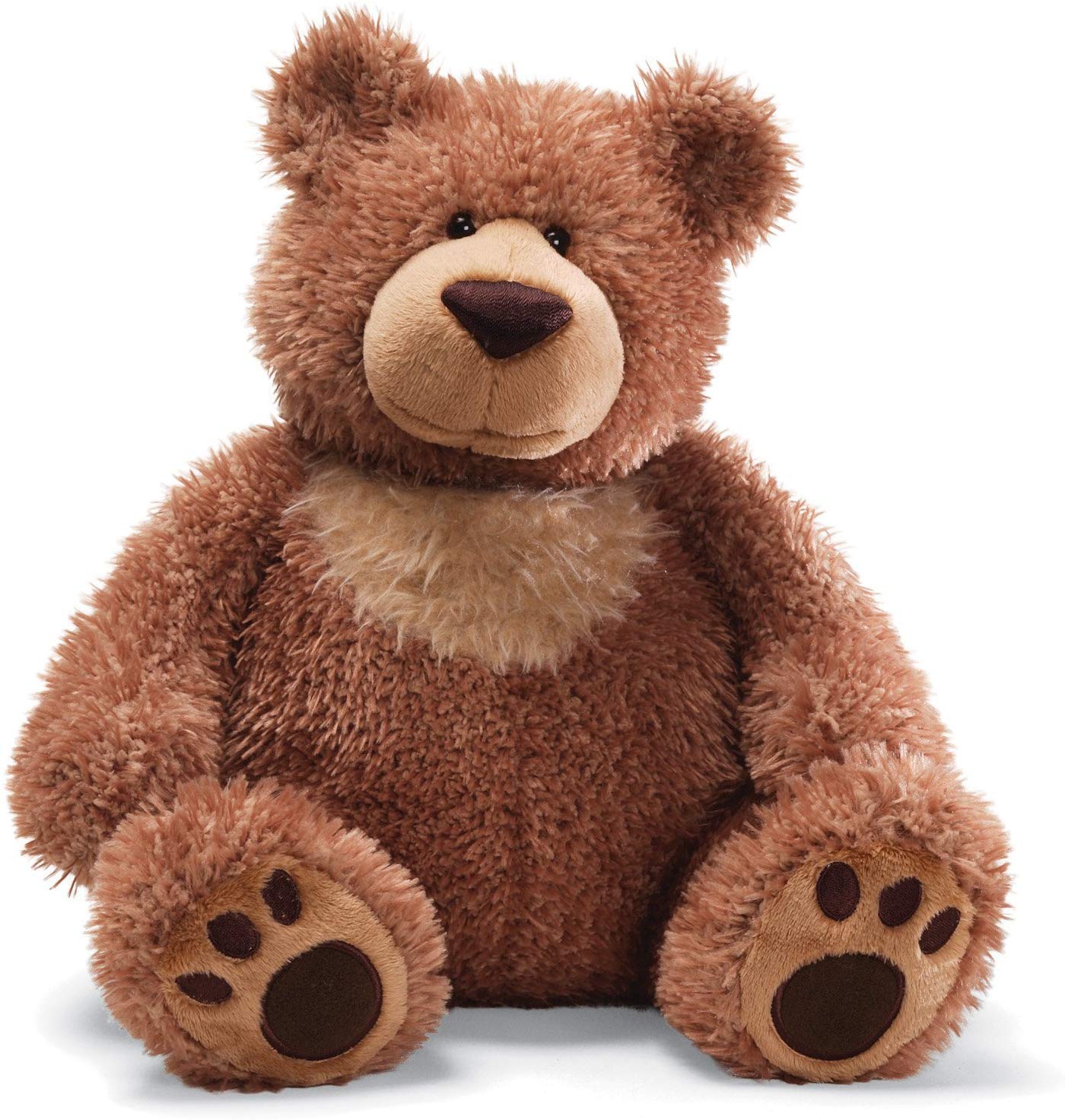}
}
\subfloat[drawing]{
	\label{fig:target_tasks:bear_drawing}
	\includegraphics[width=0.23\linewidth]{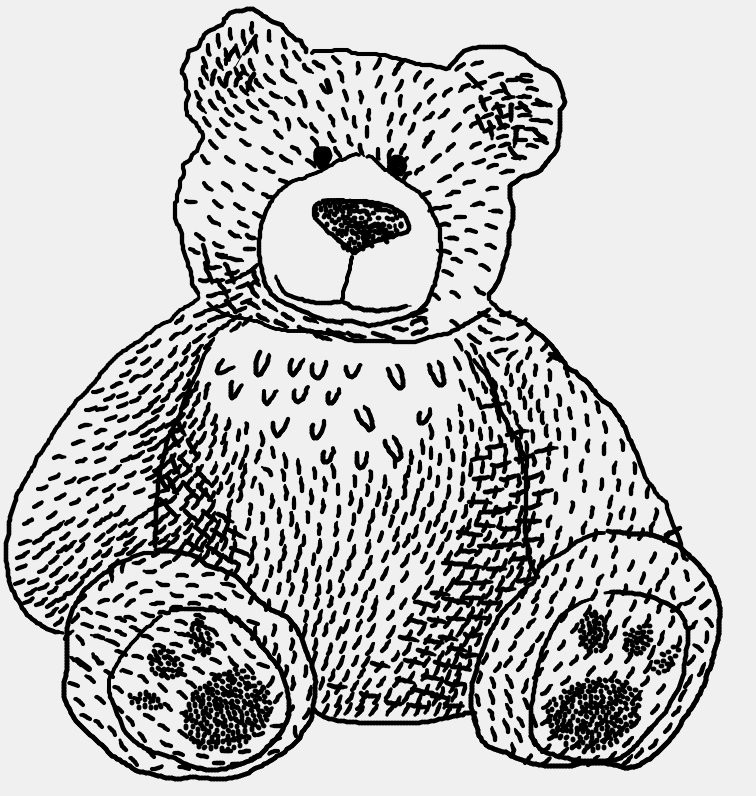}
}
\subfloat[segmentation]{
	\label{fig:target_tasks:bear_seg}
	\includegraphics[width=0.23\linewidth]{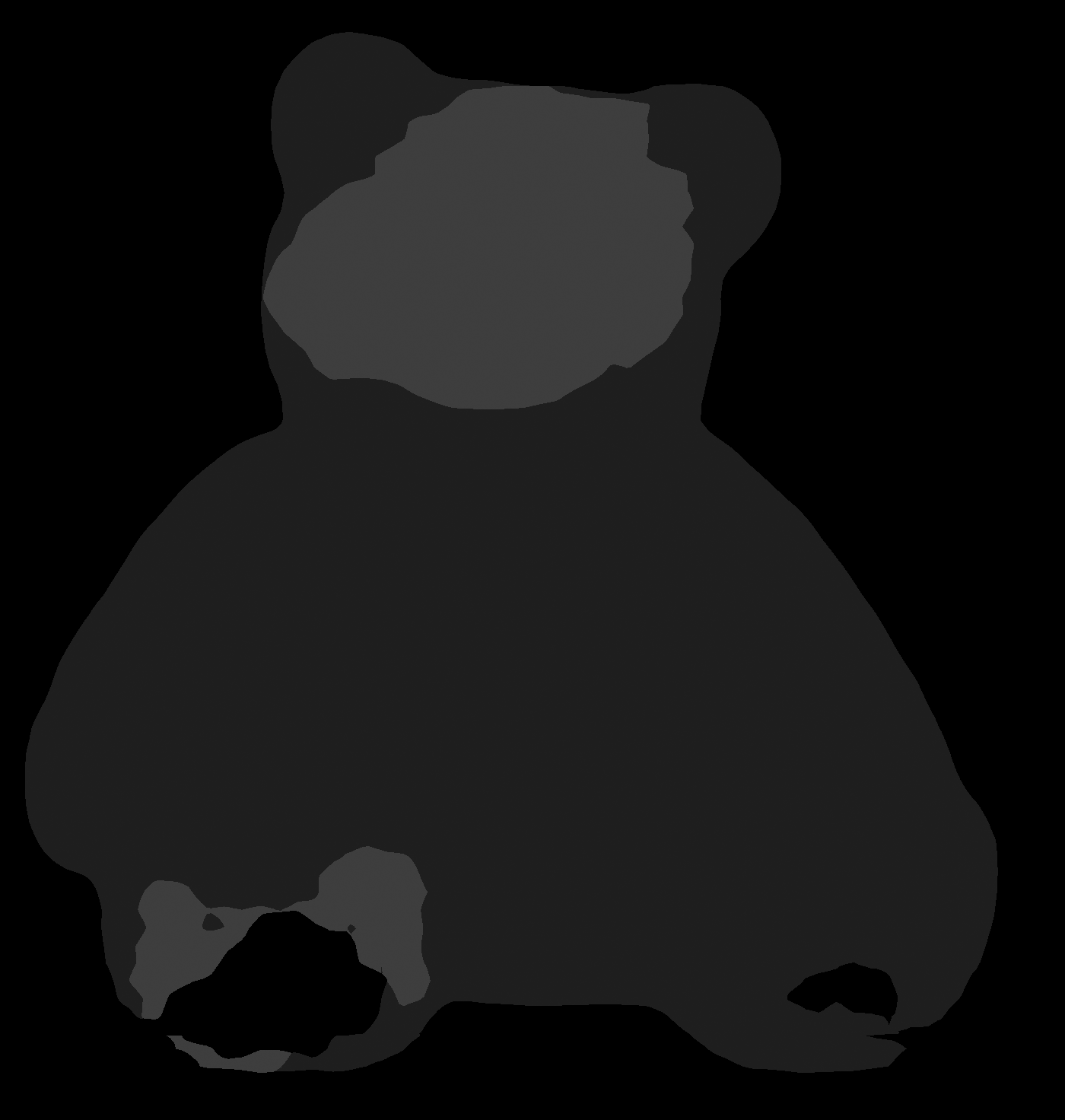}
}
\subfloat[orientation]{
	\label{fig:target_tasks:bear_orient}
	\includegraphics[width=0.23\linewidth]{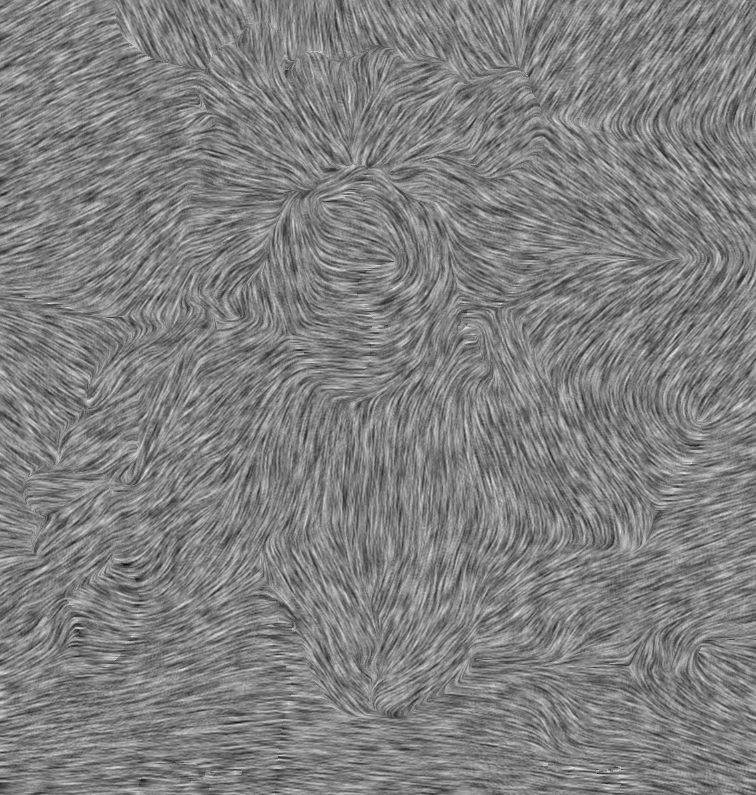}
}\\
\subfloat[beach]{
	\label{fig:target_tasks:beach}
	\includegraphics[width=0.23\linewidth]{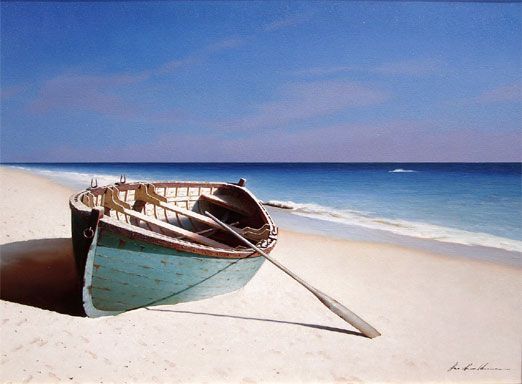}
}
\subfloat[drawing]{
	\label{fig:target_tasks:beach_drawing}
	\includegraphics[width=0.23\linewidth]{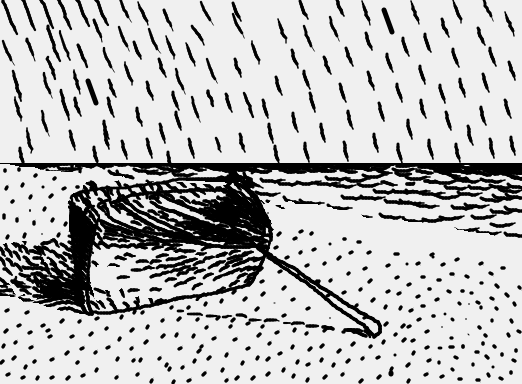}
}
\subfloat[segmentation]{
	\label{fig:target_tasks:beach_seg}
	\includegraphics[width=0.23\linewidth]{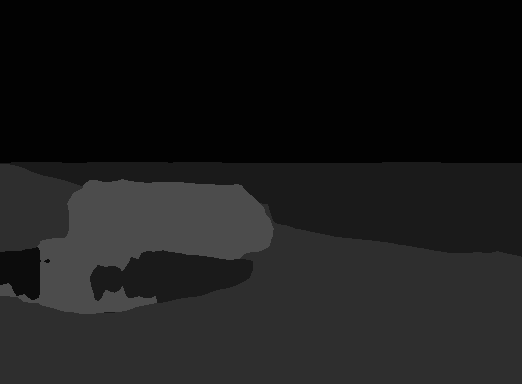}
}
\subfloat[orientation]{
	\label{fig:target_tasks:beach_orient}
	\includegraphics[width=0.23\linewidth]{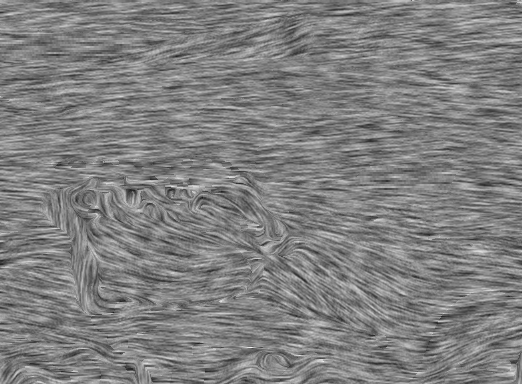}
}
\Caption{Target session tasks.}
{%
Reference photos in \subref{fig:target_tasks:bear} and \subref{fig:target_tasks:beach}, and the corresponding sample outputs in \subref{fig:target_tasks:bear_drawing} and \subref{fig:target_tasks:beach_drawing}.
}
\label{fig:target_tasks}
\end{figure}

%% file: target_session_data.tex
\begin{figure*}[t]
\centering
\ifdefined\noimage
\else
\resizebox{\linewidth}{!}{
\fi
\subfloat[NASA-TLX]{
	\label{fig:compare_study:nasa}
	\includegraphics[height=3cm]{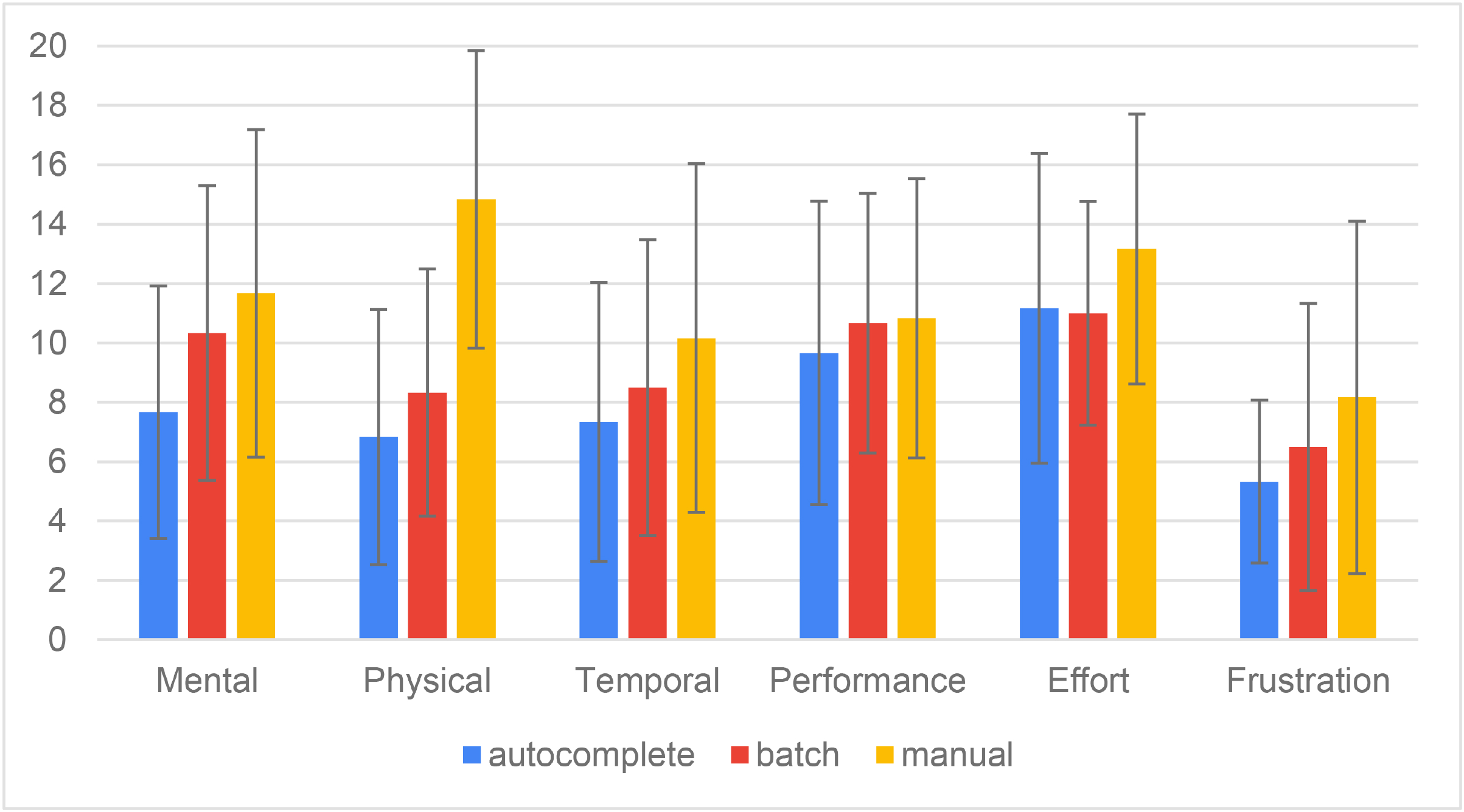}
}
\subfloat[time]{
	\label{fig:target_data:time}
	\includegraphics[height=3cm]{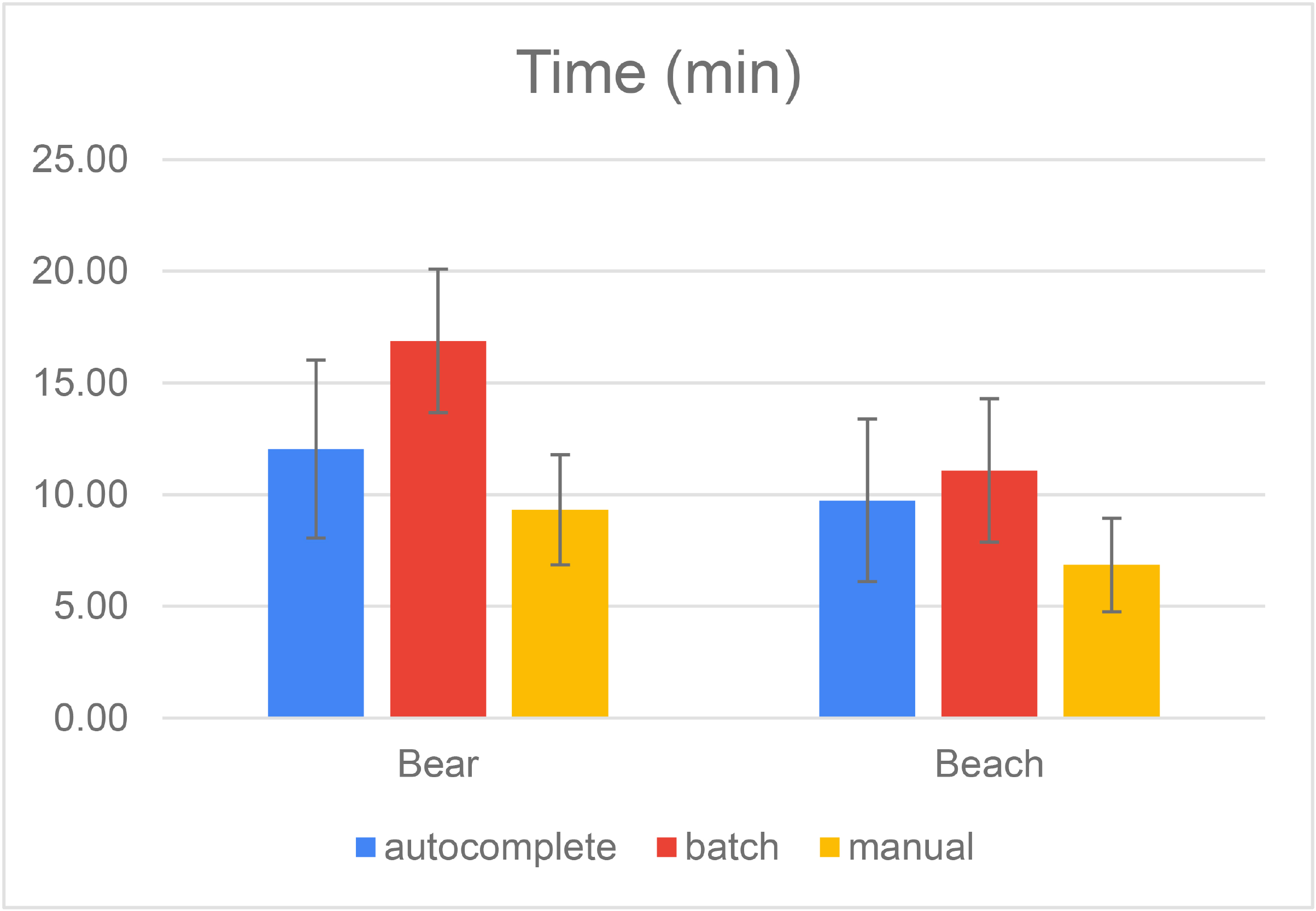}
}
\subfloat[\# strokes]{
	\label{fig:target_data:stroke}
	\includegraphics[height=3cm]{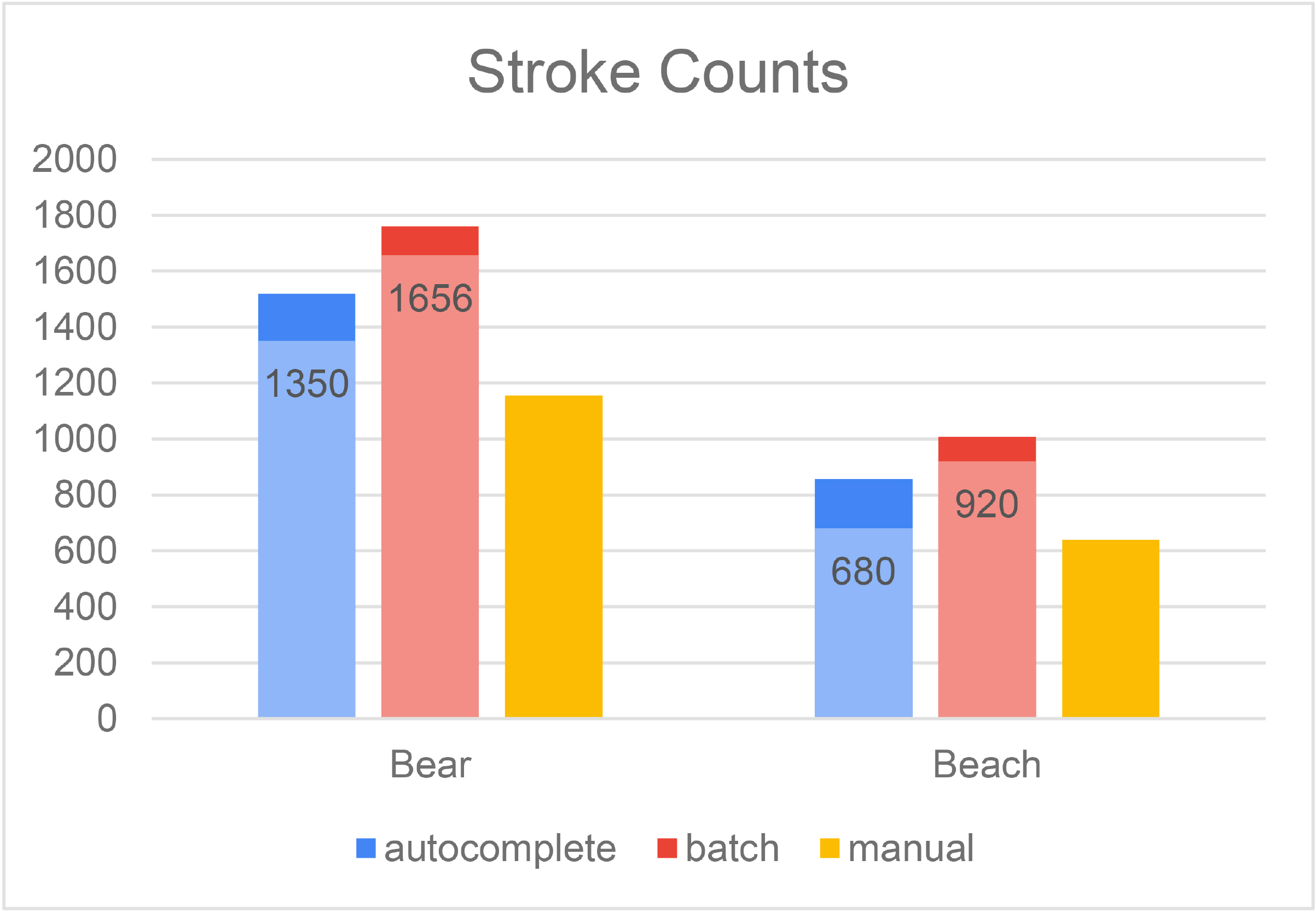}
}
\ifdefined\noimage
\else
}%
\fi
\Caption{Target sessions results.}
{%
	\subref{fig:compare_study:nasa} Average NASA-TLX scores from 12 participants. The lower the better. 
	\subref{fig:target_data:time} Average completion time.
	\subref{fig:target_data:stroke} Average stroke counts. The number of system-generated strokes is labeled in each column.
}
\end{figure*}

%% file: open_session_fig.tex
\begin{figure}[tp!]
\setlength{\belowcaptionskip}{-10pt} 
  \centering

  \subfloat[reference]{
    \label{fig:open_result:reference}
    \includegraphics[width=0.23\linewidth]{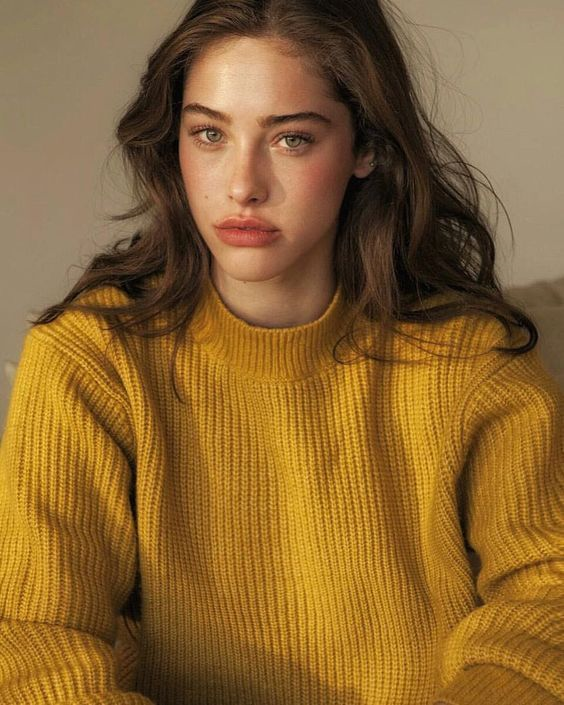}
  }%
  \subfloat[\manualstrokes{81}/\autostrokes{1563}]{
    \includegraphics[width=0.23\linewidth]{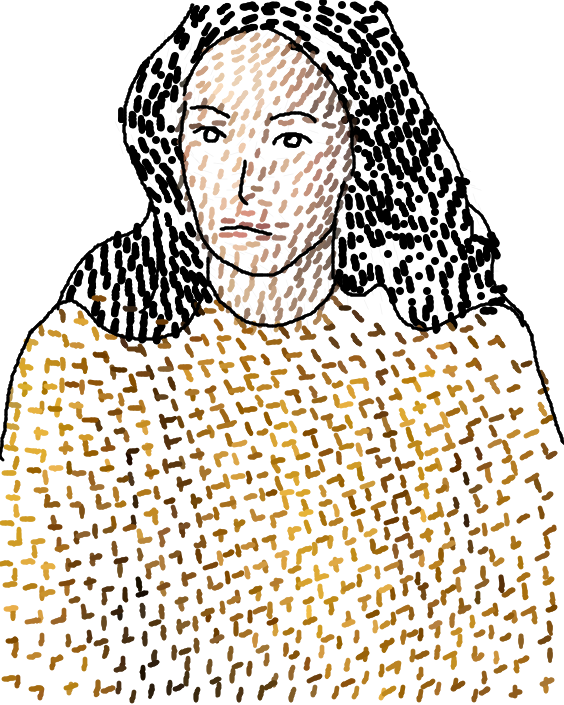}
  }%
  \subfloat[\manualstrokes{428}/\autostrokes{4593}]{
    \includegraphics[width=0.23\linewidth]{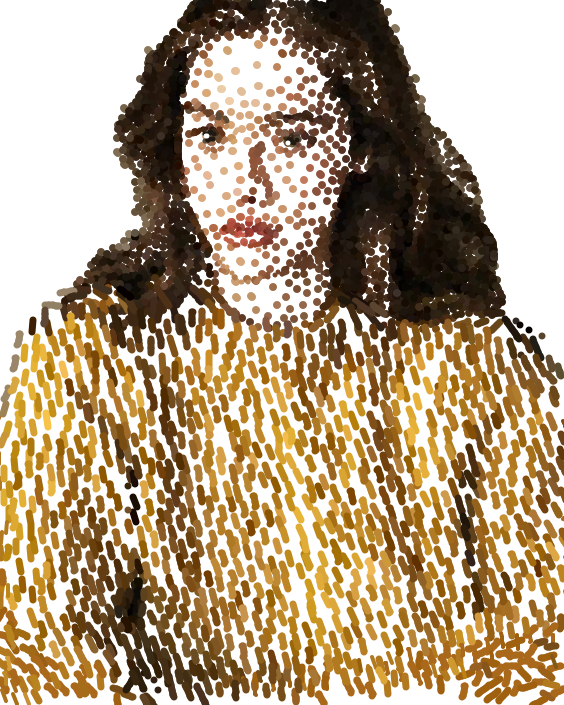}
  }%
  \subfloat[\manualstrokes{272}/\autostrokes{1266}]{
    \includegraphics[width=0.23\linewidth]{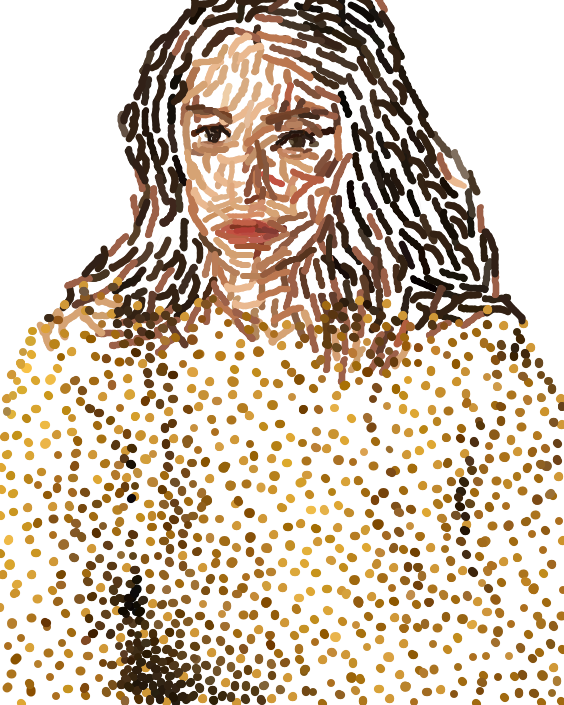}
  }%
\\
  \subfloat[\manualstrokes{68}/\autostrokes{8356}]{
    \includegraphics[width=0.23\linewidth]{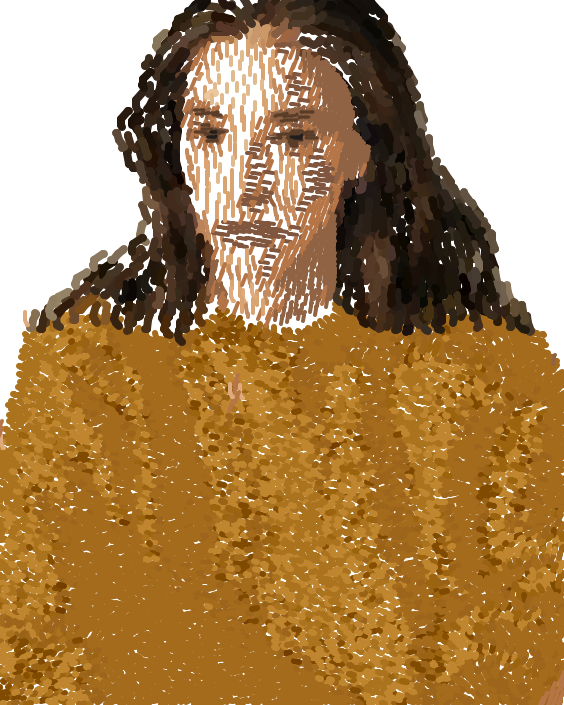}
  }%
  \subfloat[\manualstrokes{165}/\autostrokes{17111}]{
    \includegraphics[width=0.23\linewidth]{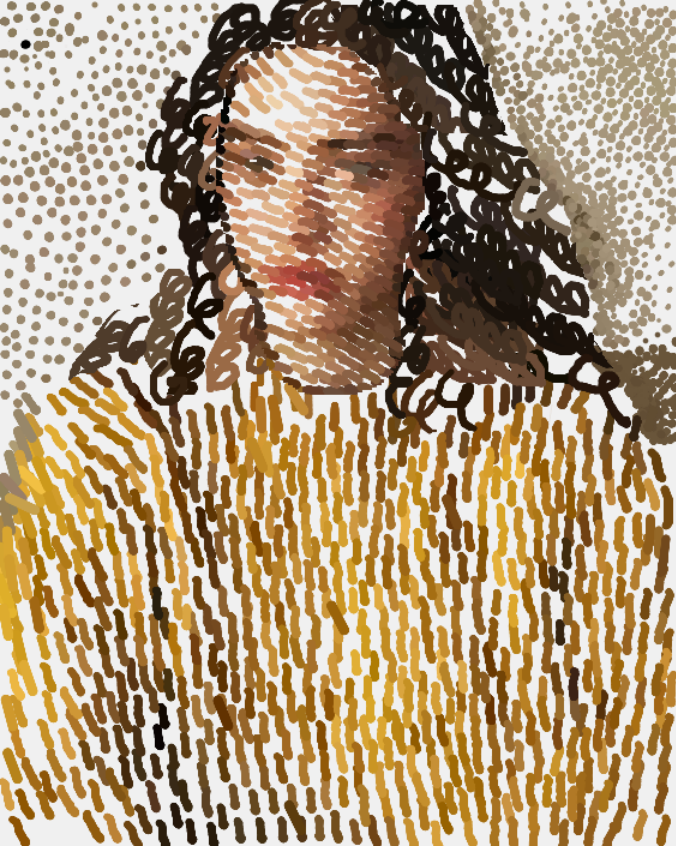}
  }%
\subfloat[\manualstrokes{443}/\autostrokes{2931}]{
	\includegraphics[width=0.23\linewidth]{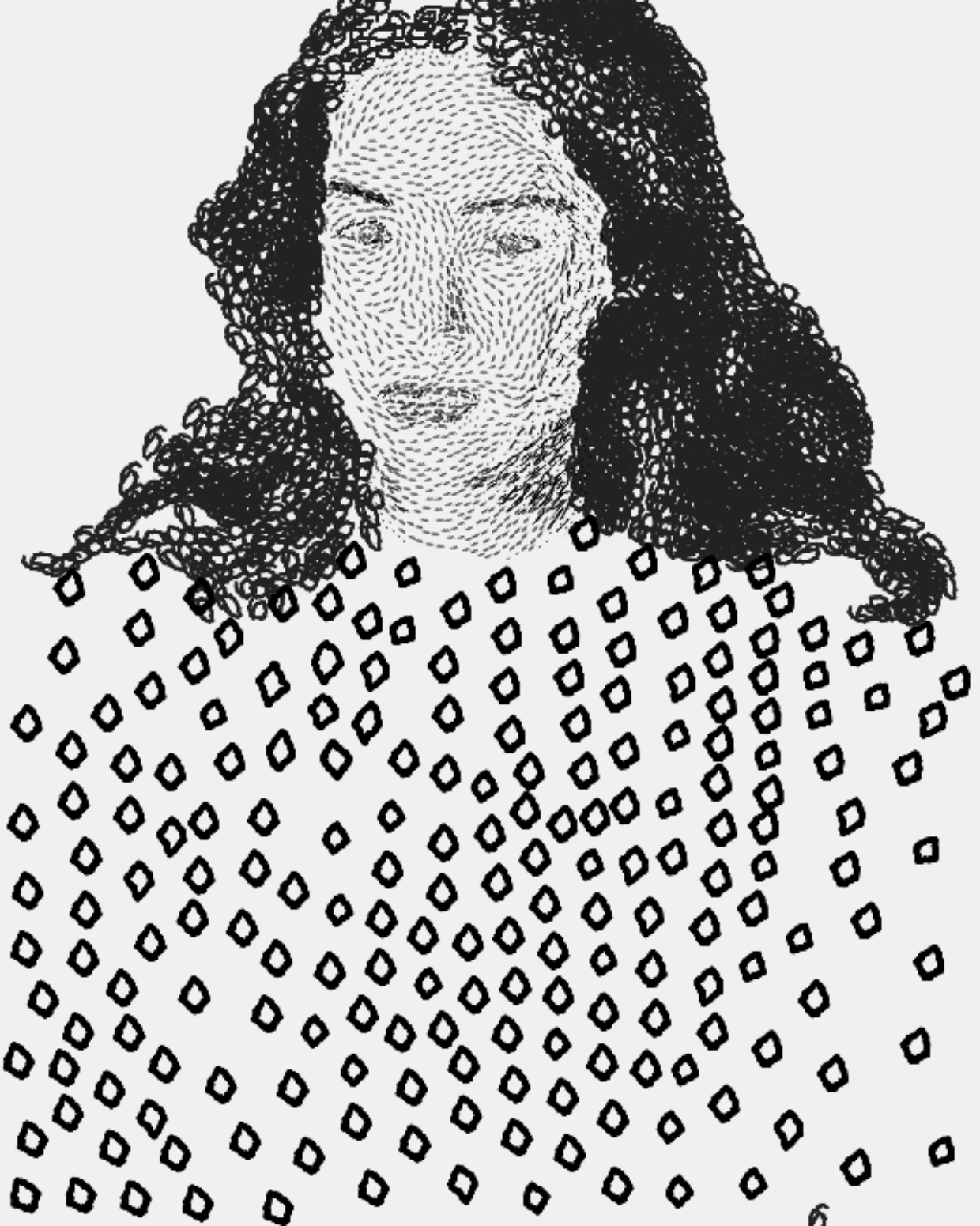}
}
\subfloat[\manualstrokes{261}/\autostrokes{6018}]{
	\includegraphics[width=0.23\linewidth]{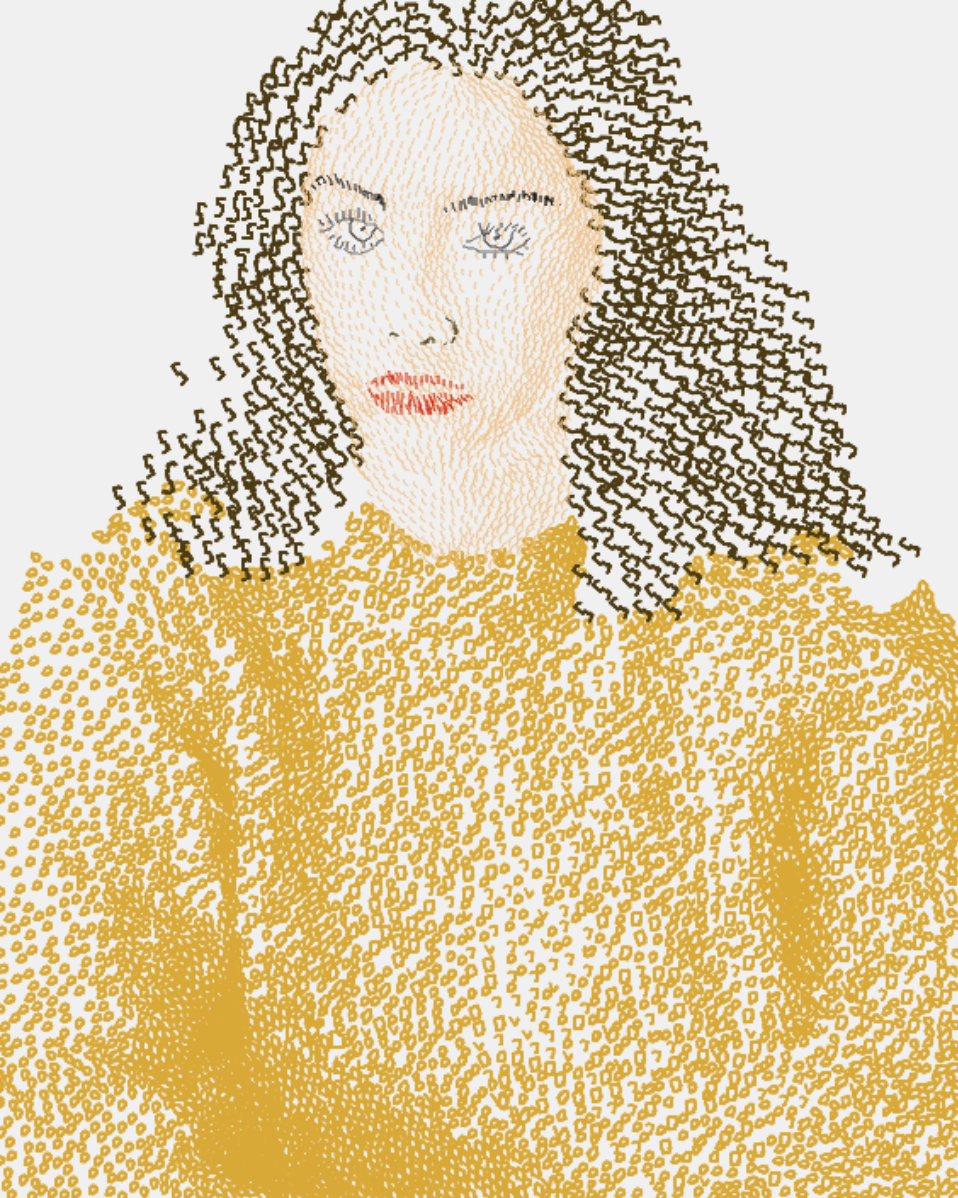}
}

\Caption{Example drawing results from the open session.}
{%
Each case is marked with the \# of \manualstrokes{manual} and \autostrokes{autocompleted} strokes.
}
\label{fig:open_result}
\end{figure}

%% file: results_fig.tex
\begin{figure*}[t]
	\centering
	
	\subfloat[]{\includegraphics[width=0.28\linewidth]{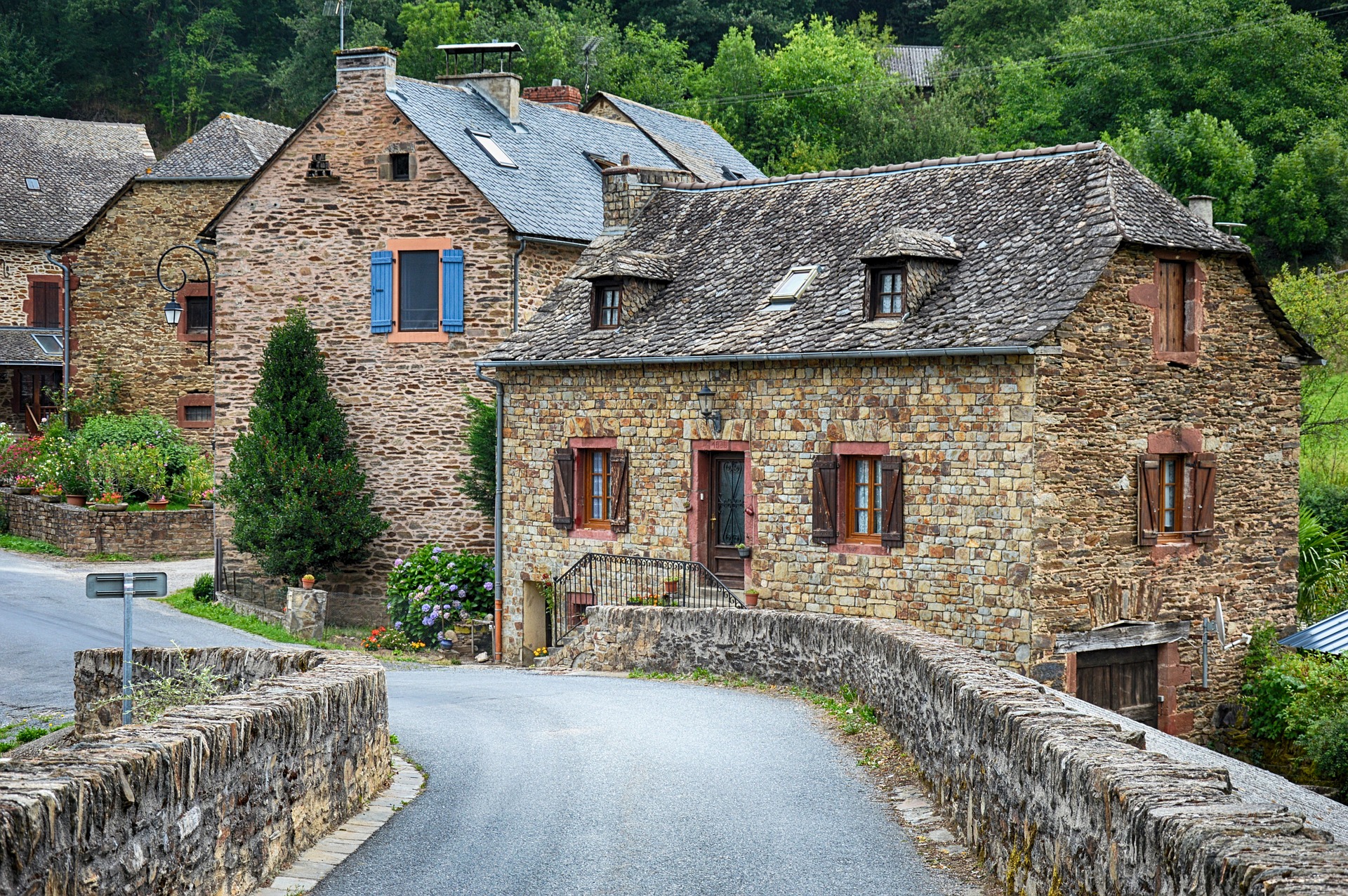}}
	\subfloat[\manualstrokes{446}/\autostrokes{9617} strokes]{\includegraphics[width=0.28\linewidth]{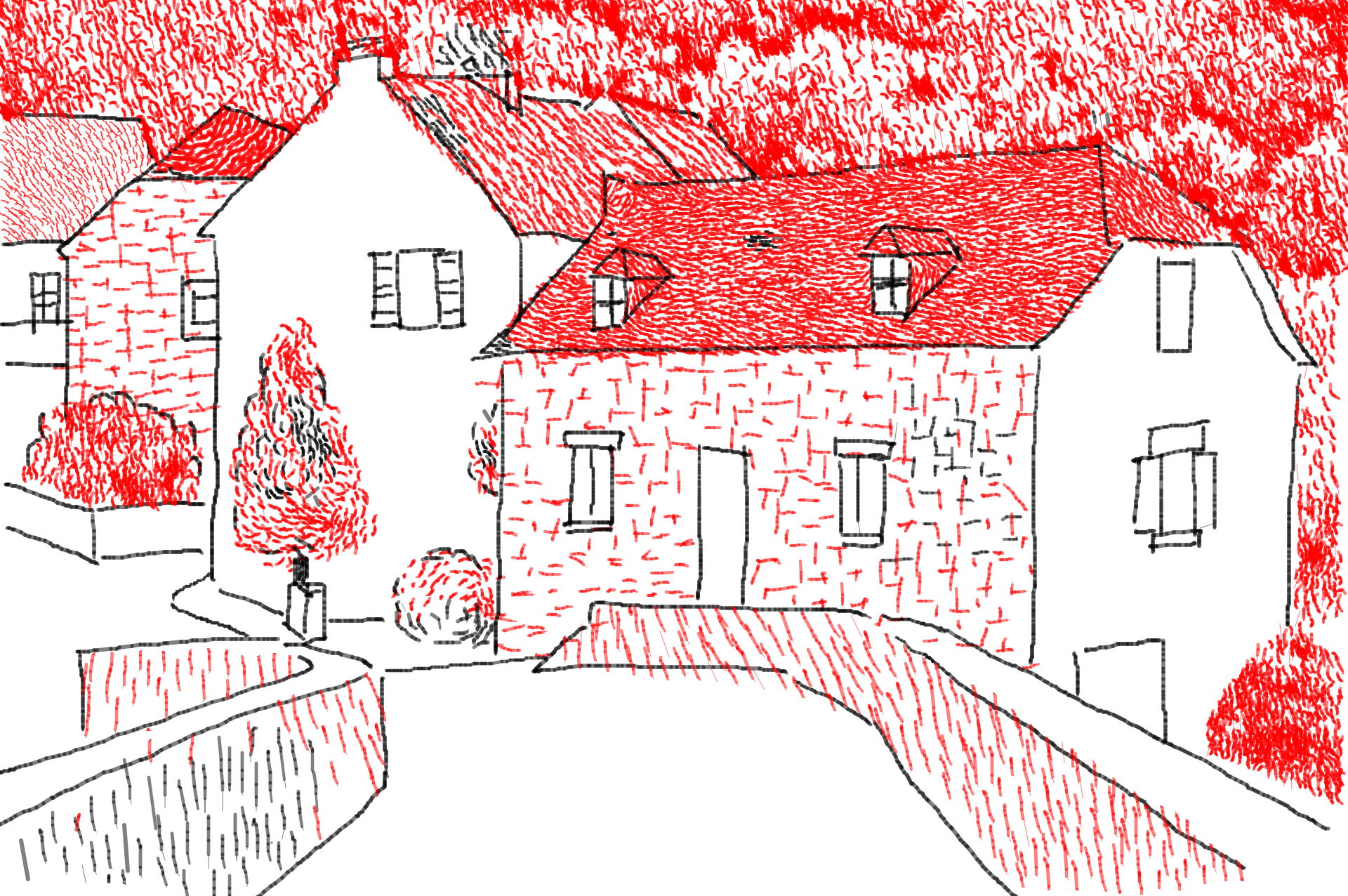}}
	\subfloat[]{\includegraphics[width=0.28\linewidth]{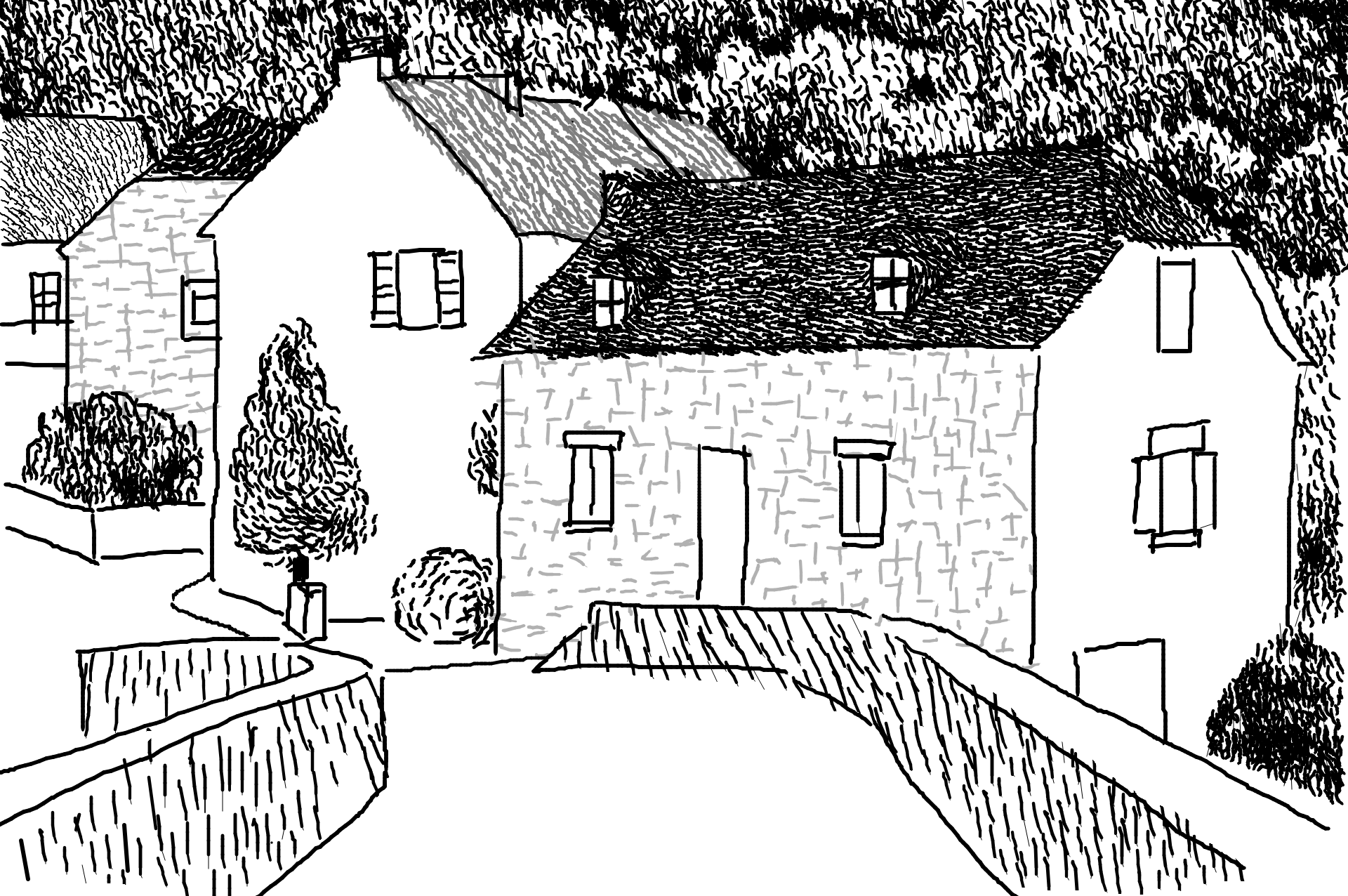}}
	
	\subfloat[]{\includegraphics[width=0.28\linewidth]{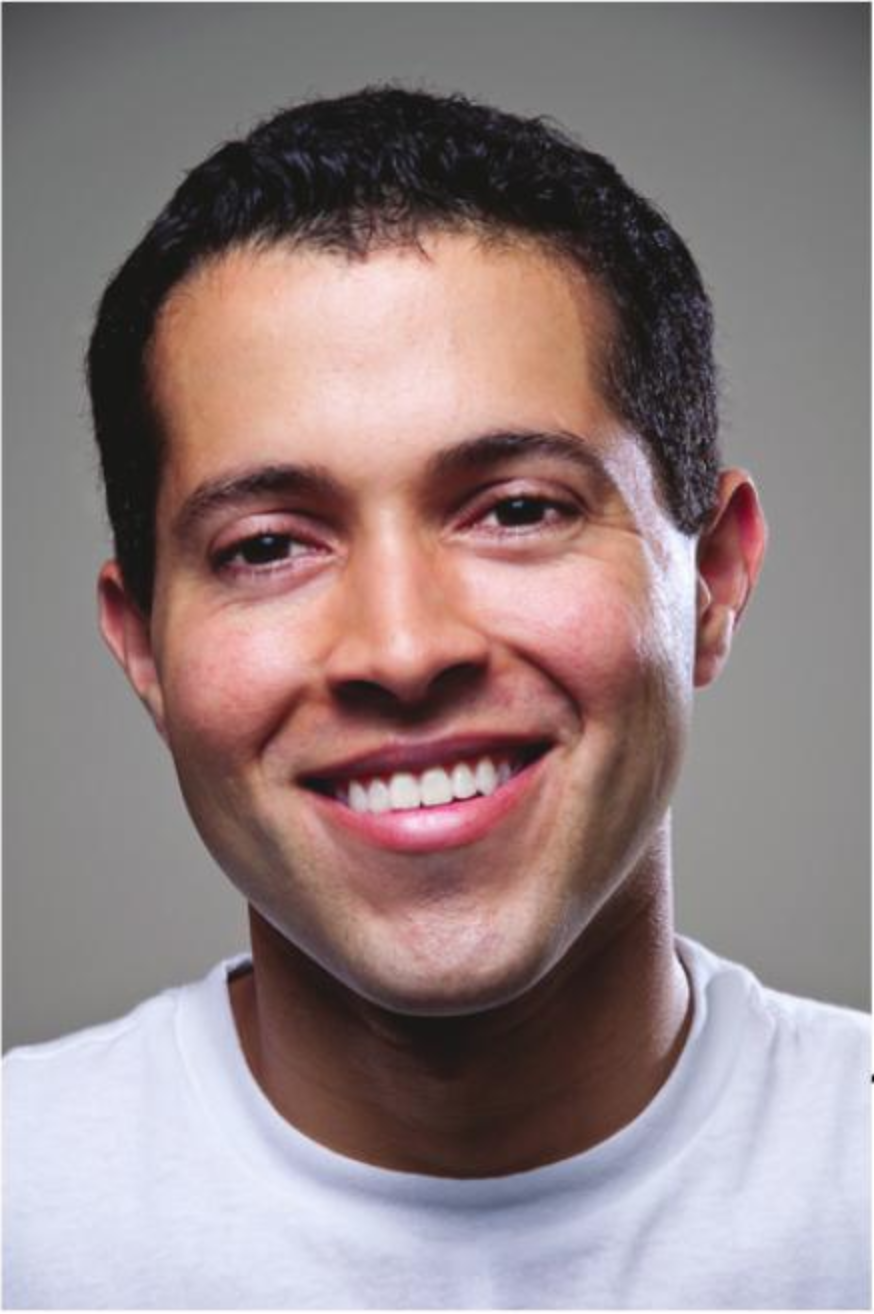}}%
	\subfloat[\manualstrokes{264}/\autostrokes{840} strokes]{\includegraphics[width=0.28\linewidth]{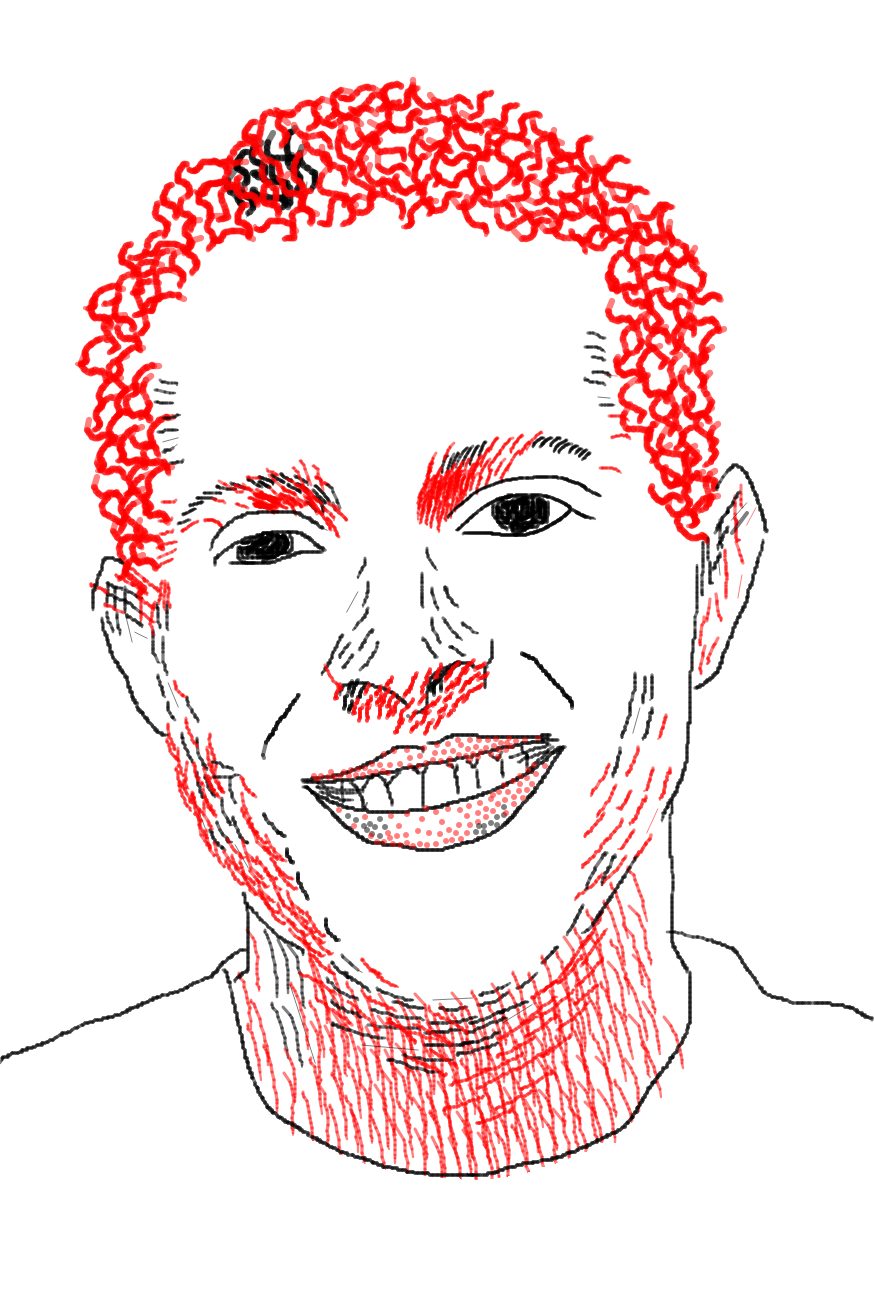}}%
	\subfloat[]{\includegraphics[width=0.28\linewidth]{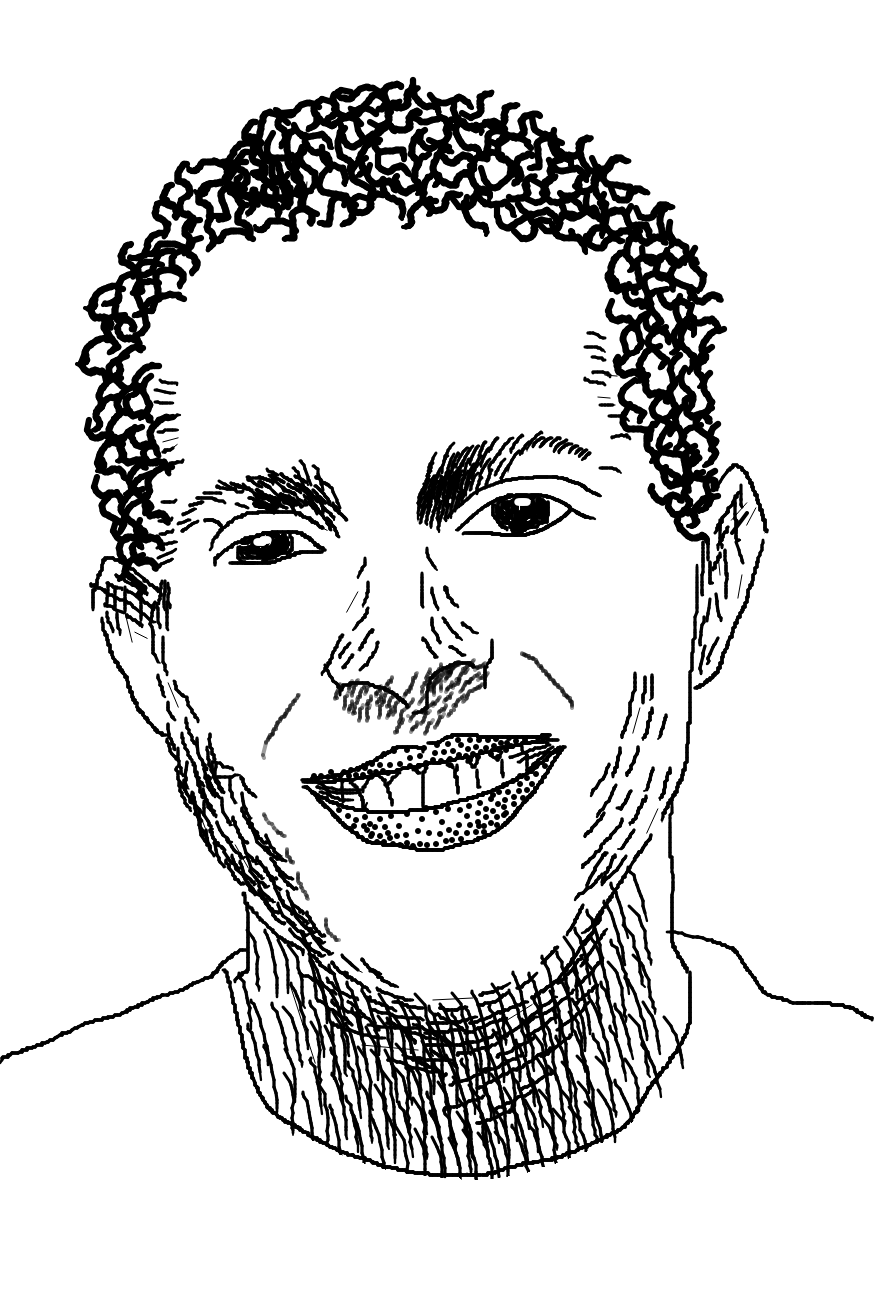}}%

	\subfloat[]{\includegraphics[width=0.28\linewidth]{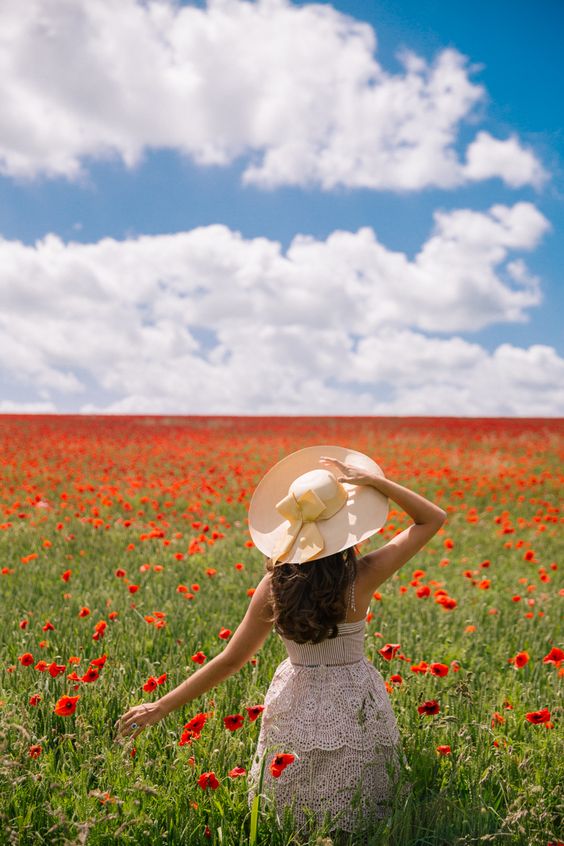}}
	\subfloat[\manualstrokes{654}/\autostrokes{1971} strokes]{\includegraphics[width=0.28\linewidth]{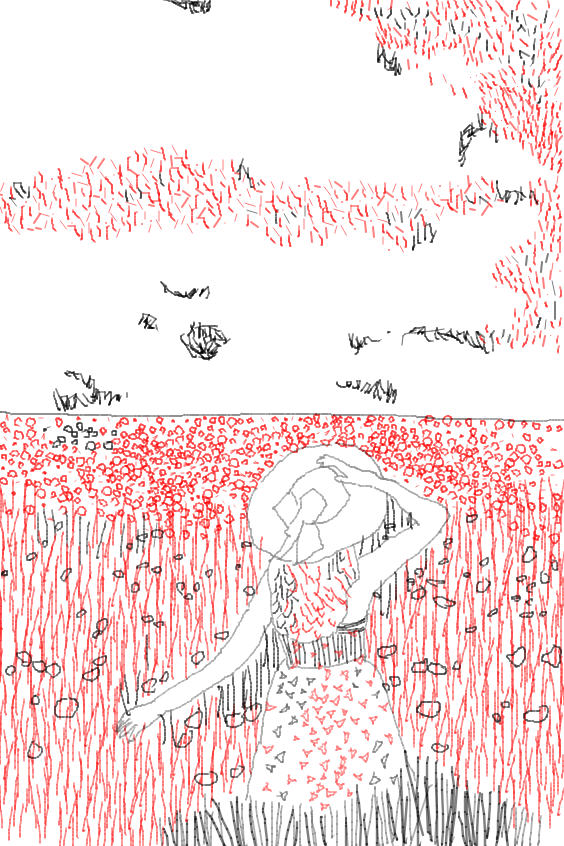}}
	\subfloat[]{\includegraphics[width=0.28\linewidth]{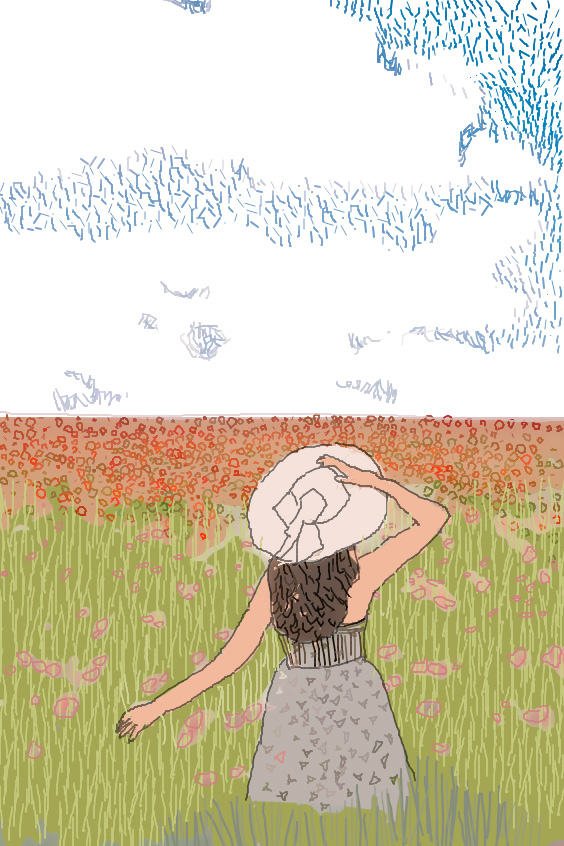}
	\label{fig:results:ladyhat}	
	}

	\Caption{Sample results.}
{%
	In each example, the left column shows the reference images, the middle column visualizes the manual strokes (\manualstrokes{black}) and autocompleted results (\autostrokes{red}) of the final drawings on the right column.
	In the last example, the strokes are created with our system first and then imported into Photoshop for background coloring.
}
\label{fig:results}
\end{figure*}

%% file: results_supp_fig.tex
\begin{figure*}[t]
  \centering

  \subfloat[]{\includegraphics[width=0.28\linewidth]{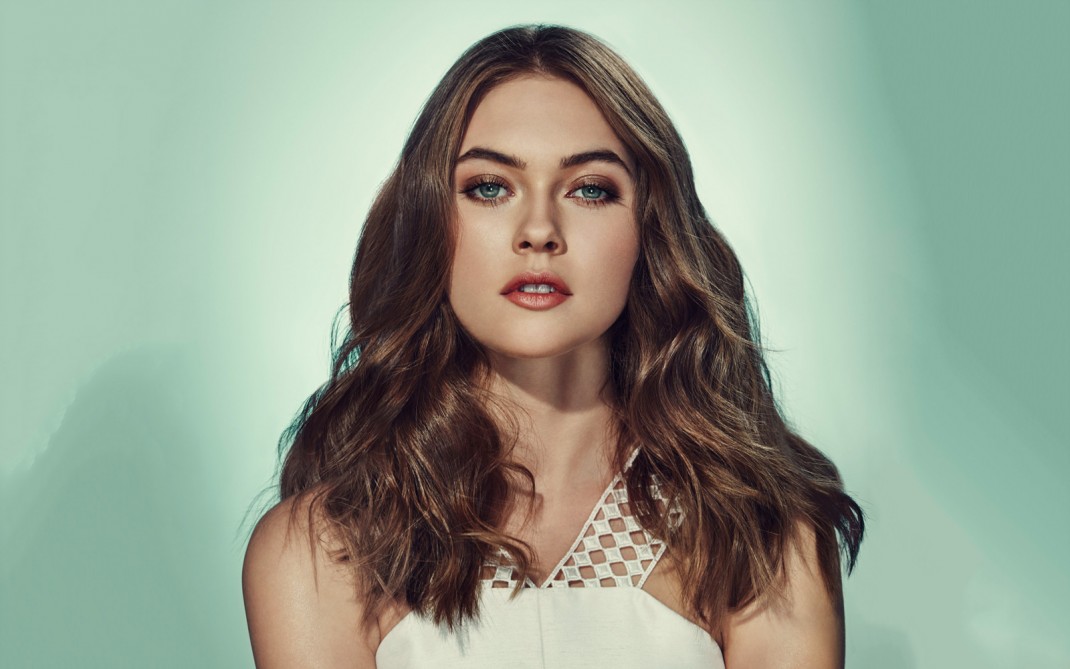}}
  \subfloat[\manualstrokes{151}/\autostrokes{1590} strokes]{\includegraphics[width=0.28\linewidth]{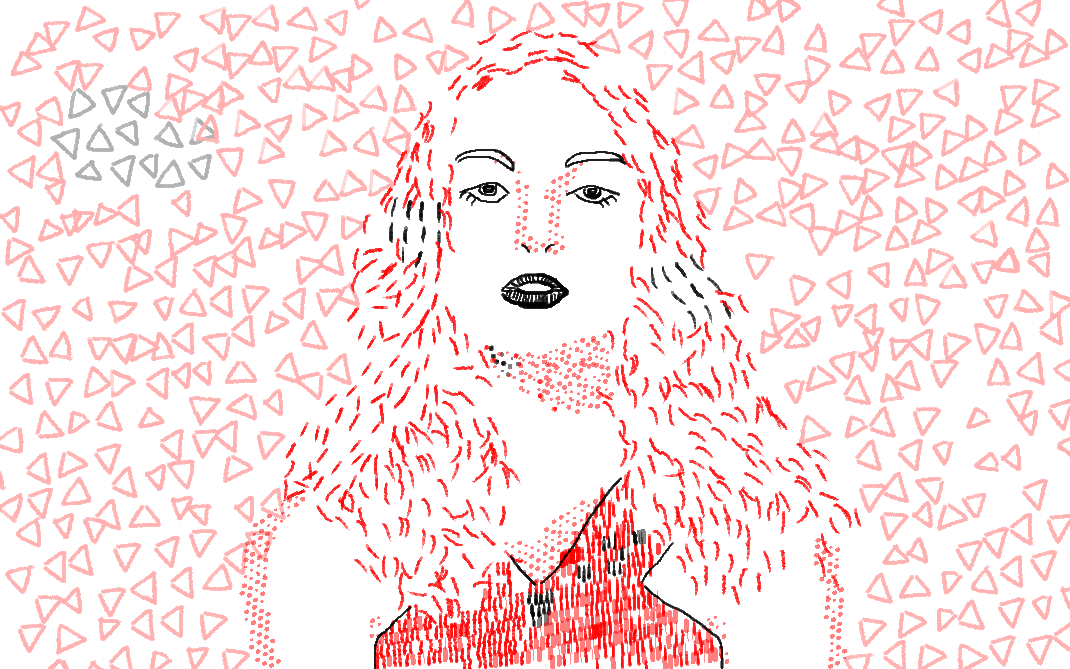}}
  \subfloat[]{\includegraphics[width=0.28\linewidth]{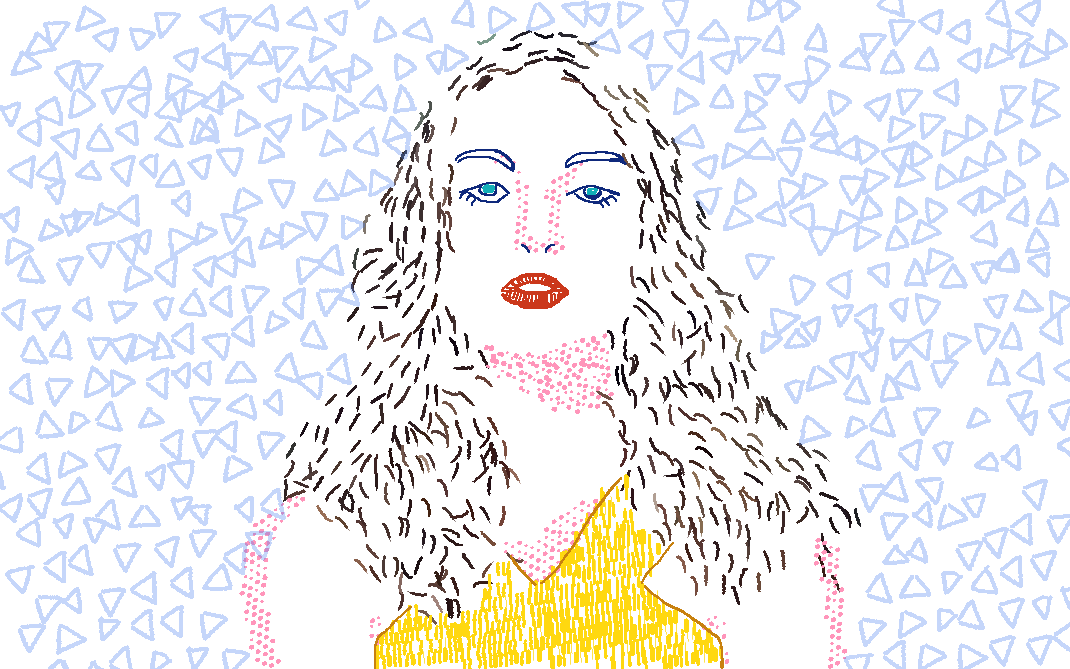}}
  
  \subfloat[]{\includegraphics[width=0.28\linewidth]{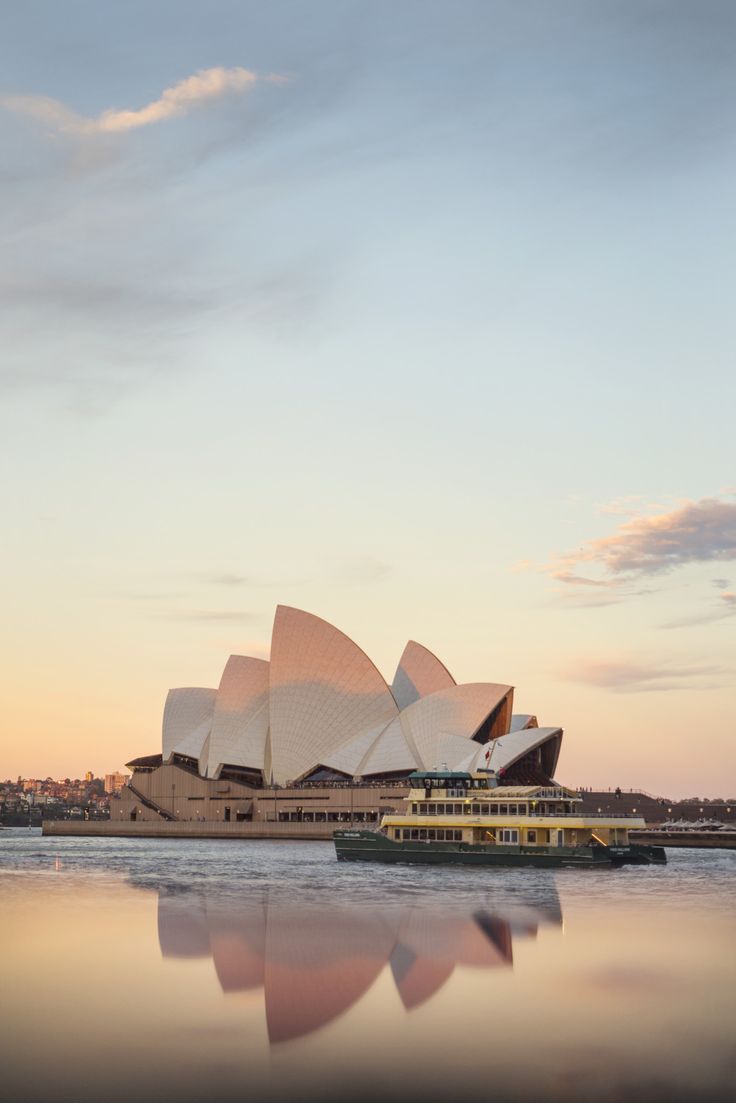}}
  \subfloat[\manualstrokes{470}/\autostrokes{551} strokes]{\includegraphics[width=0.28\linewidth]{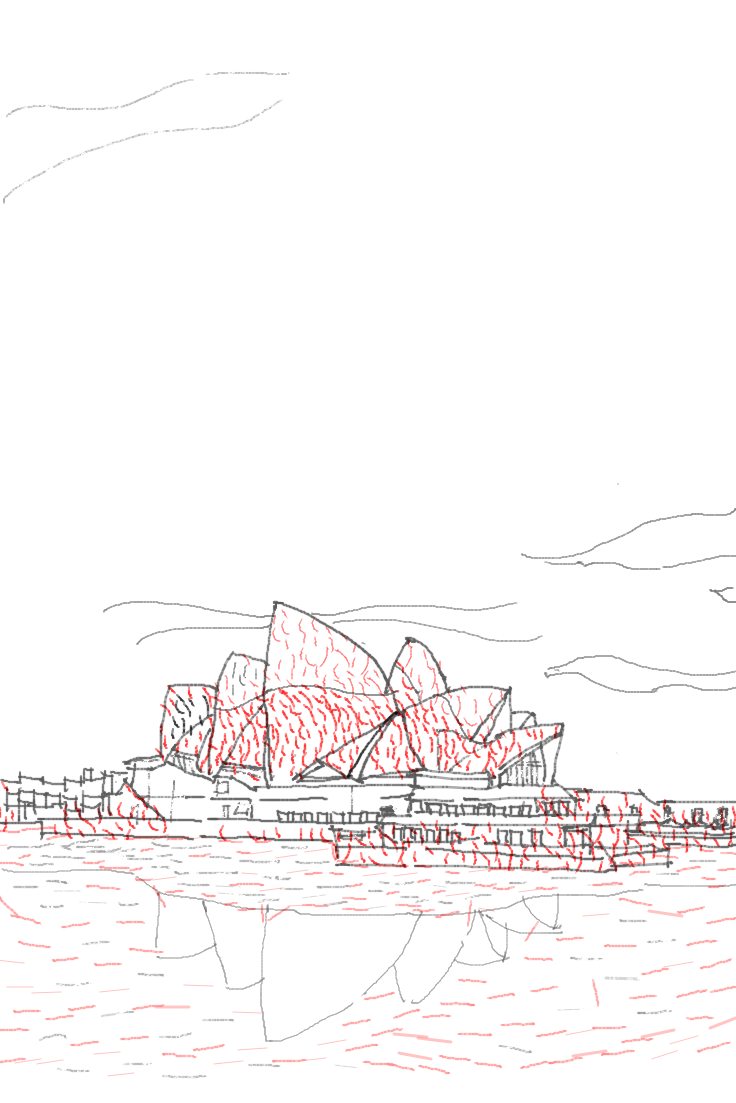}}
  \subfloat[]{\includegraphics[width=0.28\linewidth]{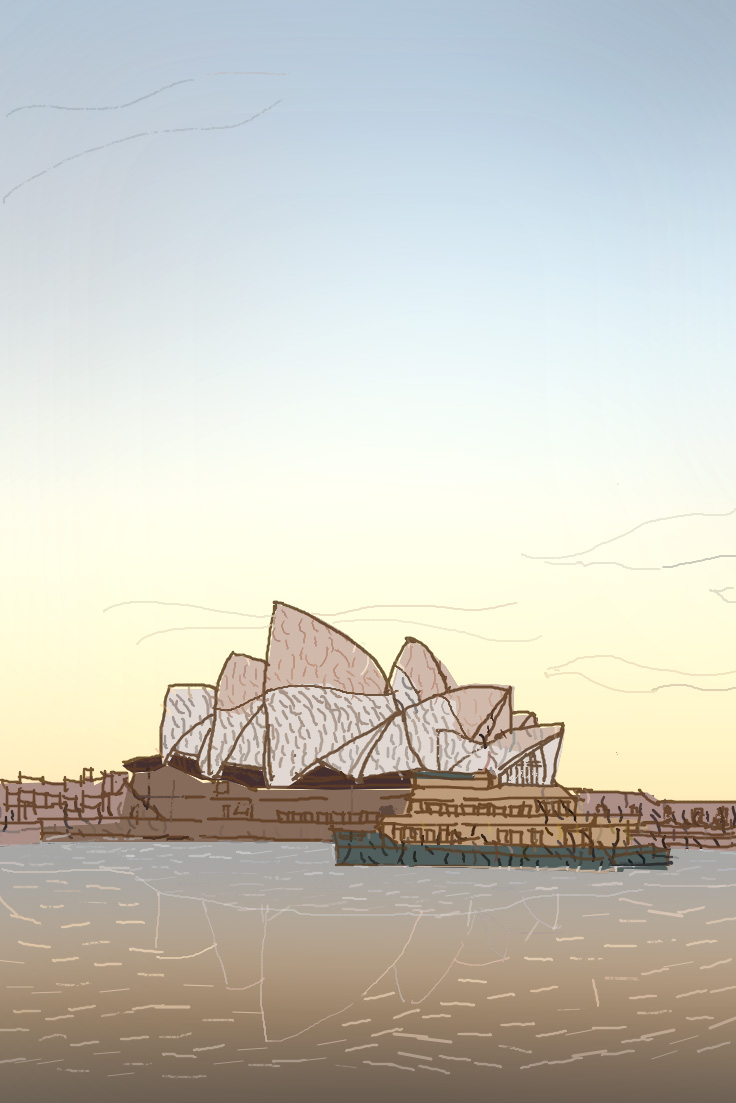}}

  \subfloat[]{\includegraphics[width=0.28\linewidth]{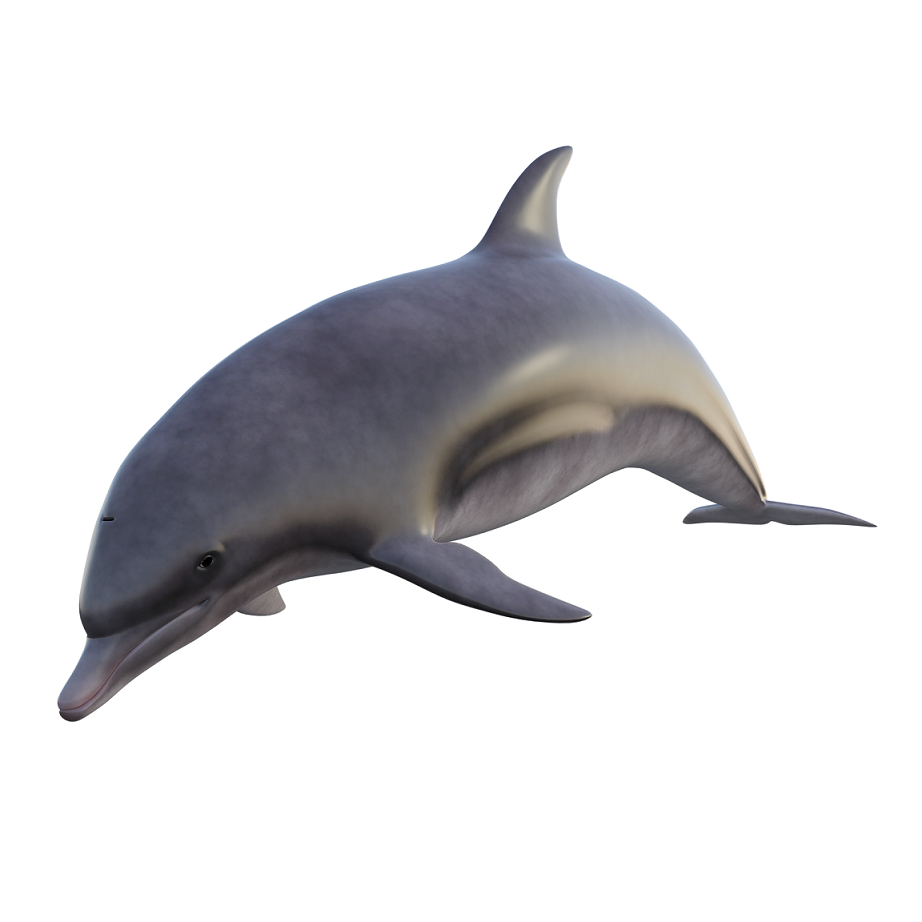}}
  \subfloat[\manualstrokes{39}/\autostrokes{1250} strokes]{\includegraphics[width=0.28\linewidth]{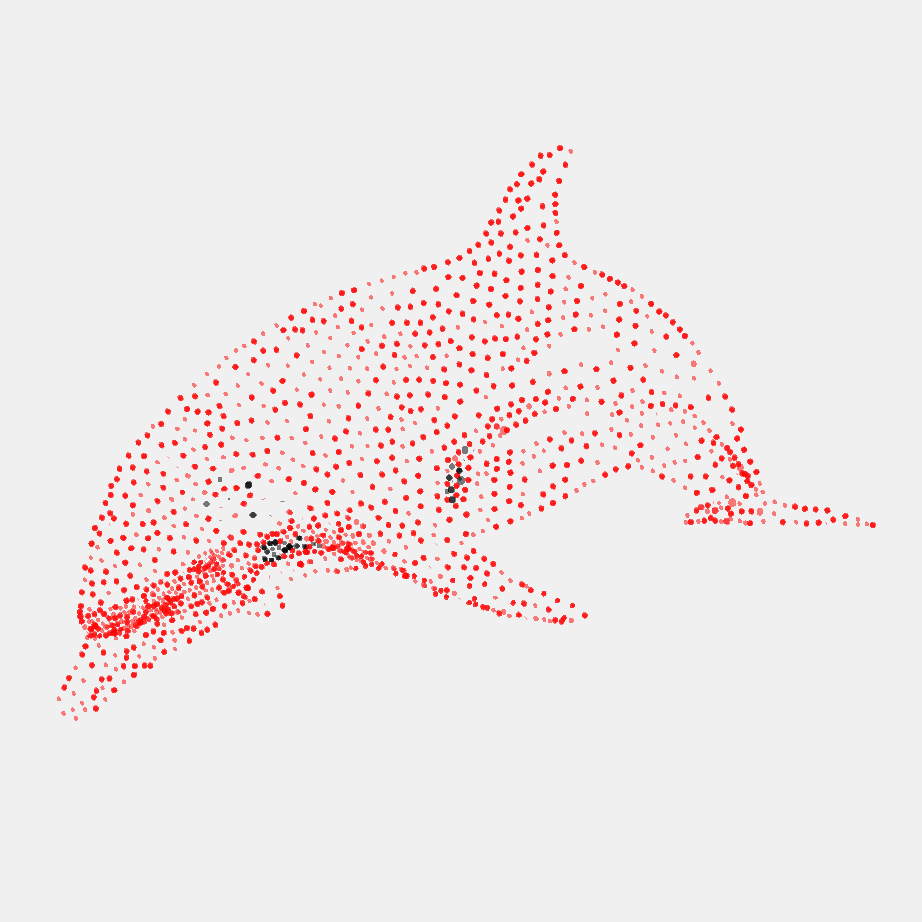}}
  \subfloat[]{\includegraphics[width=0.28\linewidth]{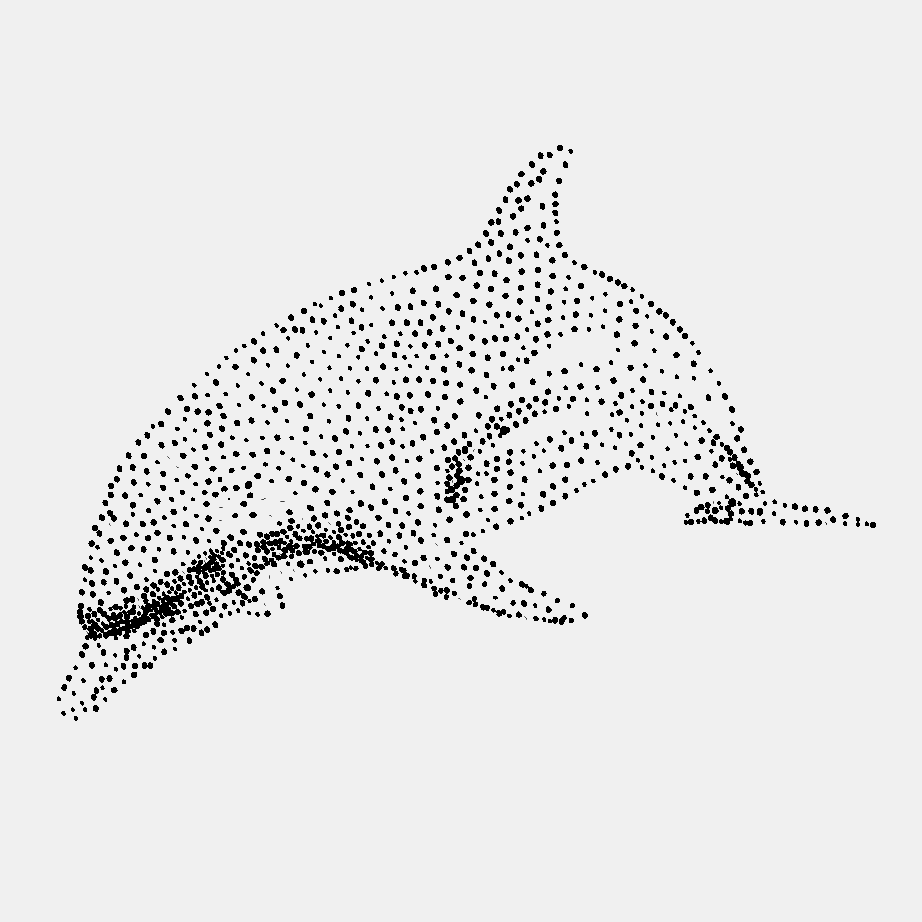}}

	\subfloat[]{\includegraphics[width=0.28\linewidth]{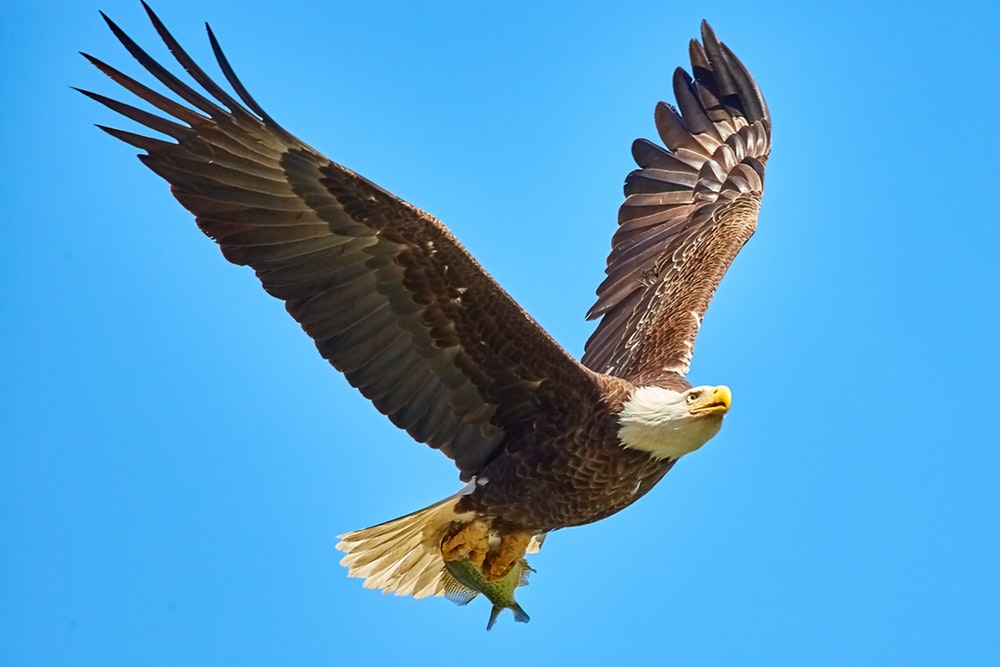}}
	\subfloat[\manualstrokes{88}/\autostrokes{604} strokes]{\includegraphics[width=0.28\linewidth]{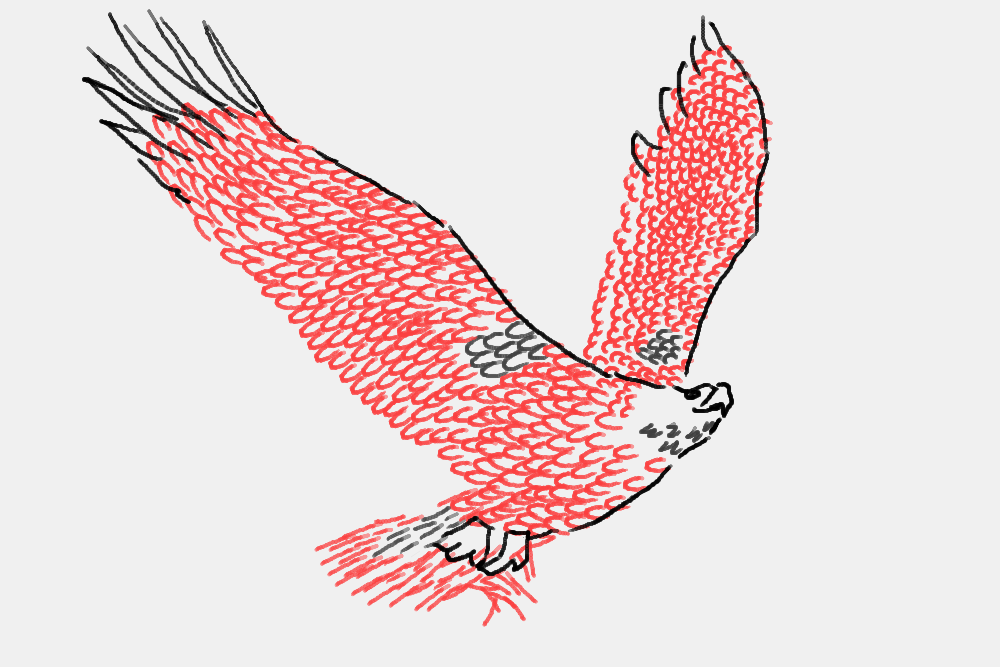}}
	\subfloat[]{\includegraphics[width=0.28\linewidth]{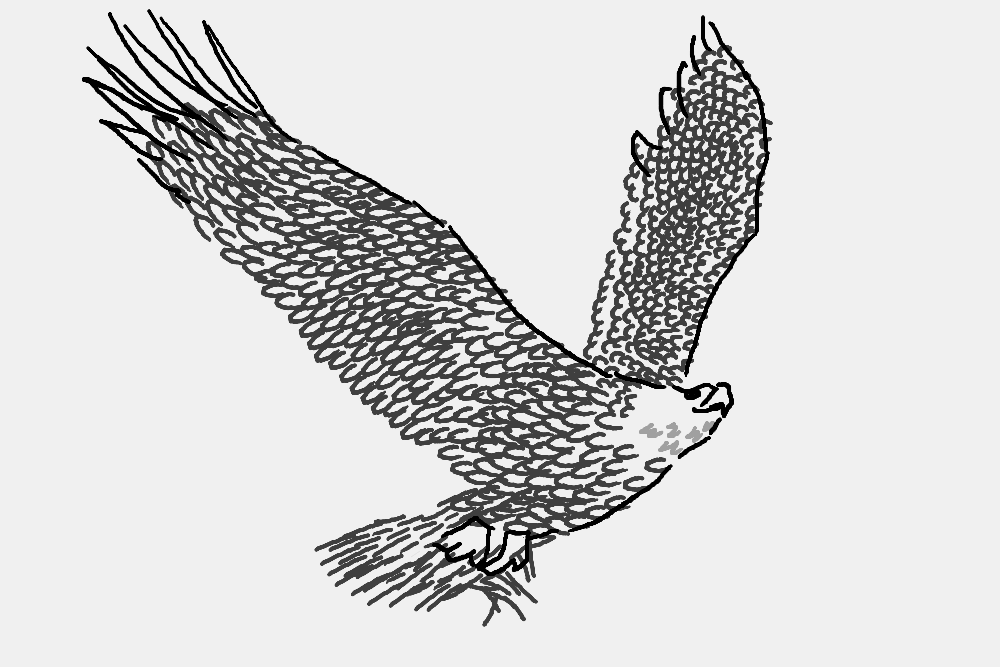}}

\phantomcaption
\end{figure*}

\begin{figure*}[t]\ContinuedFloat
\centering

\subfloat[]{\includegraphics[width=0.28\linewidth]{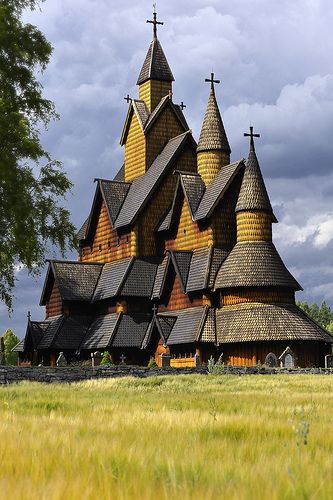}}%
\subfloat[\manualstrokes{163}/\autostrokes{939} strokes]{\includegraphics[width=0.28\linewidth]{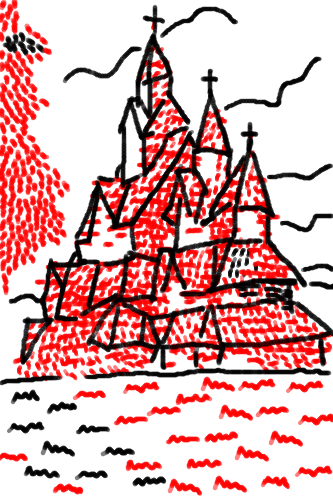}}%
\subfloat[]{\includegraphics[width=0.28\linewidth]{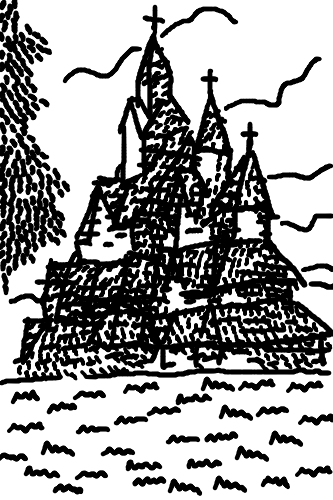}}%

  \subfloat[]{\includegraphics[width=0.28\linewidth]{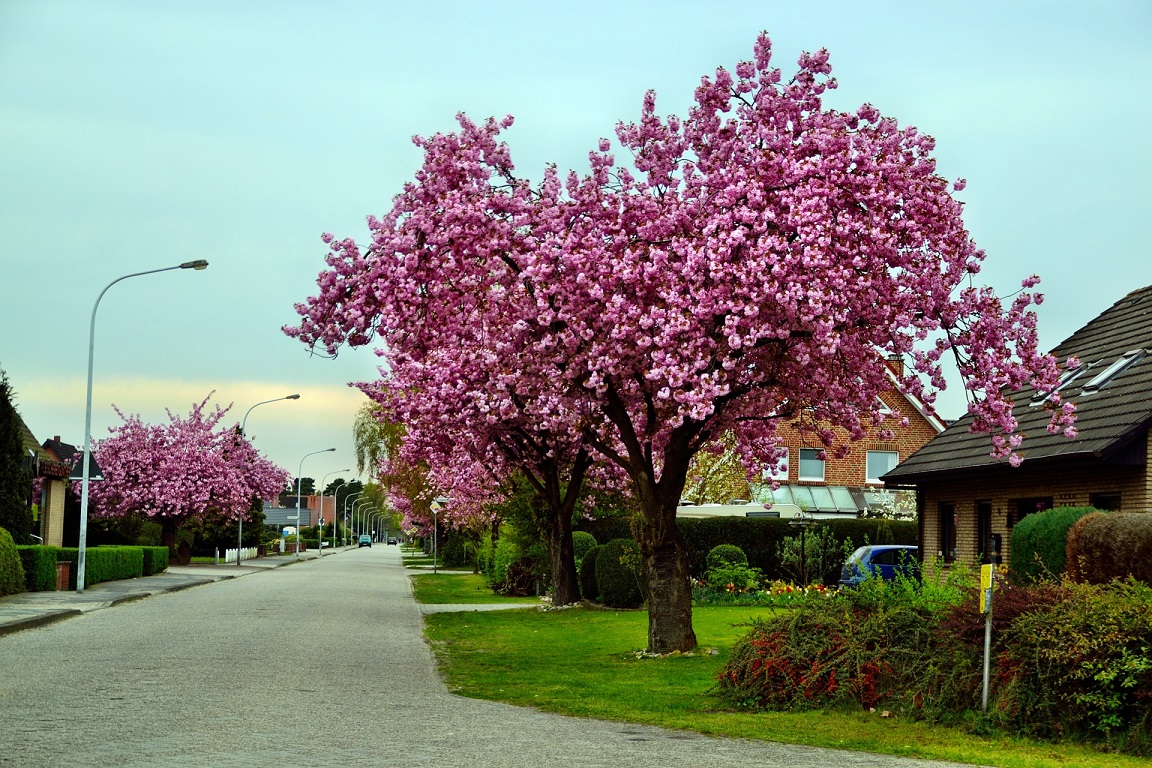}}
  \subfloat[\manualstrokes{322}/\autostrokes{2832} strokes]{\includegraphics[width=0.28\linewidth]{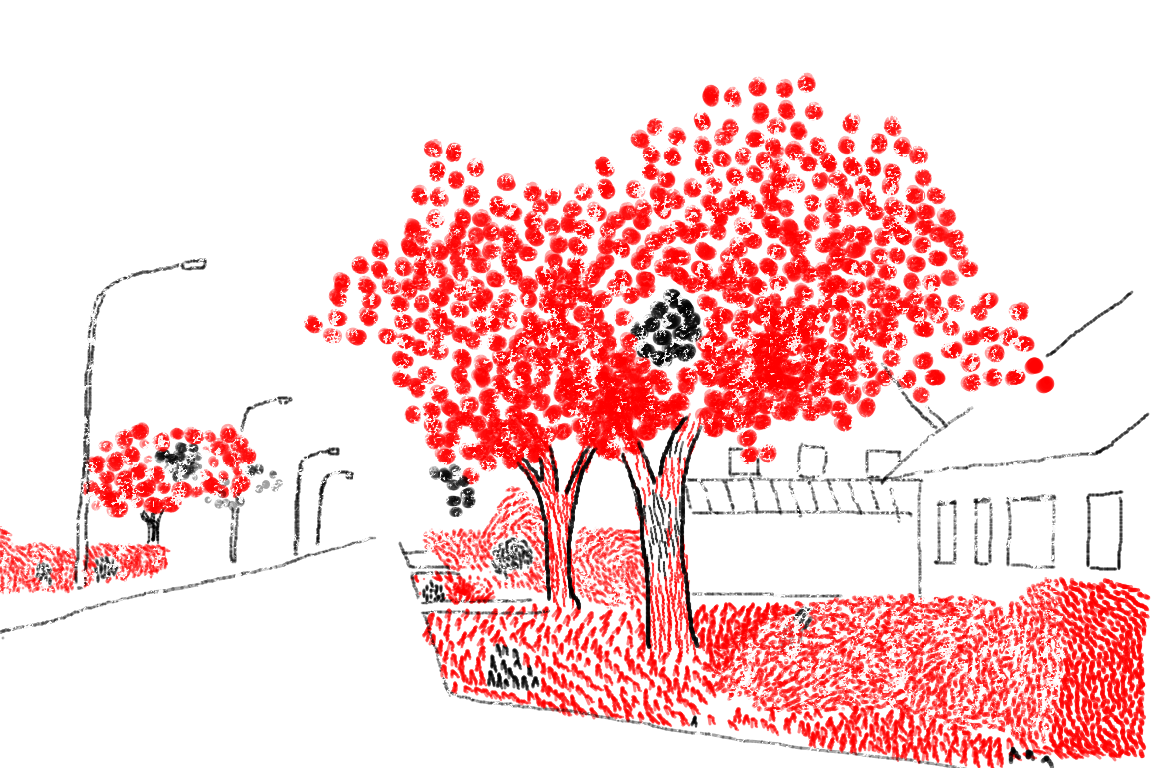}}
  \subfloat[]{\includegraphics[width=0.28\linewidth]{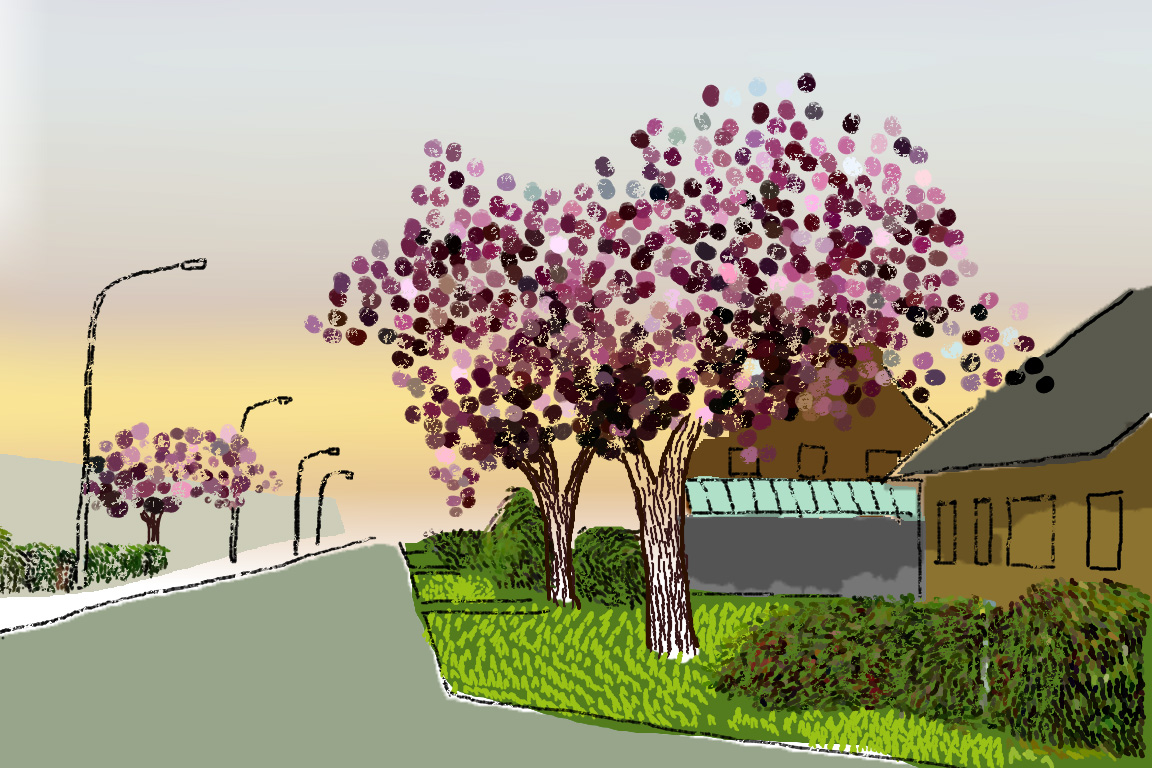}}
  
  \subfloat[]{\includegraphics[width=0.28\linewidth]{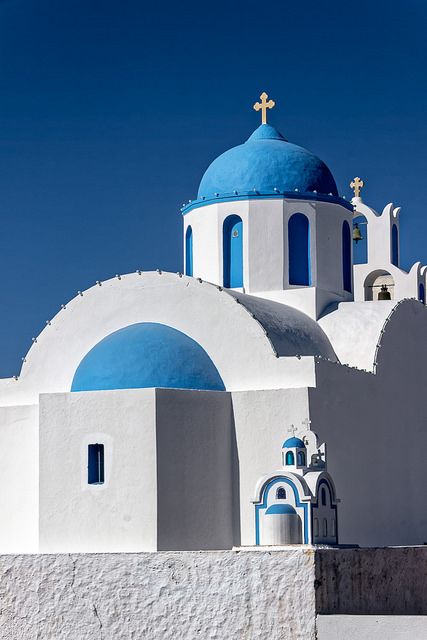}}
  \subfloat[\manualstrokes{134}/\autostrokes{1039} strokes]{\includegraphics[width=0.28\linewidth]{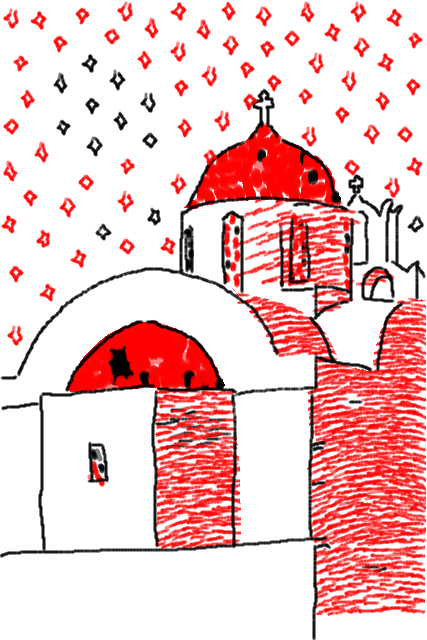}}
  \subfloat[]{\includegraphics[width=0.28\linewidth]{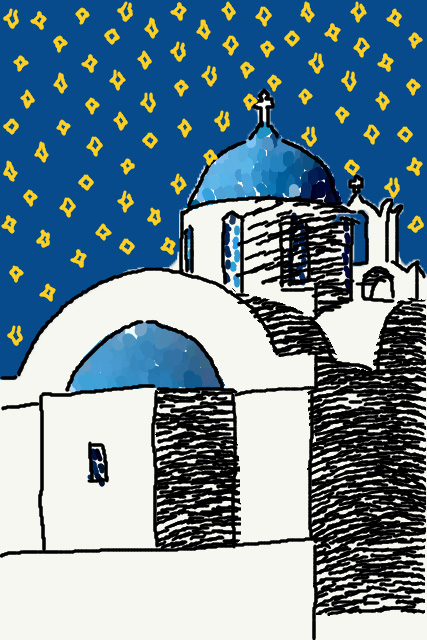}}

  \Caption{Additional results\ifdefined\resultsupp~for \Cref{fig:results}\fi.}
  {%
\ifdefined\resultsupp
\else
  	In each example, the left column shows the reference images, the middle column visualizes the manual strokes (\manualstrokes{black}) and autocompleted results (\autostrokes{red}) of the final drawings on the right column.
\fi
  }
  \label{fig:results:supp}
\end{figure*}

%% file: conclusion.tex
\section{Conclusion}
\label{sec:conclusion}

We have presented a method to help users autocomplete repetitive short strokes with guidance from reference images while maintaining the flexible control of manual drawing.
By extending operation history analysis and synthesis with image analysis, our method is able to generate results that adapt to reference images and users' prior inputs.
We conducted a pilot study to validate the usefulness of our approach and show various drawing results from the users.

%% file: limitation.tex
\section{Limitations and Future Work}
\label{sec:limit}

From our observation and users' feedback, we identified several improvement opportunities.

\paragraph{Improve accuracy of autocompletion.}
We rely on simple $Lab*$ color and semantic segmentation for region inference.
While color feature is sufficient for most cases, regions with similar colors but different semantics will require sufficient segmentation accuracy for region inference (\Cref{fig:target_tasks:bear_seg,fig:target_tasks:beach_seg}).
Since our segmentation map is precomputed, taking users' input as additional cues might help improve the segmentation accuracy (e.g., using interactive semantic segmentation methods like \cite{Ning:2017:DGO}).

\input{visual_block_fig}

\paragraph{Resolve visual blocking.}
Since the drawing and the system suggestions are overlaid on the reference image, it might be difficult for users to refer to the image when selecting parts of the suggestions (e.g., \Cref{fig:visualblock}) or adding a new layer of strokes.
Although users can switch the views via a hotkey, it might be helpful to provide some reference information, like image darkness or boundaries, through additional visual hints \cite{Xie:2014:PFS,Williford:2019:DAN}.

\paragraph{Consider relationships with higher-level image features.}
We only consider the relationships between strokes and low-level image features, like colors and flows, over regions.
By considering higher-level image features, such as elements and edges, it is possible to extend the scope of autocompletion, such as autocomplting the sparse flowers in the foreground of \Cref{fig:results:ladyhat} through the correspondences between strokes and elements.

\paragraph{Support more stroke types.}
Our method only supports short strokes, while artists also use long repetitive strokes frequently \cite{Dunn:2015:PID}.
It is worth investigating the possibility of incorporating continuous strokes \cite{Tu:2020:CCT} in our analysis and synthesis framework and extending the support for different input strokes.

%% file: visual_block_fig.tex
\begin{wrapfigure}{r}{0.4\linewidth}
	\centering
	{\includegraphics[width=0.48\linewidth]{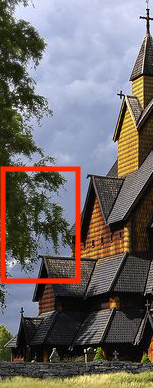}
	}%
	{\includegraphics[width=0.48\linewidth]{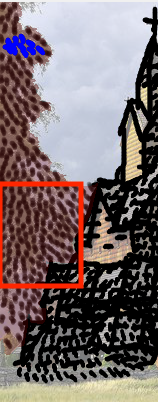}
	}
	\caption{Example of visual blocking. Left: reference image. Right: canvas view.}
	\label{fig:visualblock}
\end{wrapfigure}

%% file: supp.tex
\ifdefined\resultsupp
\else
\input{results_supp_fig}
\fi
\input{user_drawings_fig}

%% file: user_drawings_fig.tex
\begin{figure*}[tbh]
\Caption{Participants' drawings for the target session.}
{}
\label{fig:user_drawings}
\begin{tabular}{cccccc}
	\hline
	\hline
	autocomplete & batch & manual & autocomplete & batch & manual \\
	\hline
	\subfloat{\includegraphics[width = 0.15\textwidth]{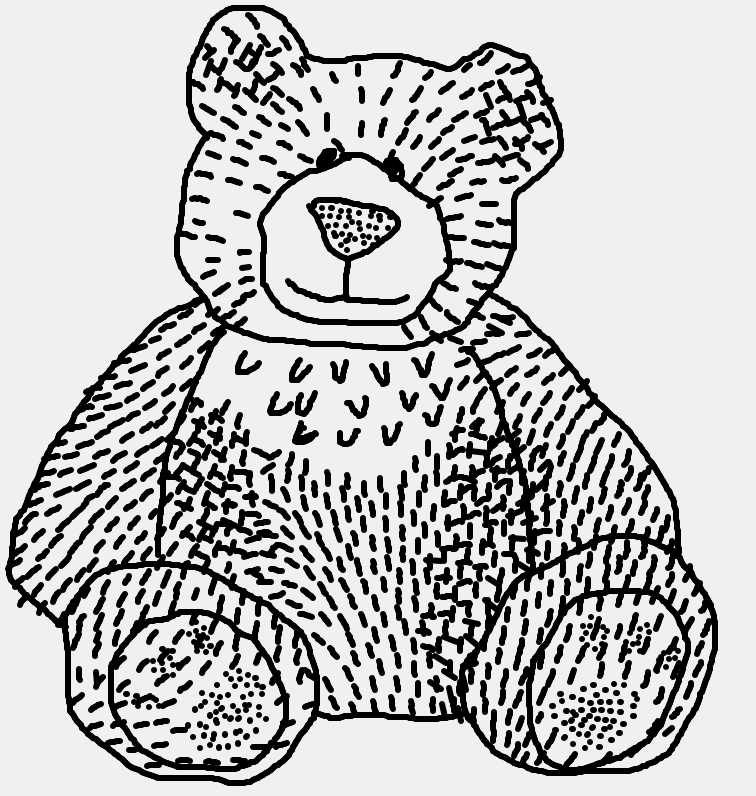}} &
	\subfloat{\includegraphics[width = 0.15\textwidth]{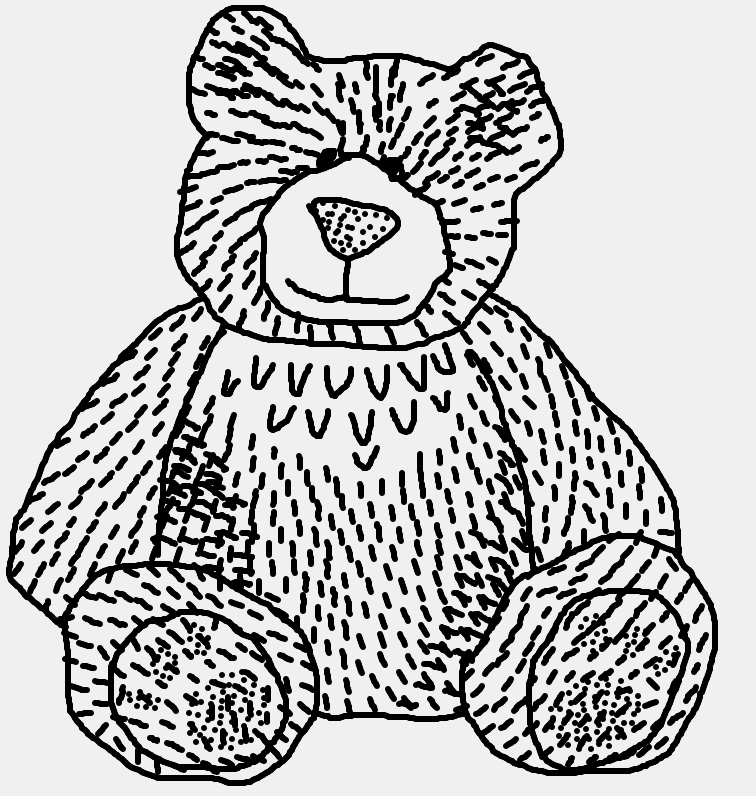}} &
	\subfloat{\includegraphics[width = 0.15\textwidth]{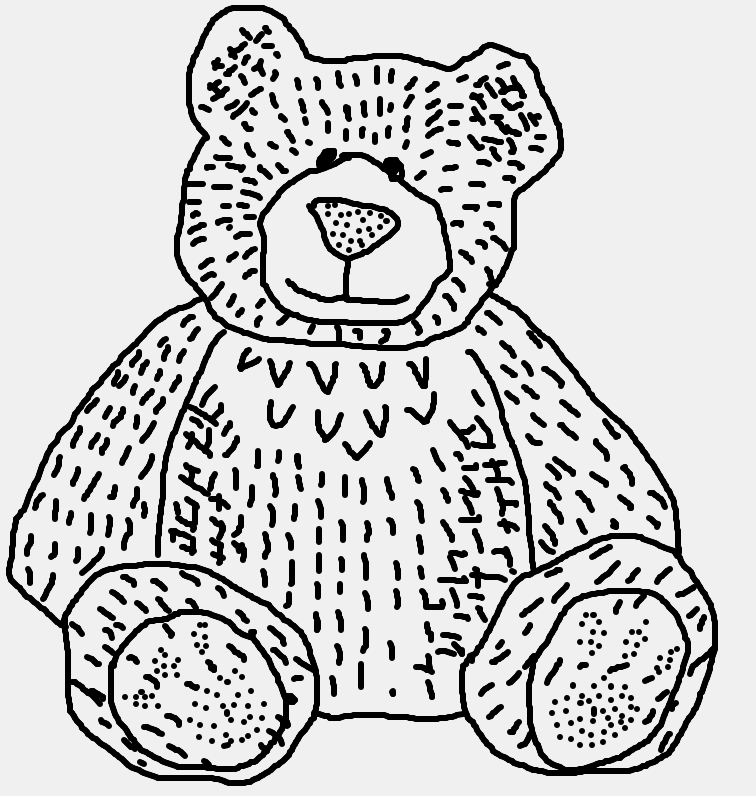}} &
	\subfloat{\includegraphics[width = 0.15\textwidth]{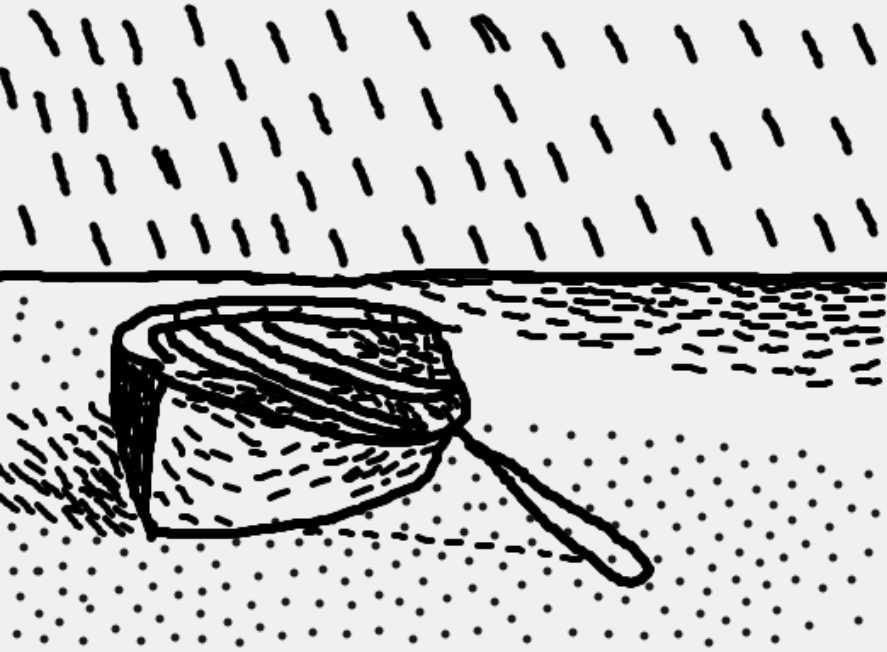}} &
	\subfloat{\includegraphics[width = 0.15\textwidth]{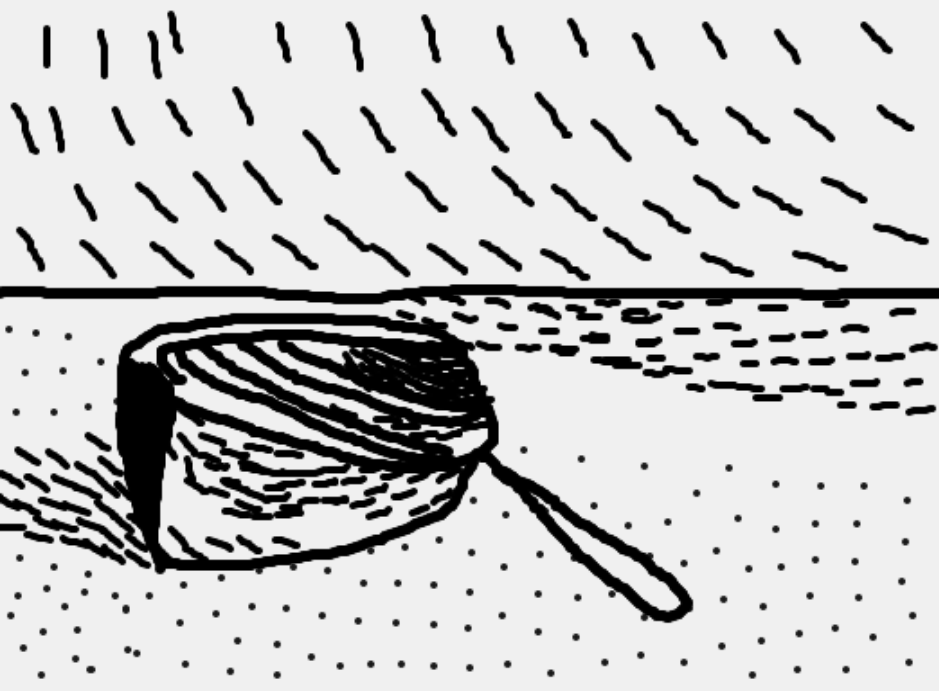}} &
	\subfloat{\includegraphics[width = 0.15\textwidth]{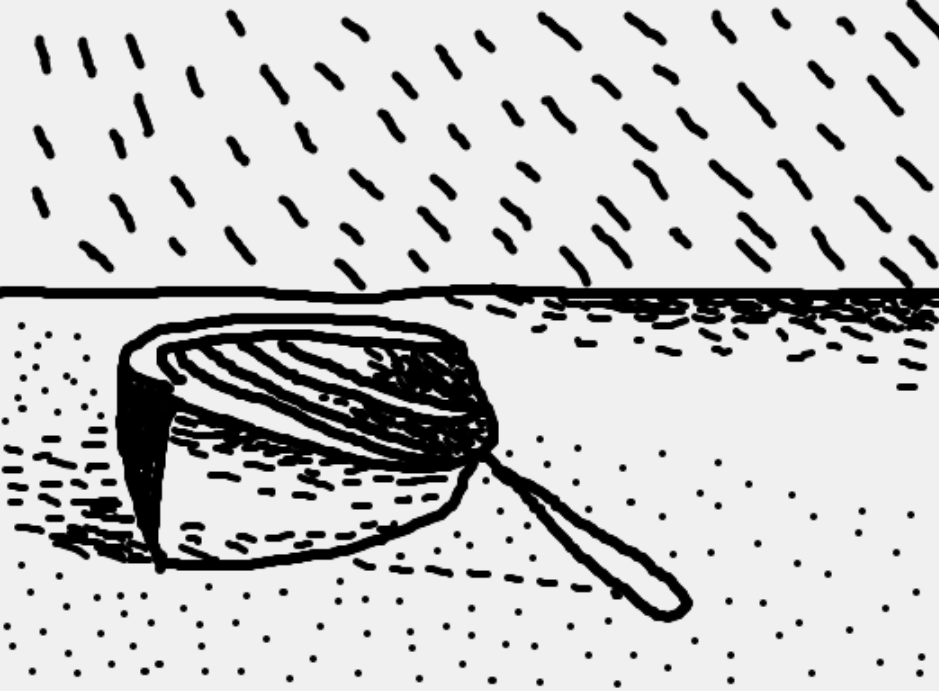}} \\
	\subfloat{\includegraphics[width = 0.15\textwidth]{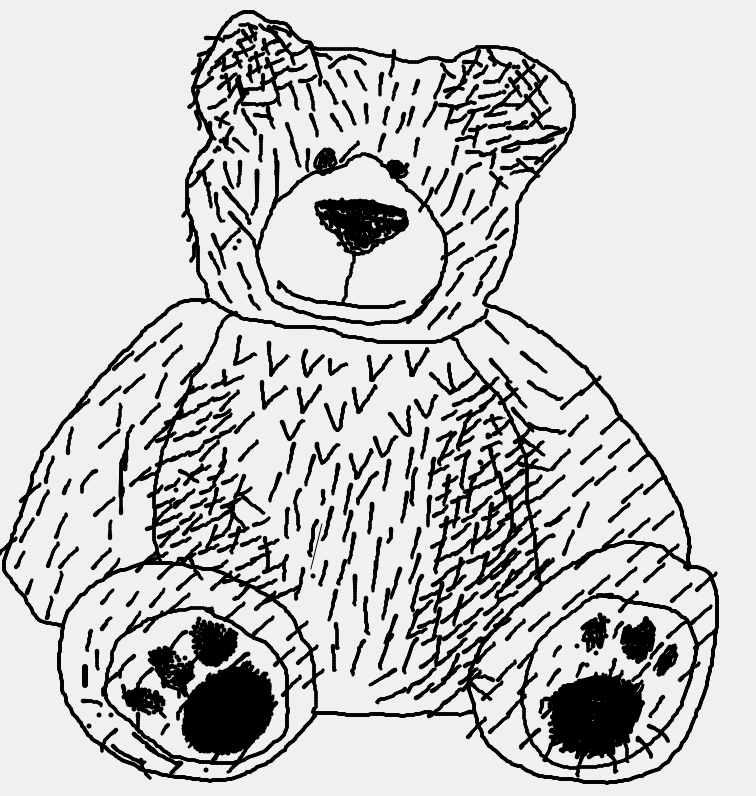}} &
	\subfloat{\includegraphics[width = 0.15\textwidth]{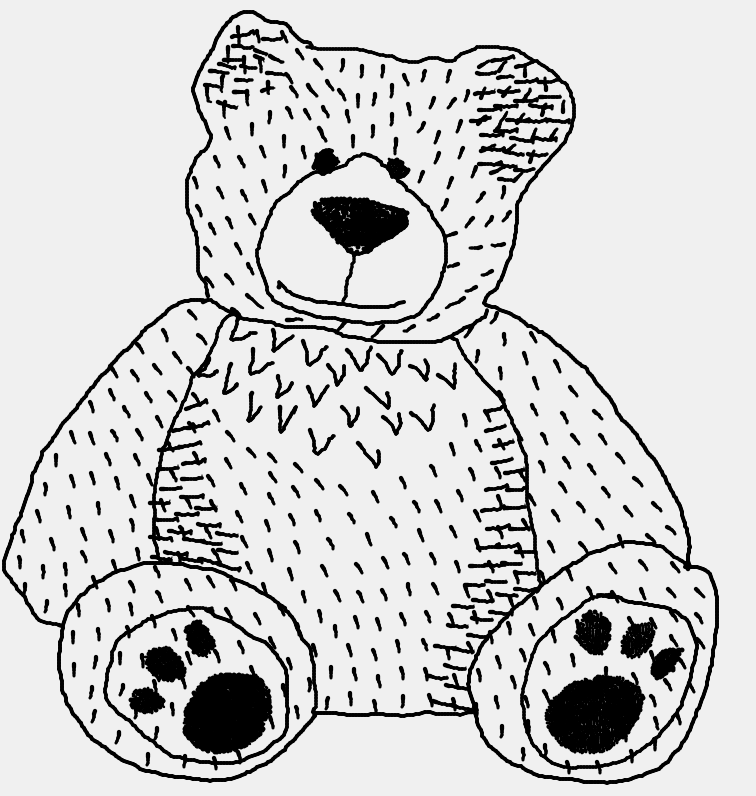}} &
	\subfloat{\includegraphics[width = 0.15\textwidth]{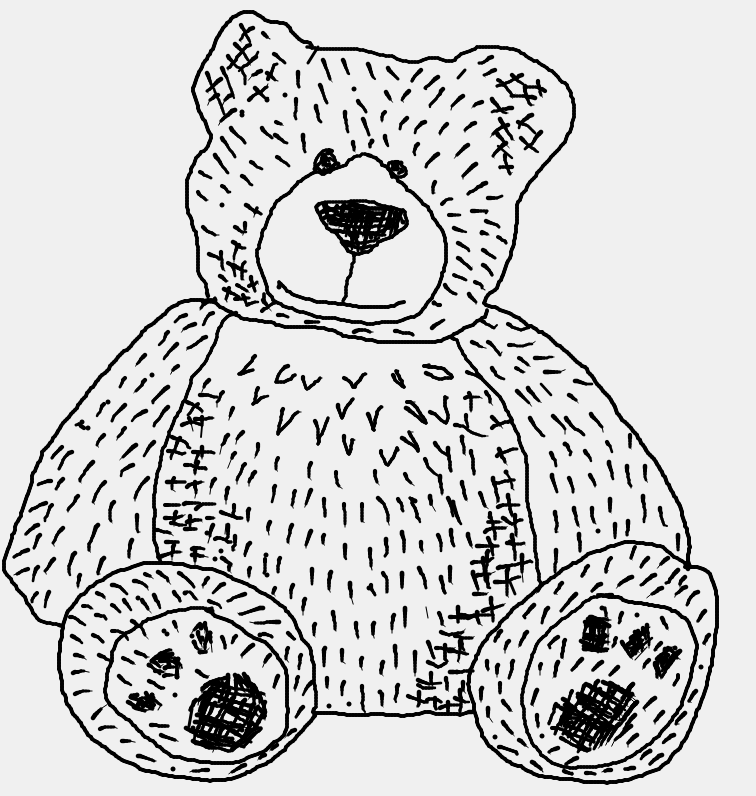}} &
	\subfloat{\includegraphics[width = 0.15\textwidth]{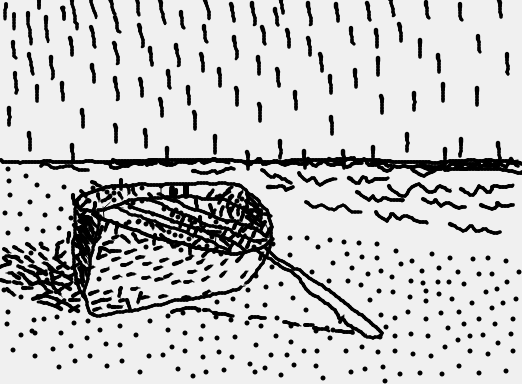}} &
	\subfloat{\includegraphics[width = 0.15\textwidth]{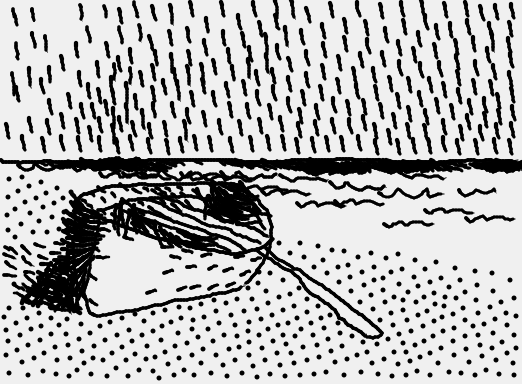}} &
	\subfloat{\includegraphics[width = 0.15\textwidth]{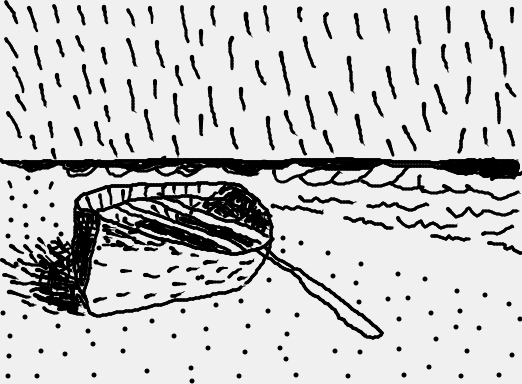}} \\
	\subfloat{\includegraphics[width = 0.15\textwidth]{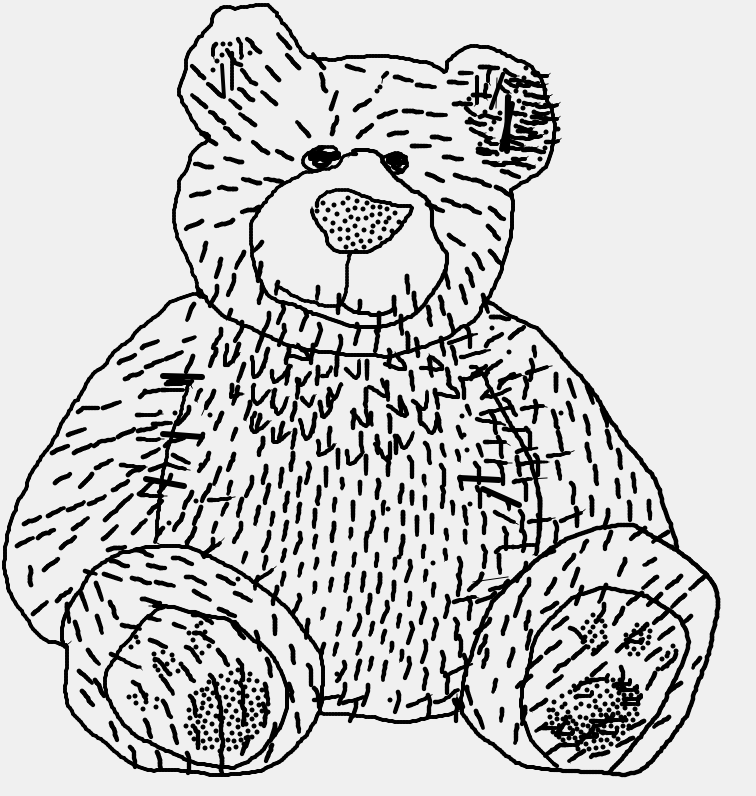}} &
	\subfloat{\includegraphics[width = 0.15\textwidth]{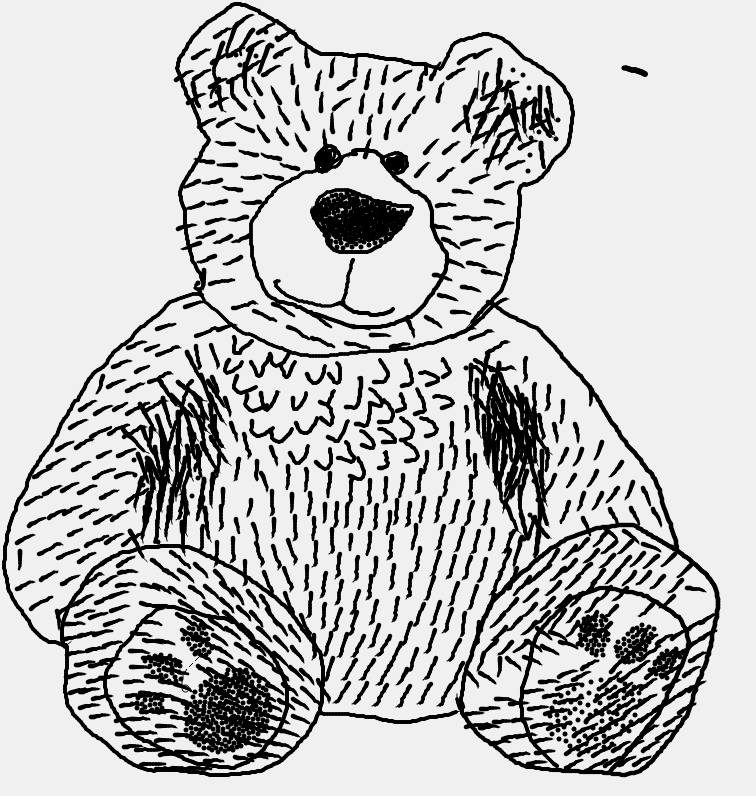}} &
	\subfloat{\includegraphics[width = 0.15\textwidth]{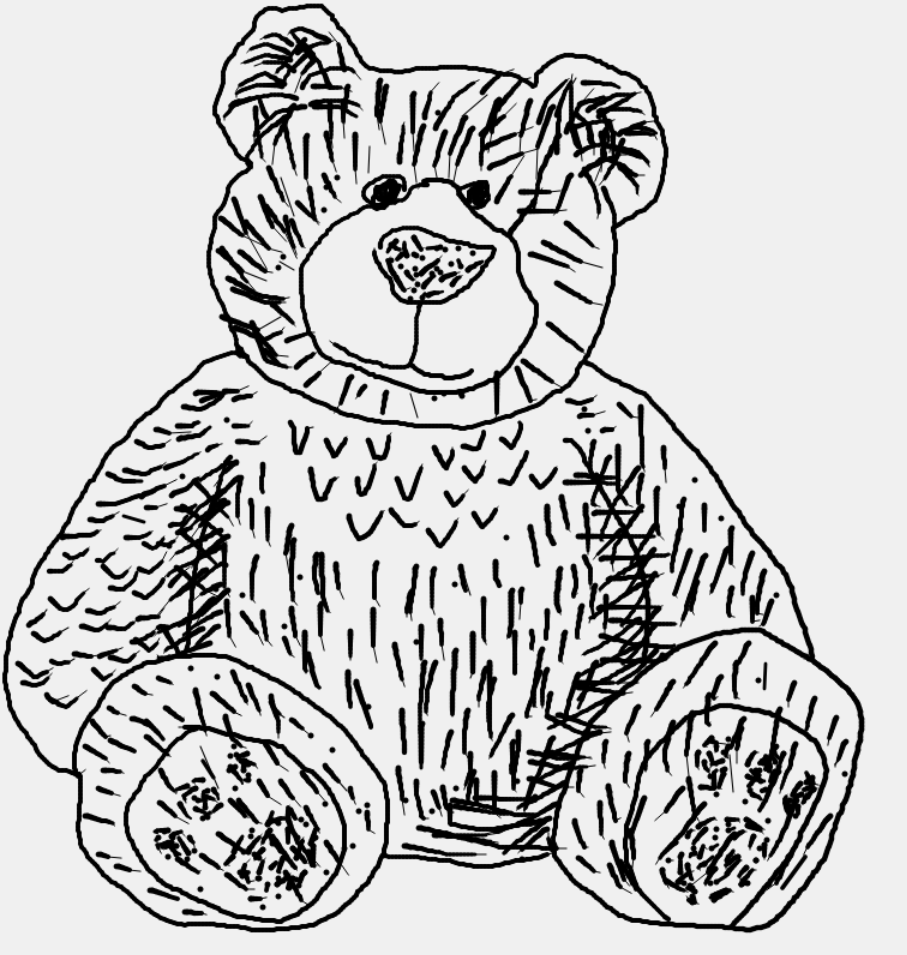}} &
	\subfloat{\includegraphics[width = 0.15\textwidth]{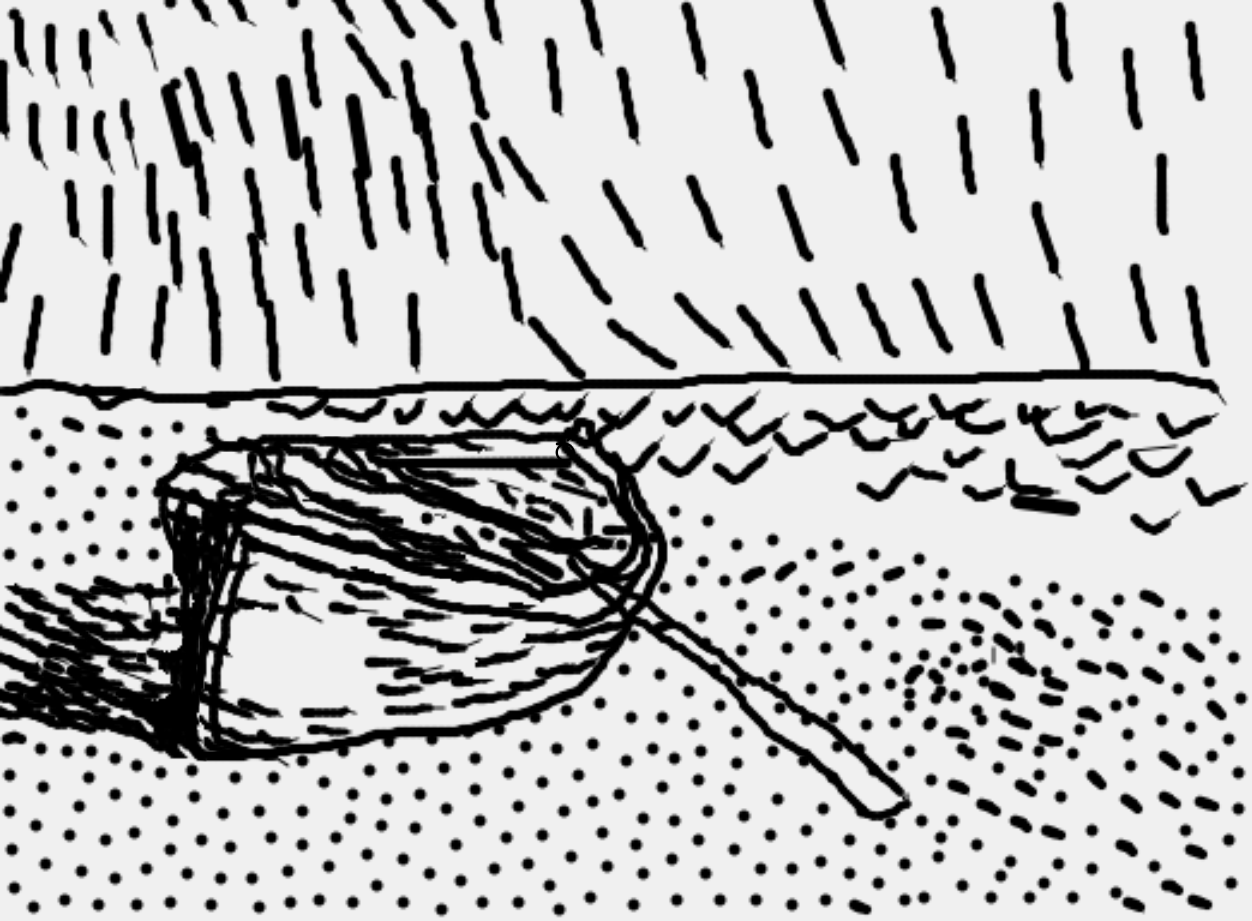}} &
	\subfloat{\includegraphics[width = 0.15\textwidth]{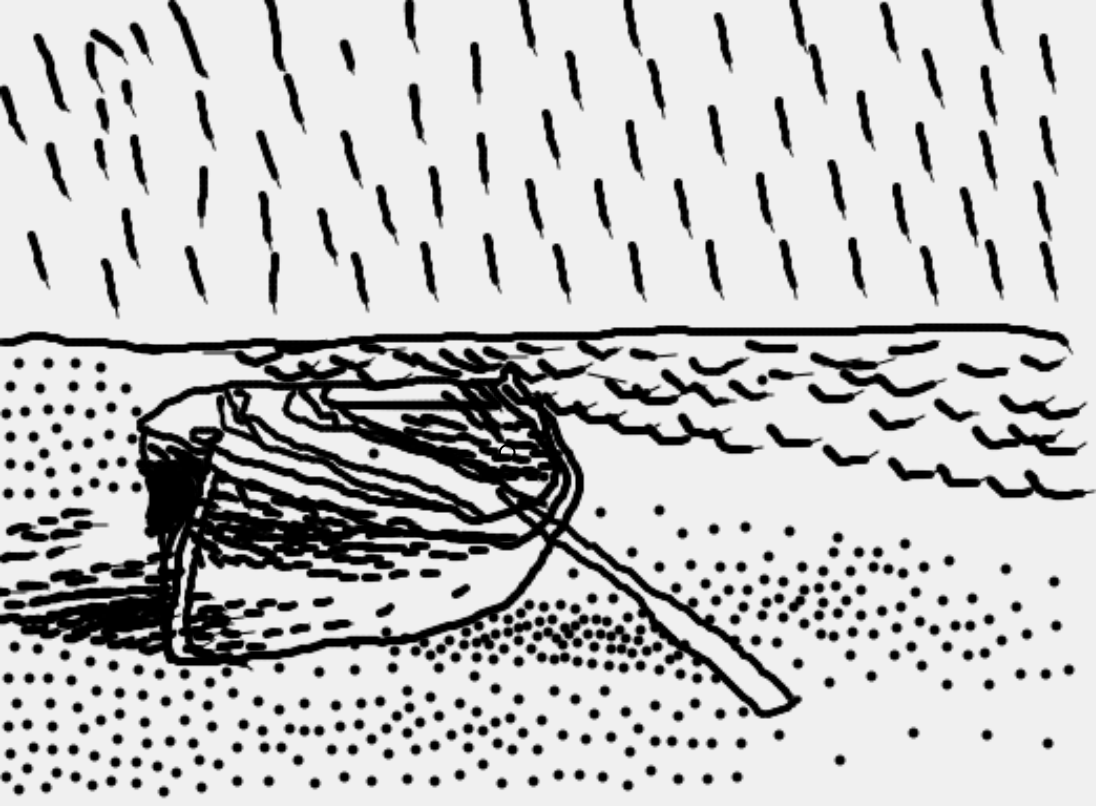}} &
	\subfloat{\includegraphics[width = 0.15\textwidth]{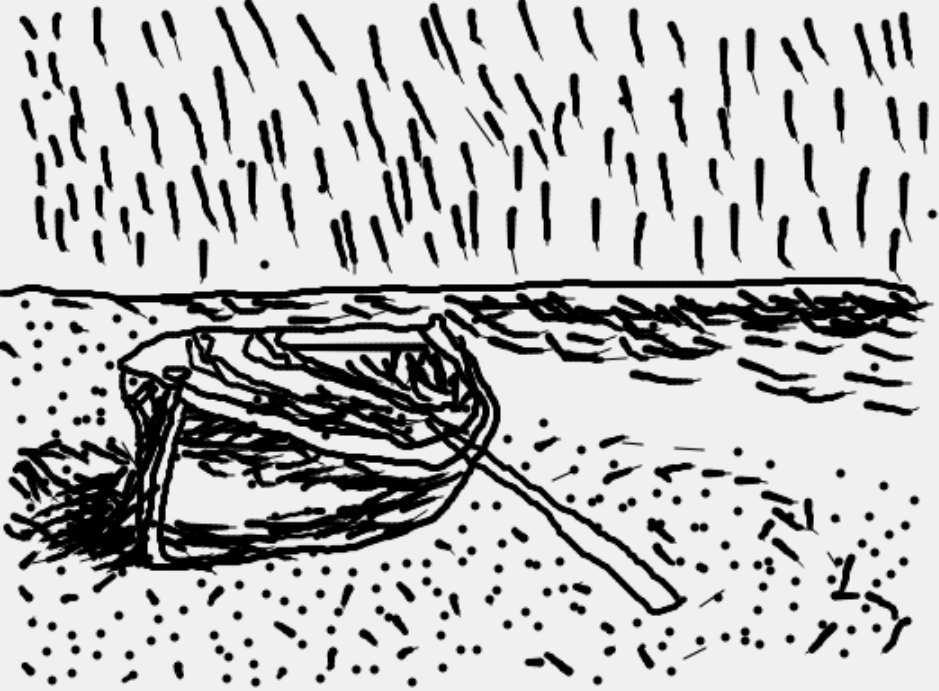}} \\
	\subfloat{\includegraphics[width = 0.15\textwidth]{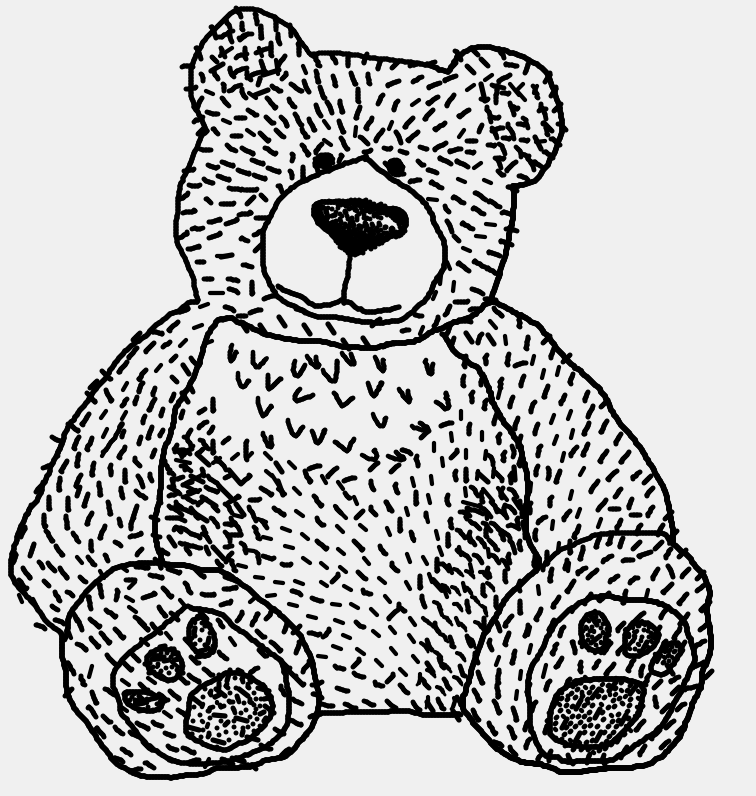}} &
	\subfloat{\includegraphics[width = 0.15\textwidth]{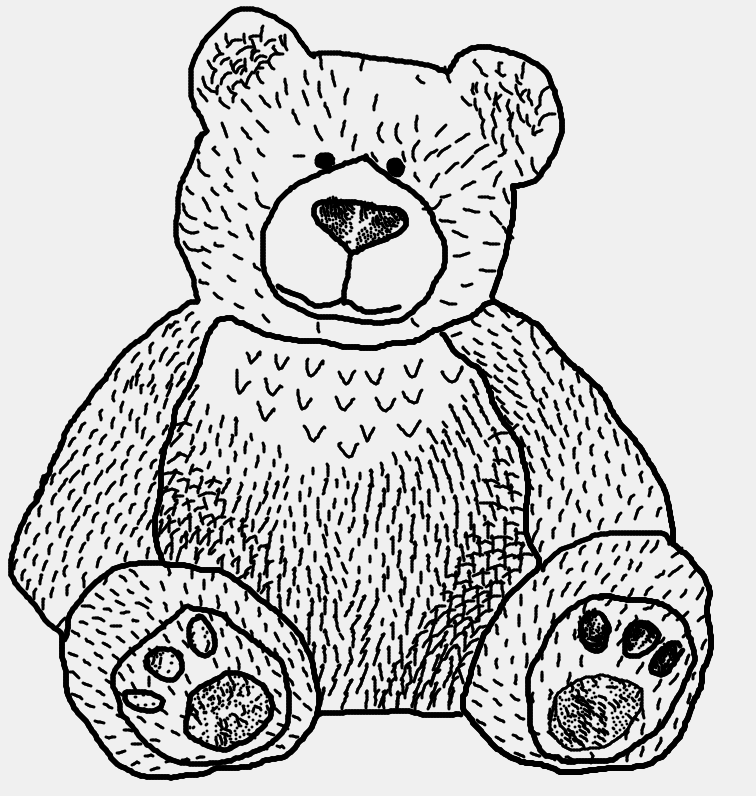}} &
	\subfloat{\includegraphics[width = 0.15\textwidth]{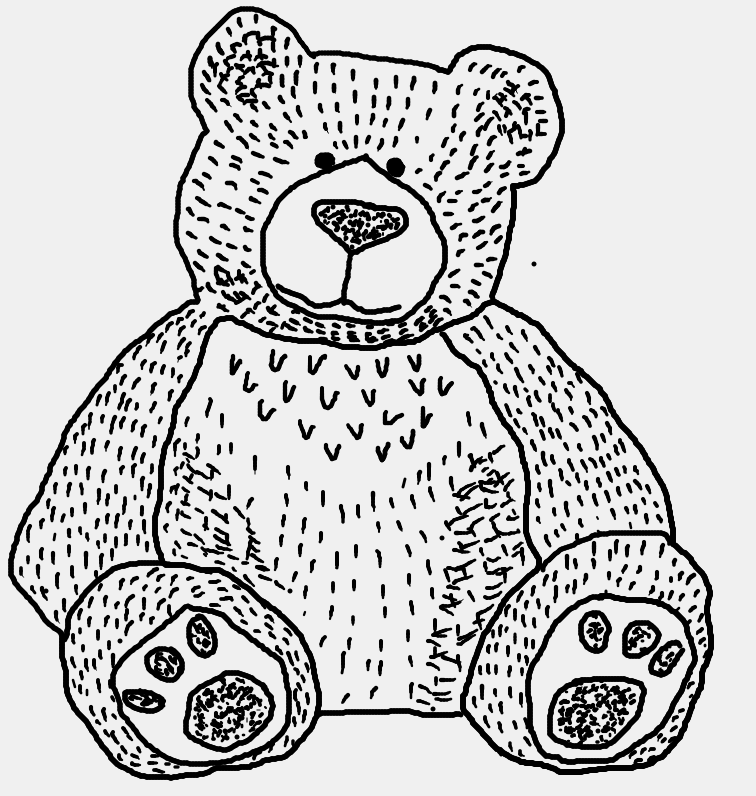}} &
	\subfloat{\includegraphics[width = 0.15\textwidth]{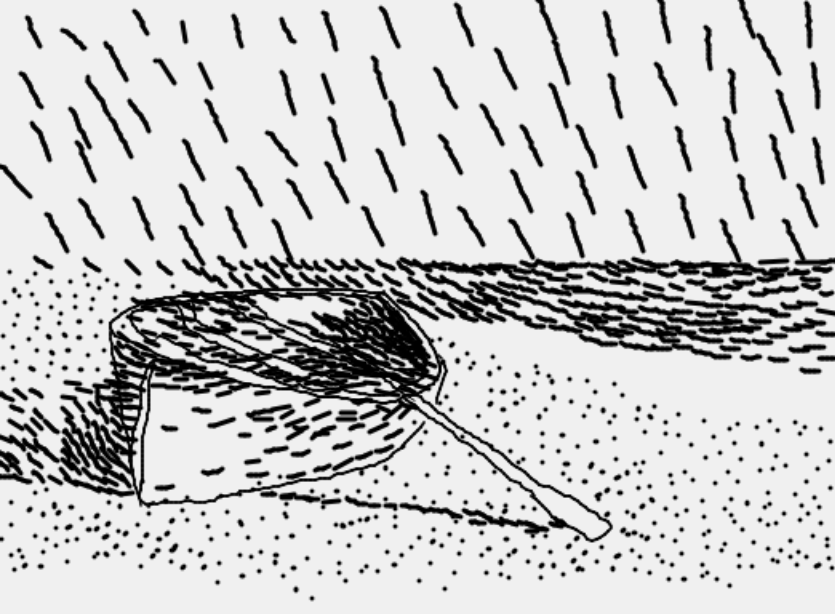}} &
	\subfloat{\includegraphics[width = 0.15\textwidth]{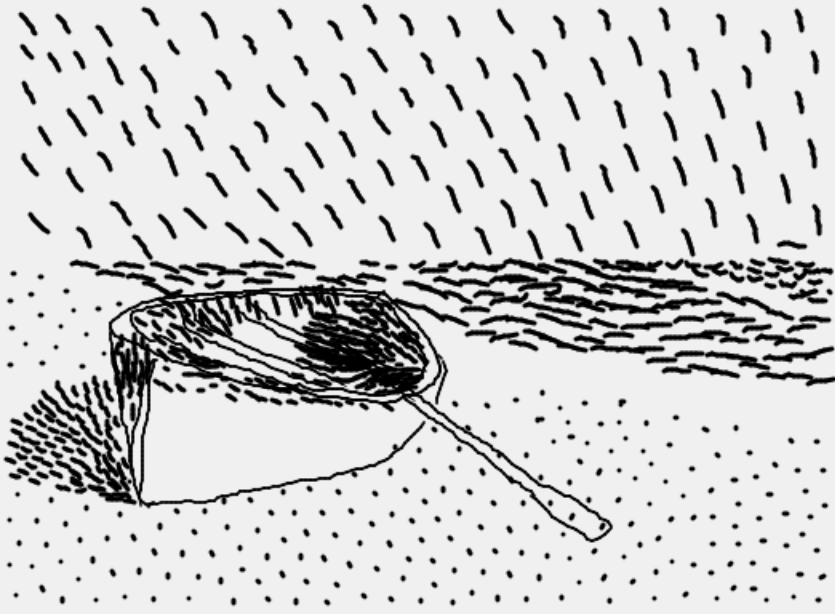}} &
	\subfloat{\includegraphics[width = 0.15\textwidth]{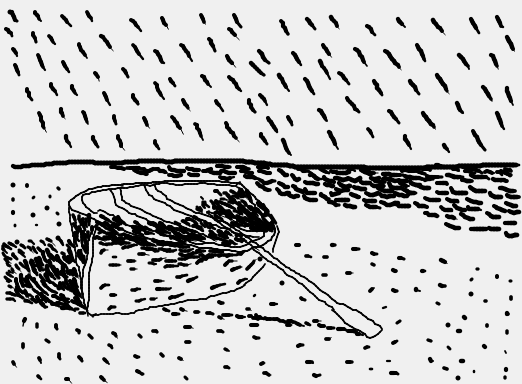}} \\
	\subfloat{\includegraphics[width = 0.15\textwidth]{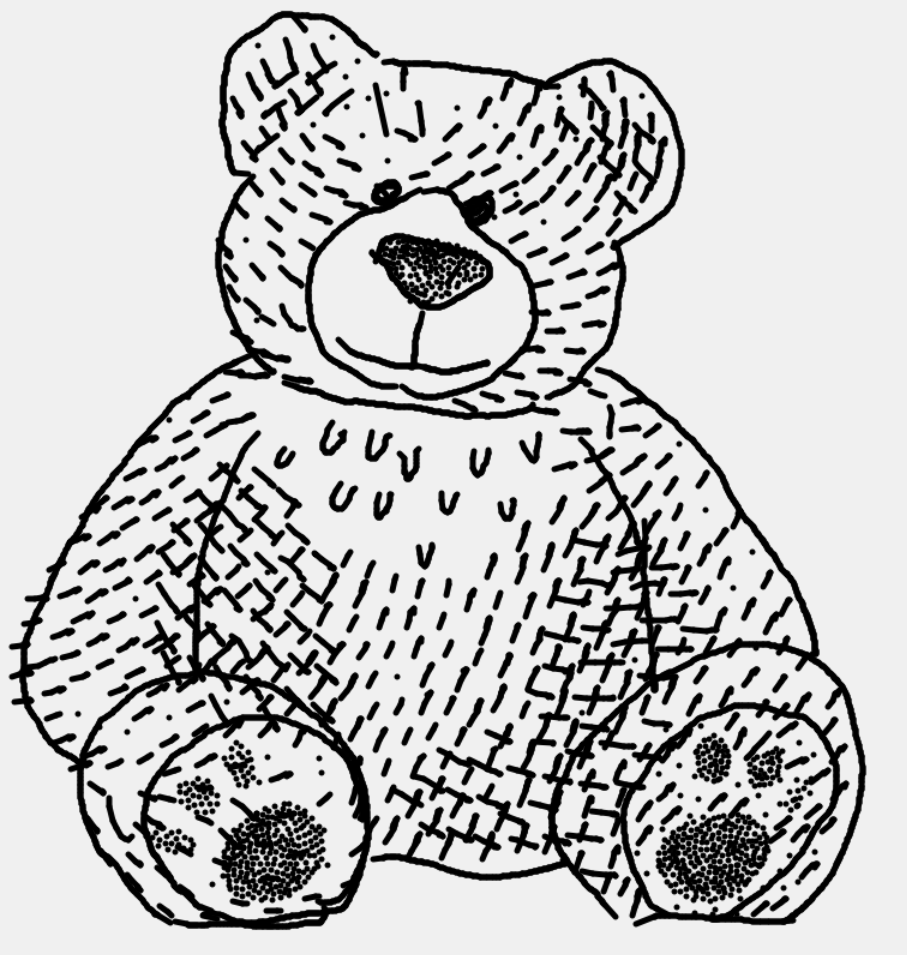}} &
	\subfloat{\includegraphics[width = 0.15\textwidth]{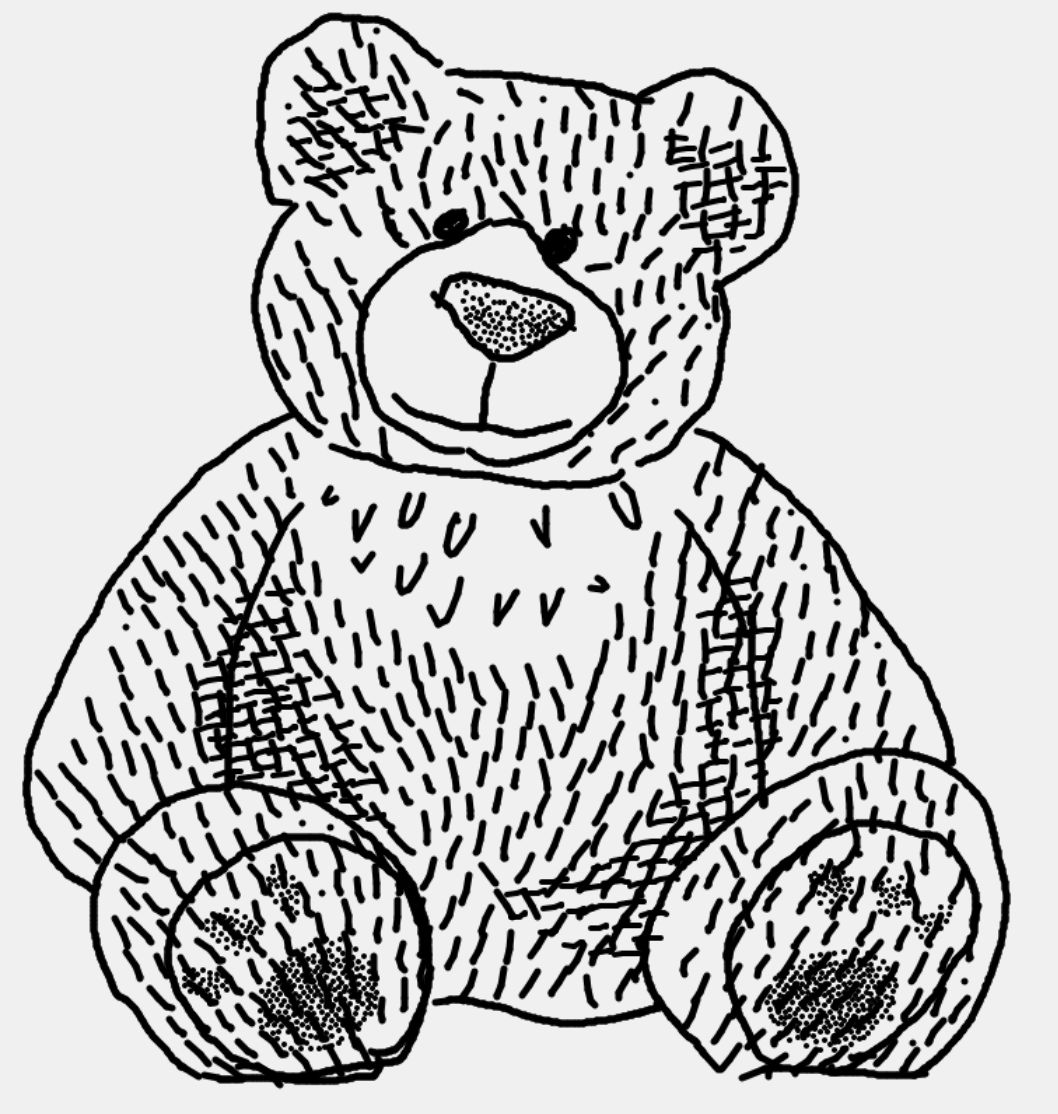}} &
	\subfloat{\includegraphics[width = 0.15\textwidth]{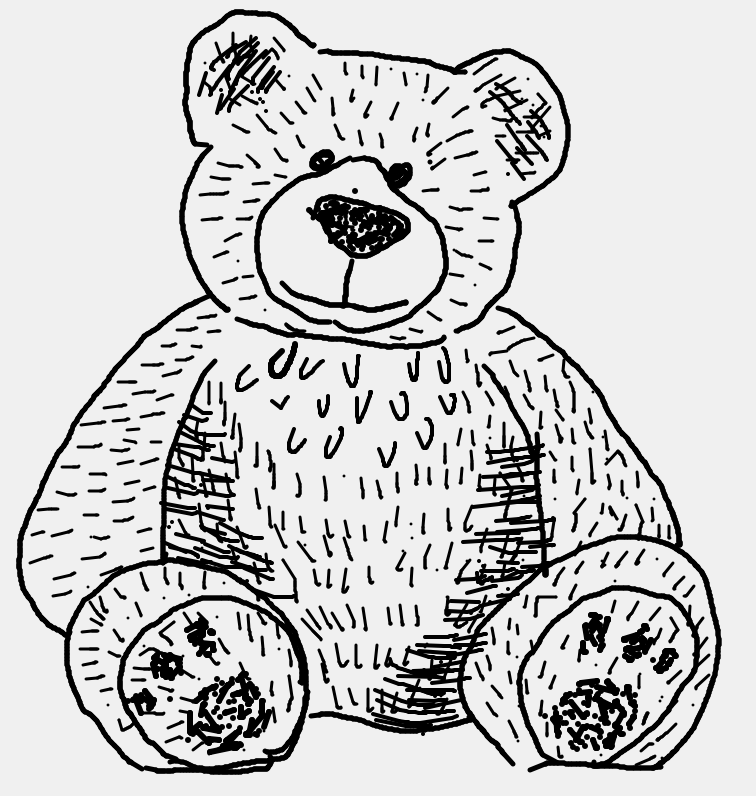}} &
	\subfloat{\includegraphics[width = 0.15\textwidth]{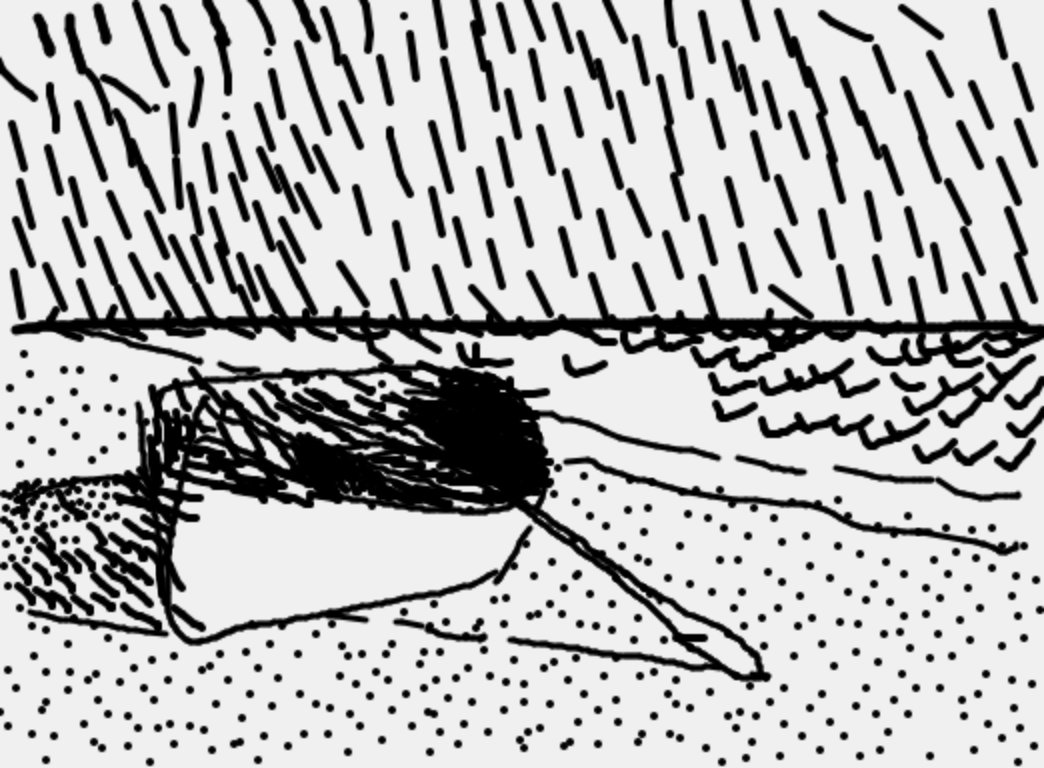}} &
	\subfloat{\includegraphics[width = 0.15\textwidth]{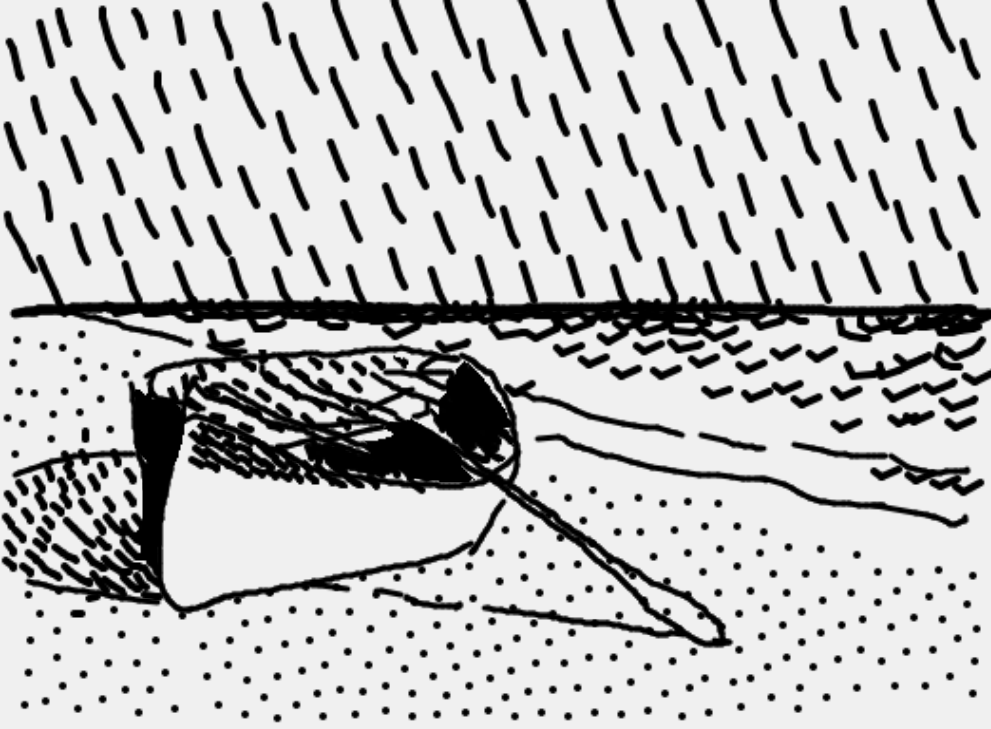}} &
	\subfloat{\includegraphics[width = 0.15\textwidth]{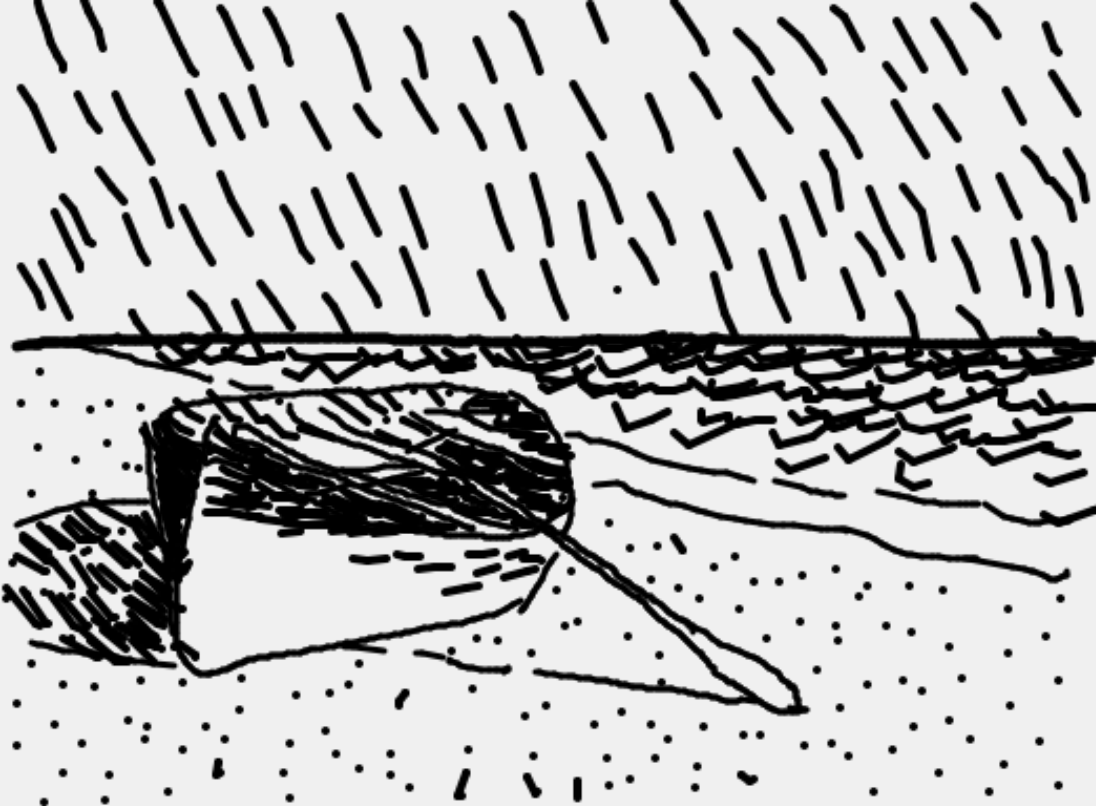}} \\
	\subfloat{\includegraphics[width = 0.15\textwidth]{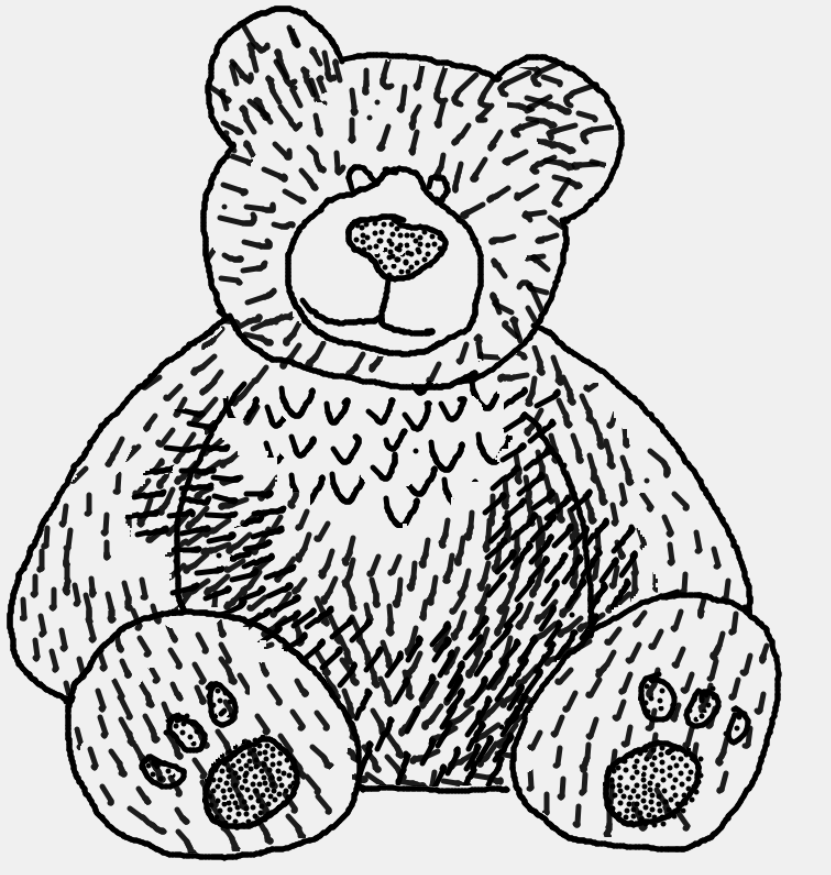}} &
	\subfloat{\includegraphics[width = 0.15\textwidth]{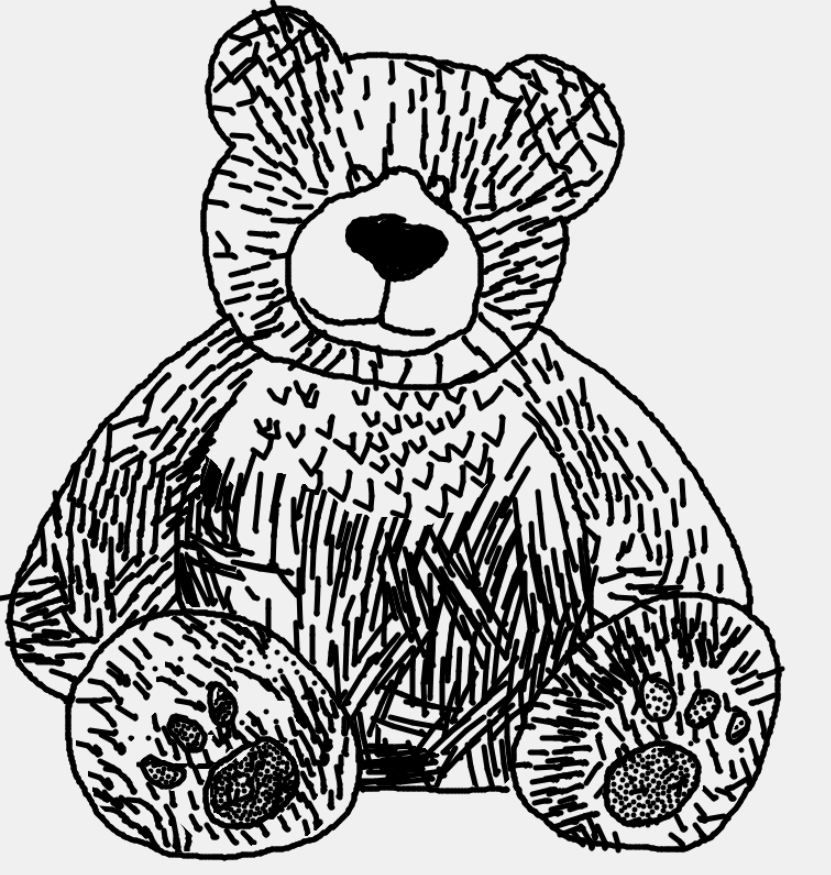}} &
	\subfloat{\includegraphics[width = 0.15\textwidth]{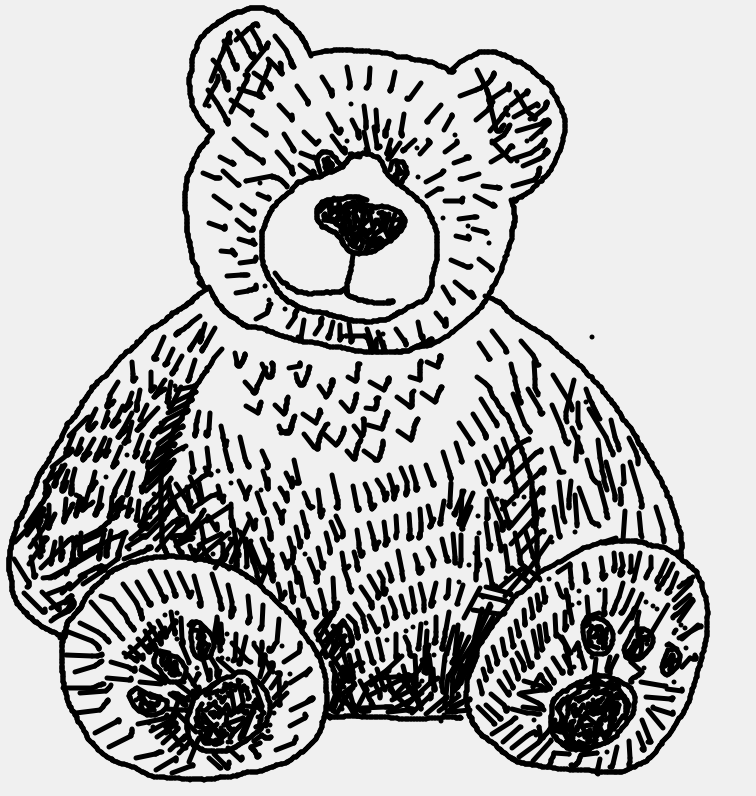}} &
	\subfloat{\includegraphics[width = 0.15\textwidth]{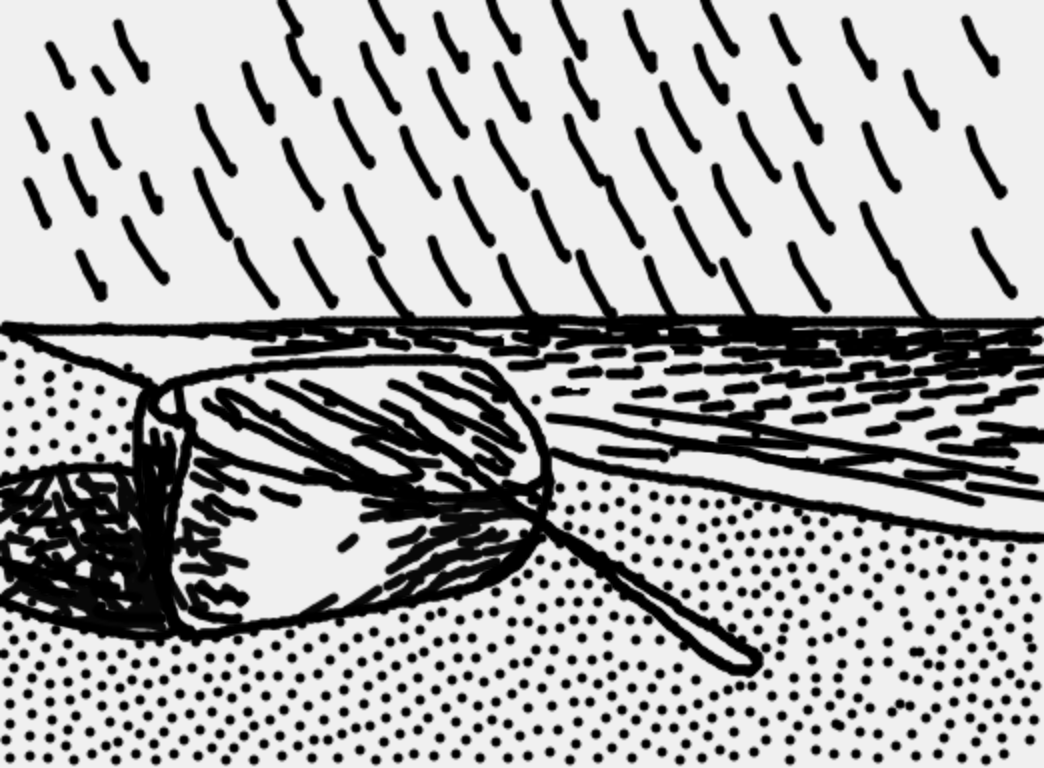}} &
	\subfloat{\includegraphics[width = 0.15\textwidth]{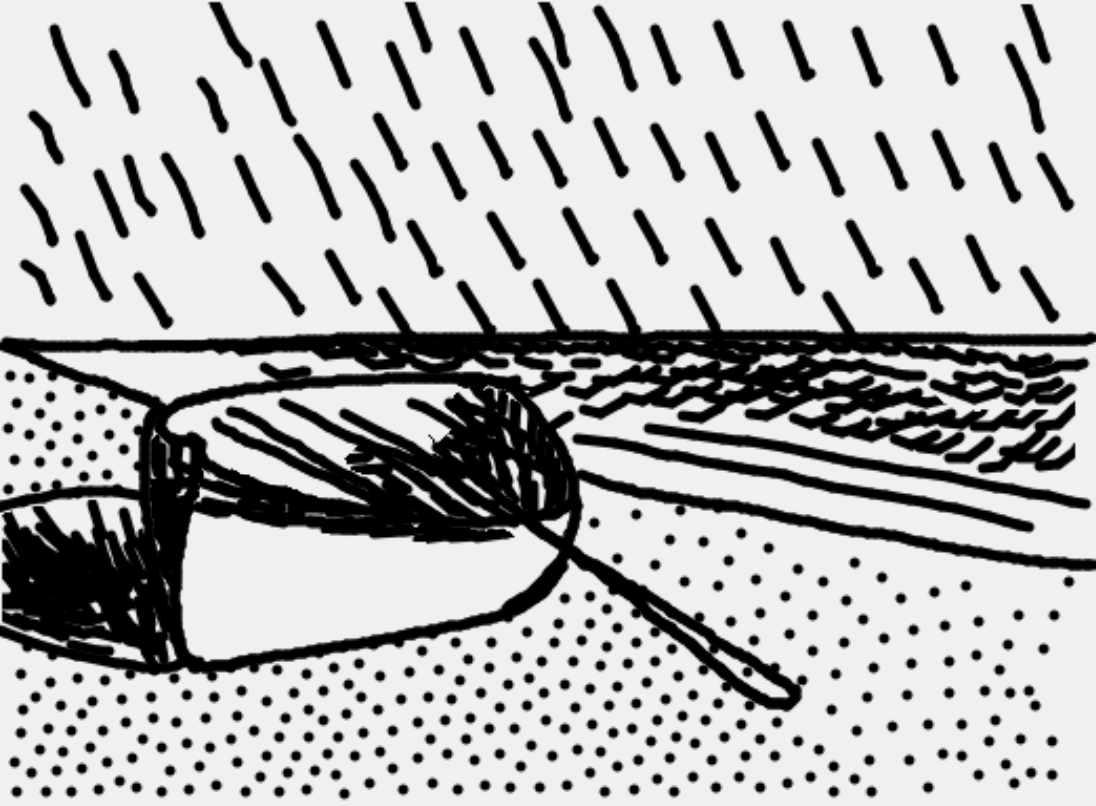}} &
	\subfloat{\includegraphics[width = 0.15\textwidth]{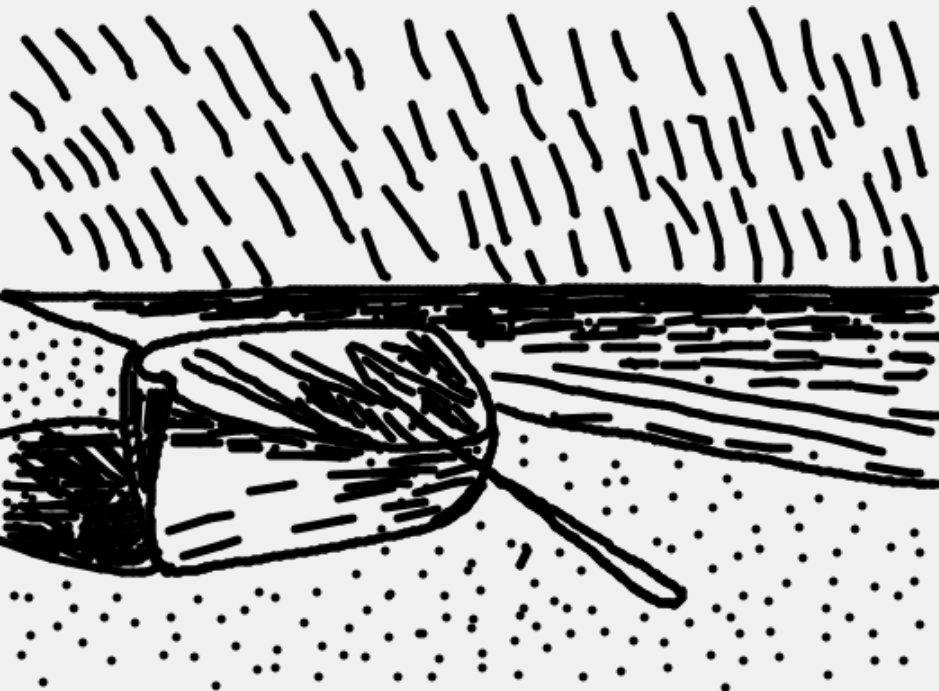}} \\
\end{tabular}
\end{figure*}

\begin{figure*}
	\begin{tabular}{cccccc}
	\subfloat{\includegraphics[width = 0.15\textwidth]{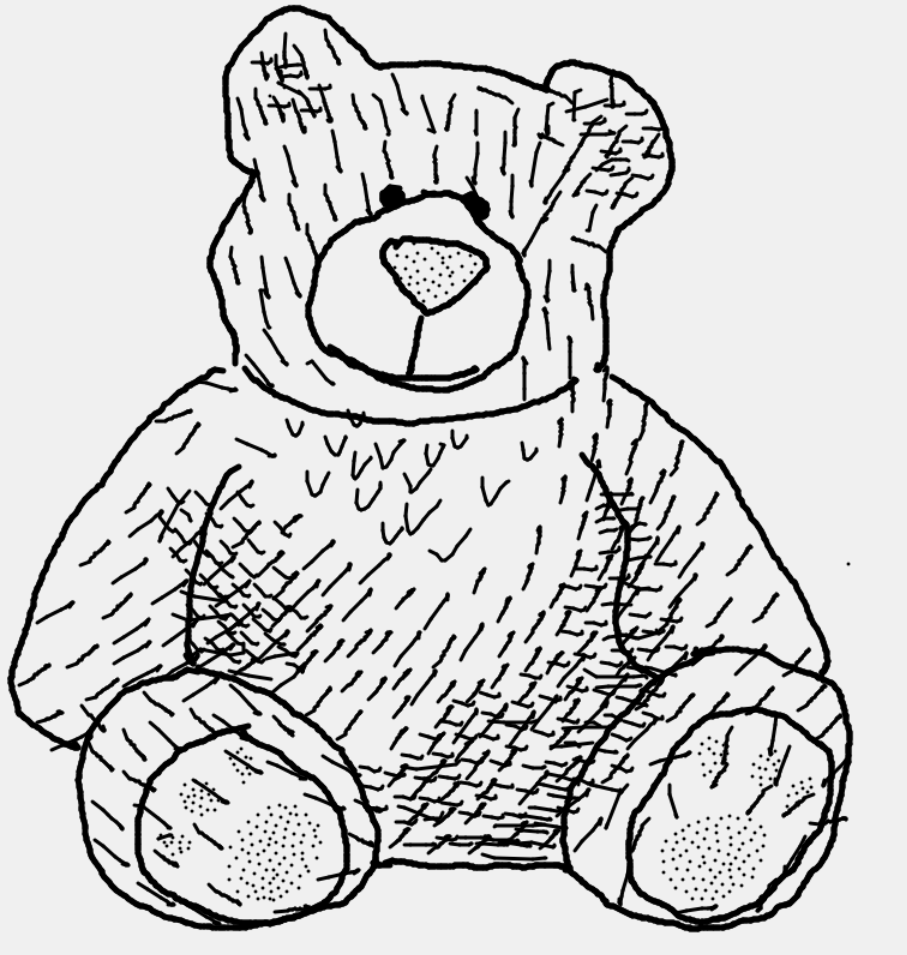}} &
	\subfloat{\includegraphics[width = 0.15\textwidth]{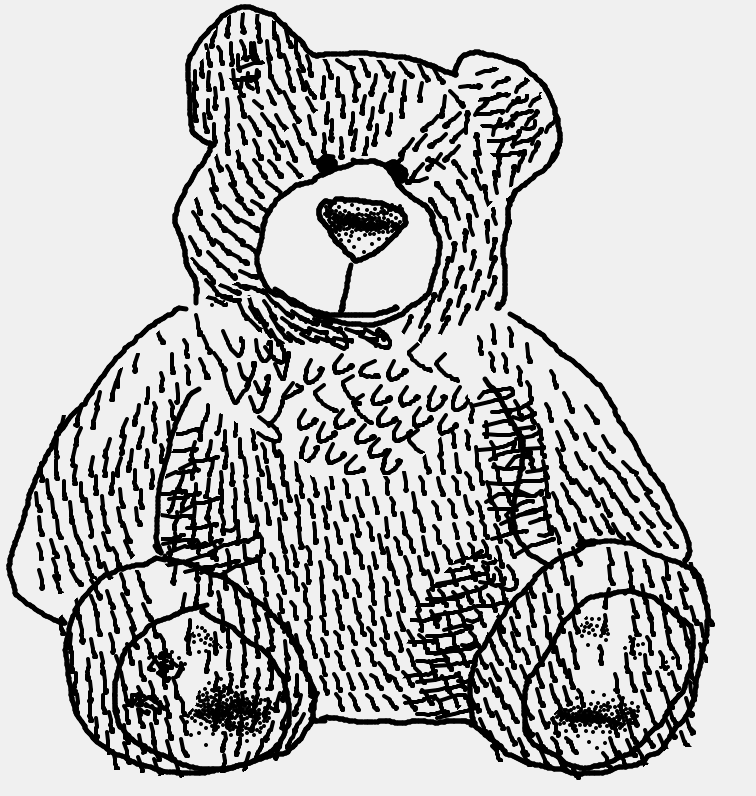}} &
	\subfloat{\includegraphics[width = 0.15\textwidth]{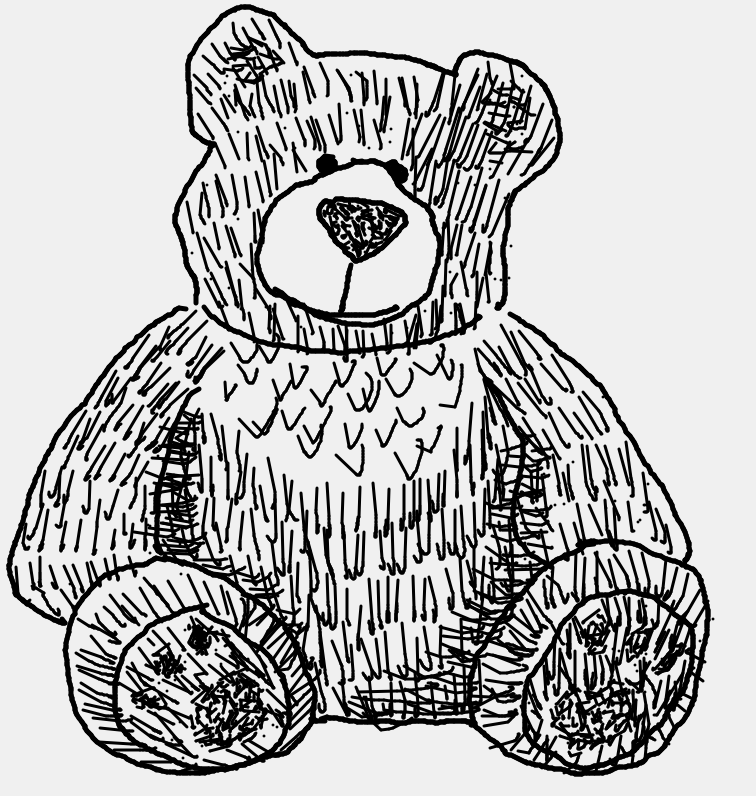}} &
	\subfloat{\includegraphics[width = 0.15\textwidth]{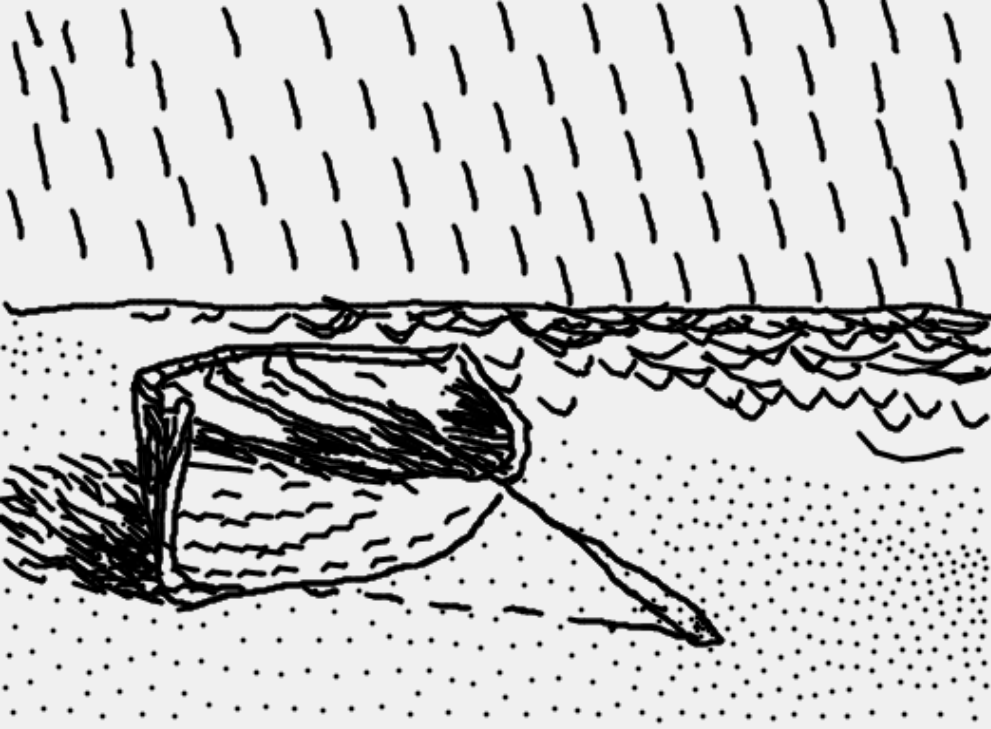}} &
	\subfloat{\includegraphics[width = 0.15\textwidth]{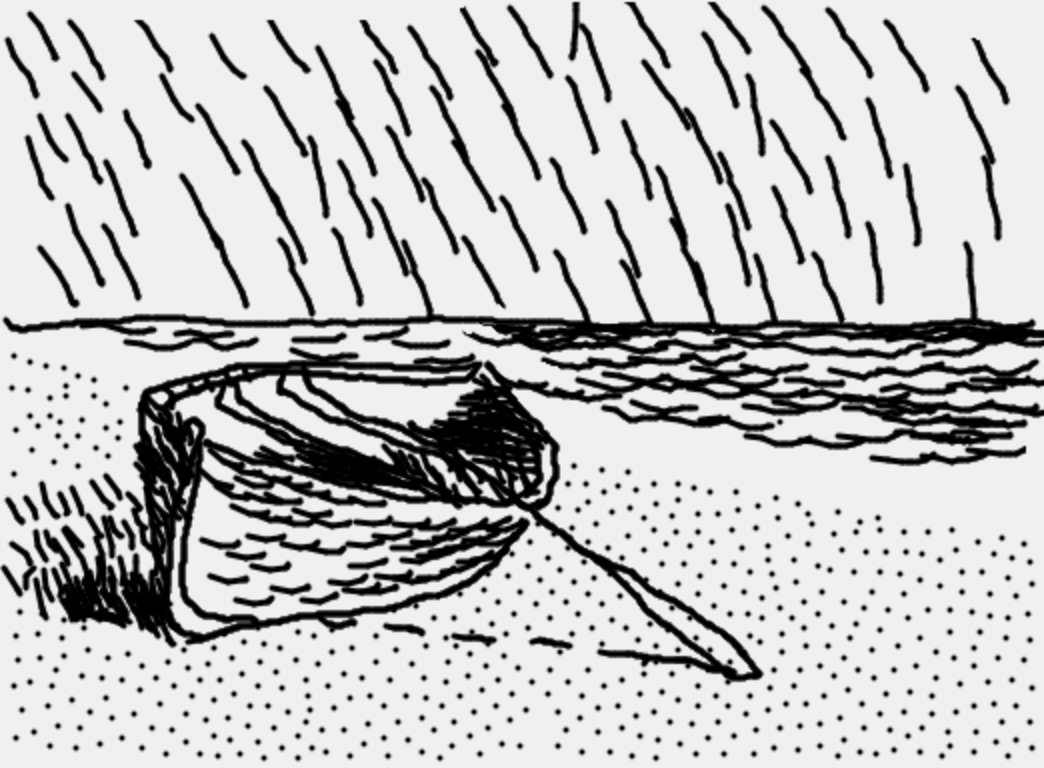}} &
	\subfloat{\includegraphics[width = 0.15\textwidth]{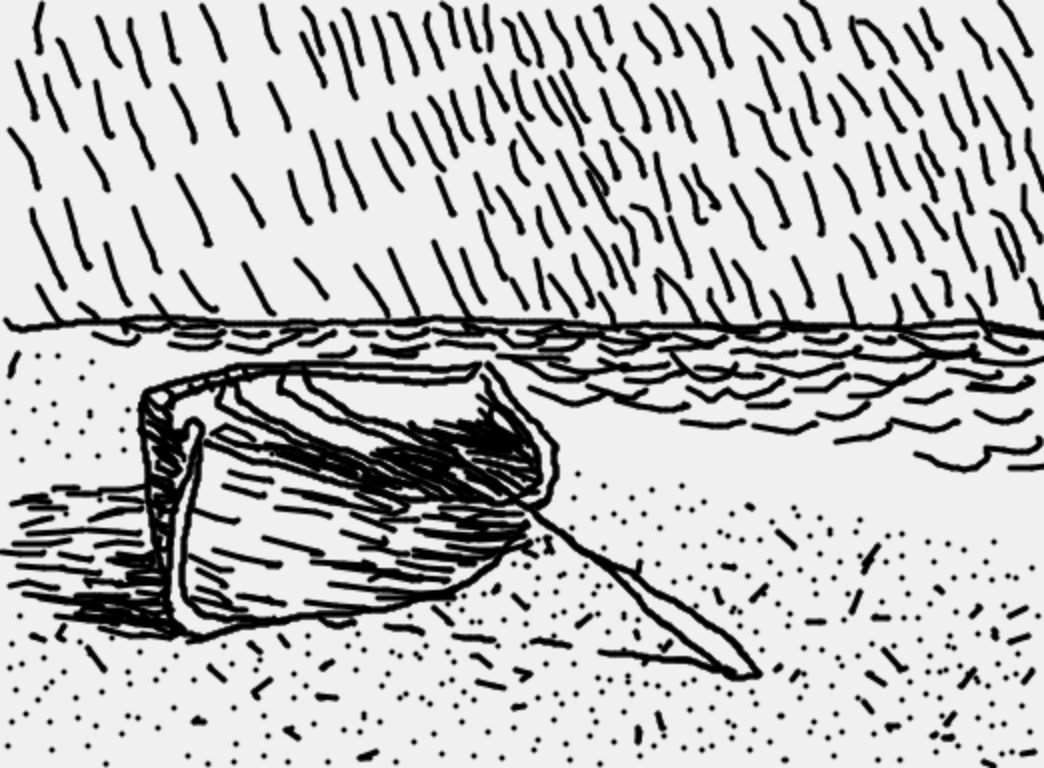}} \\
	\subfloat{\includegraphics[width = 0.15\textwidth]{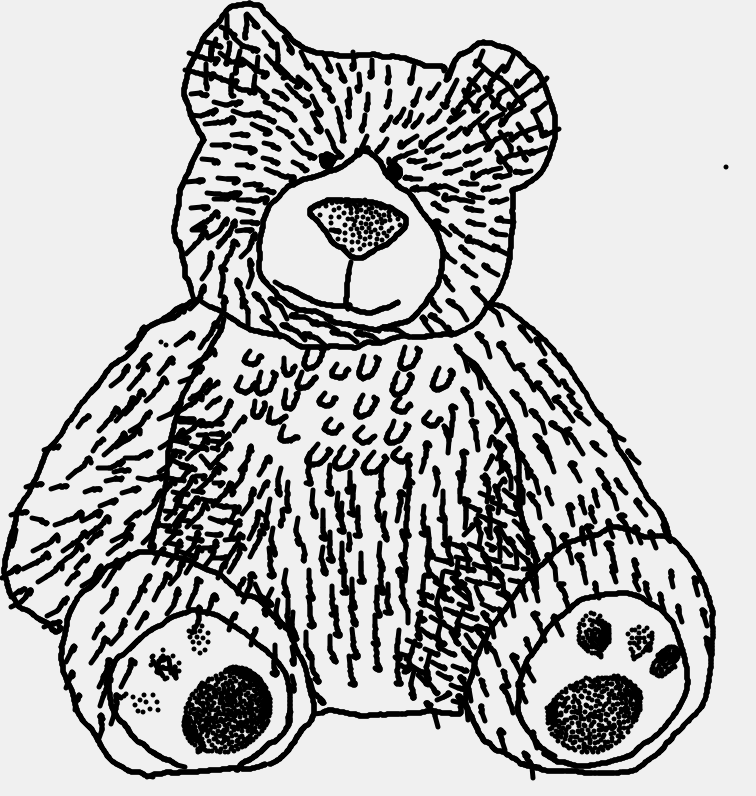}} &
	\subfloat{\includegraphics[width = 0.15\textwidth]{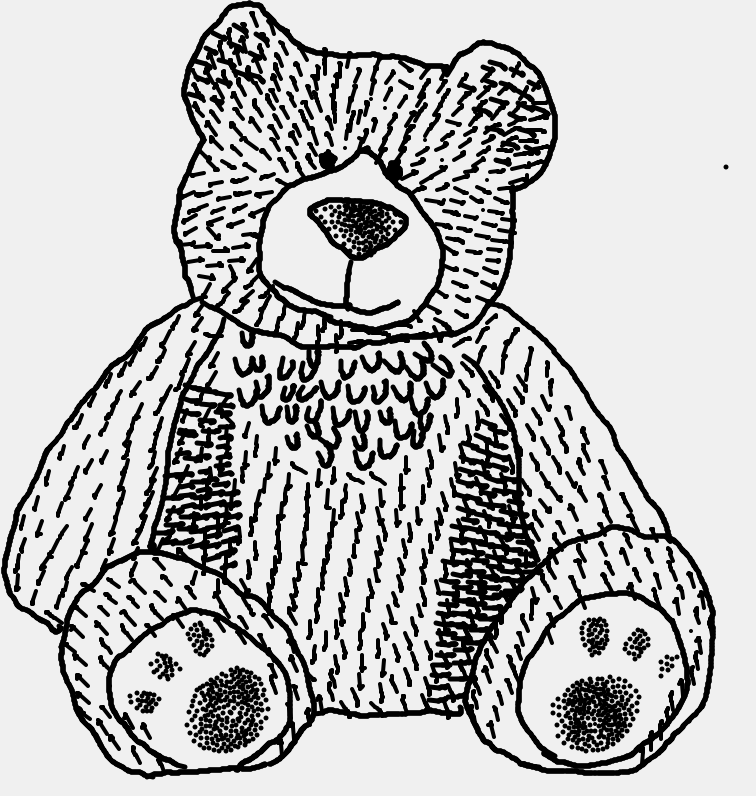}} &
	\subfloat{\includegraphics[width = 0.15\textwidth]{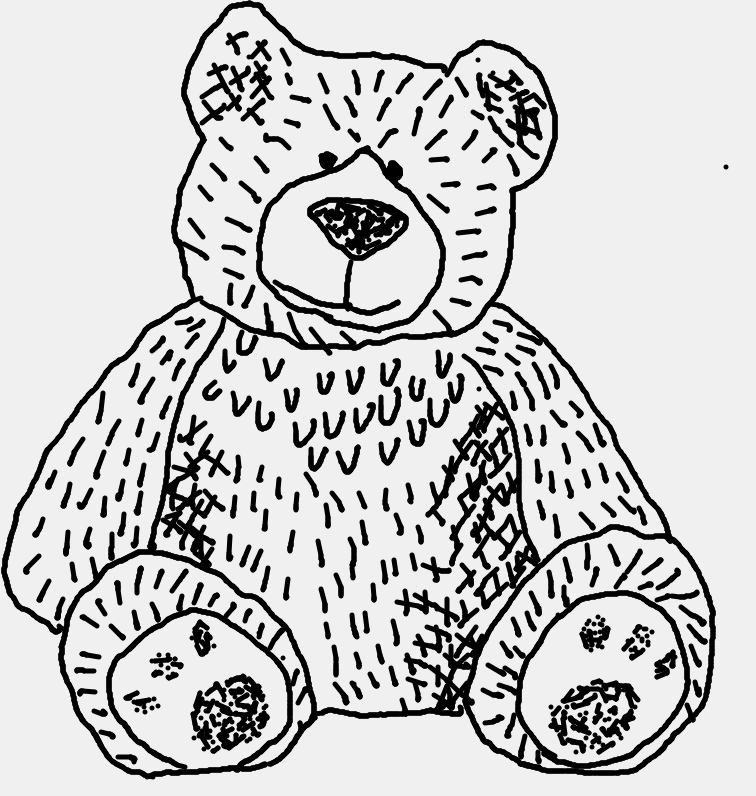}} &
	\subfloat{\includegraphics[width = 0.15\textwidth]{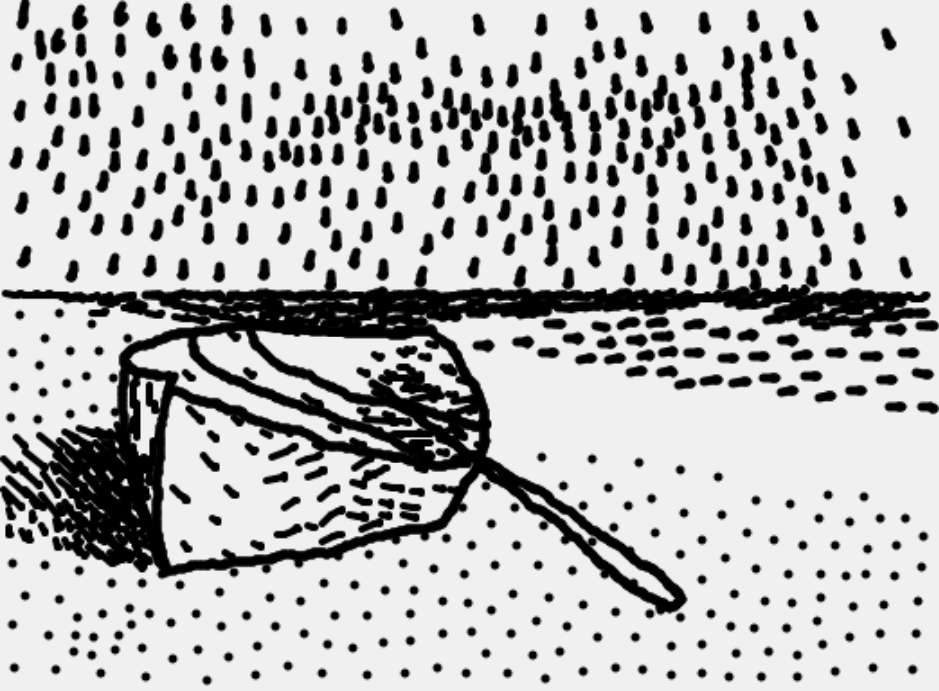}} &
	\subfloat{\includegraphics[width = 0.15\textwidth]{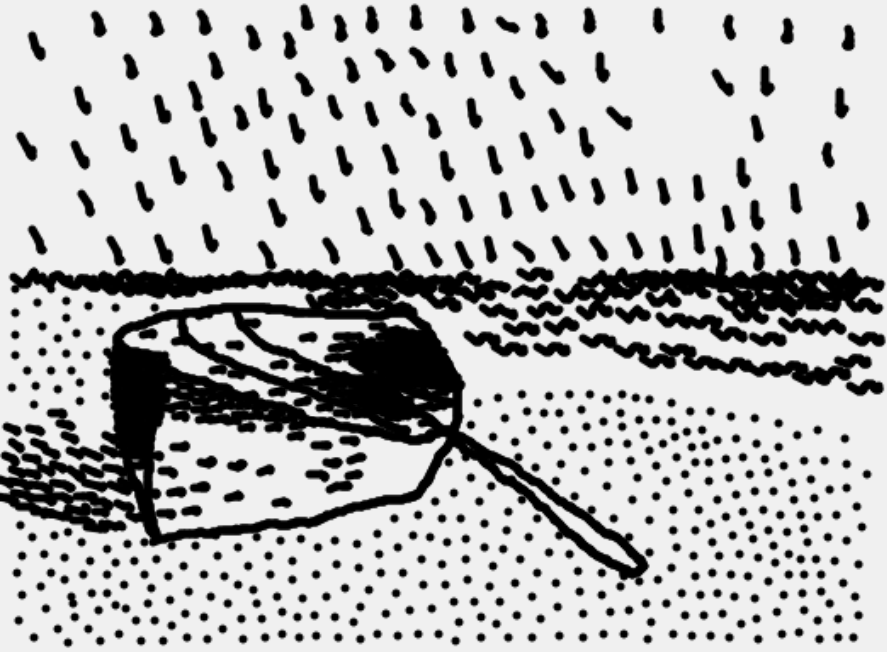}} &
	\subfloat{\includegraphics[width = 0.15\textwidth]{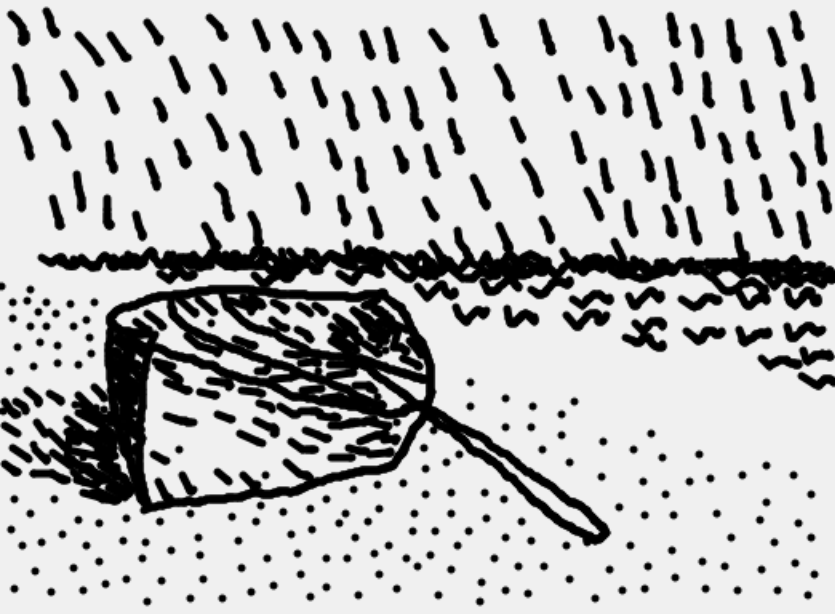}} \\
	\subfloat{\includegraphics[width = 0.15\textwidth]{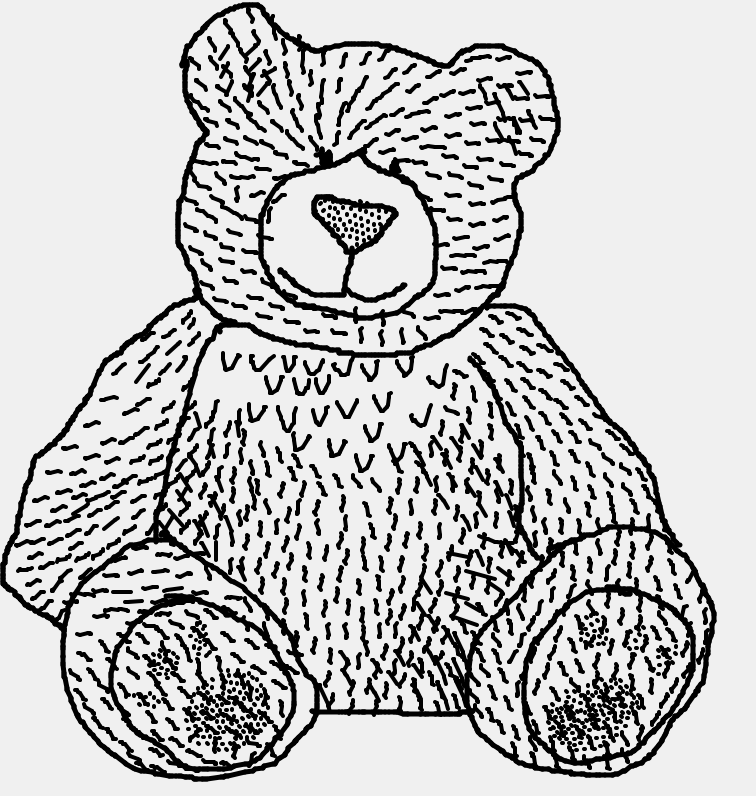}} &
	\subfloat{\includegraphics[width = 0.15\textwidth]{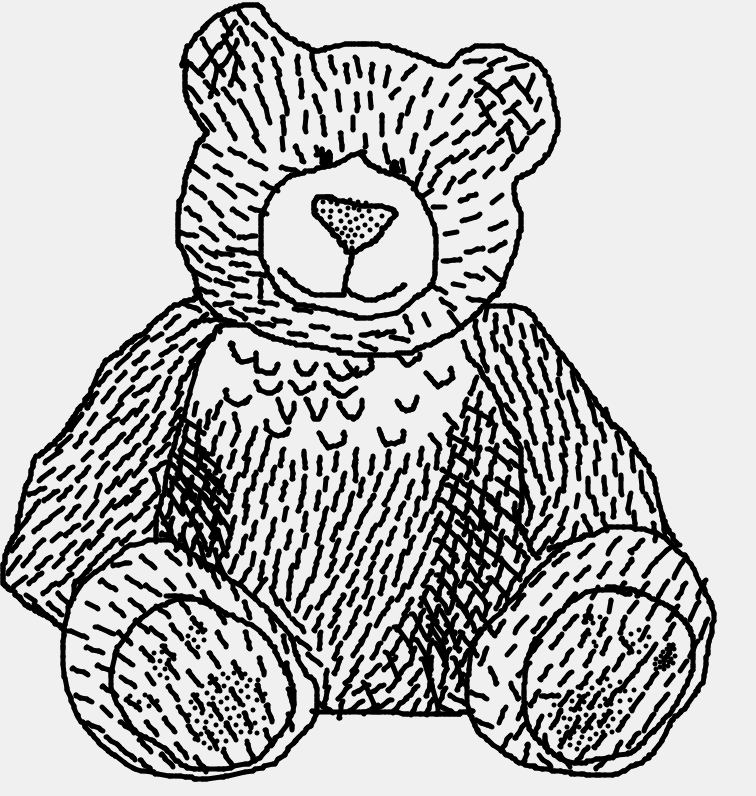}} &
	\subfloat{\includegraphics[width = 0.15\textwidth]{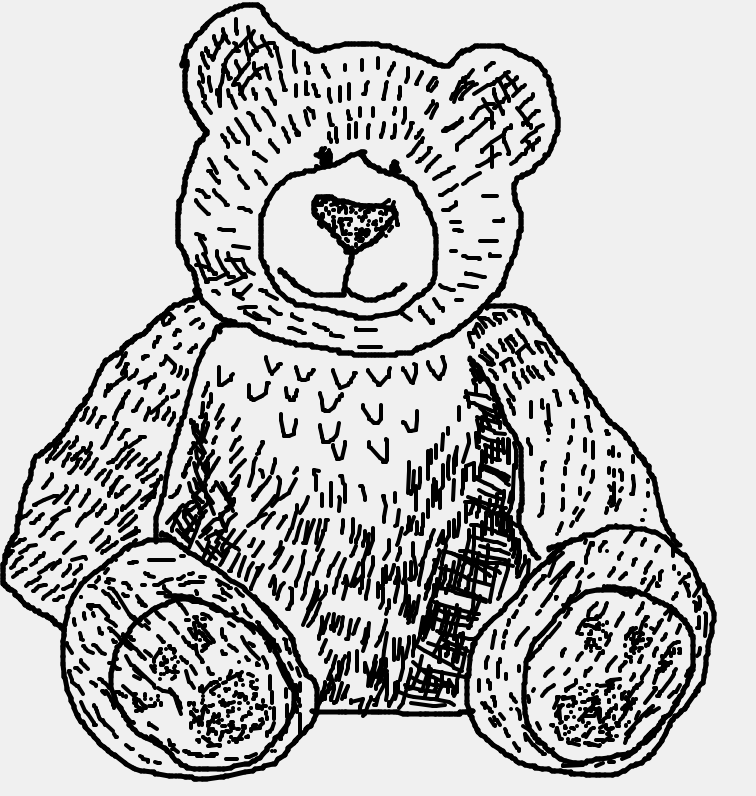}} &
	\subfloat{\includegraphics[width = 0.15\textwidth]{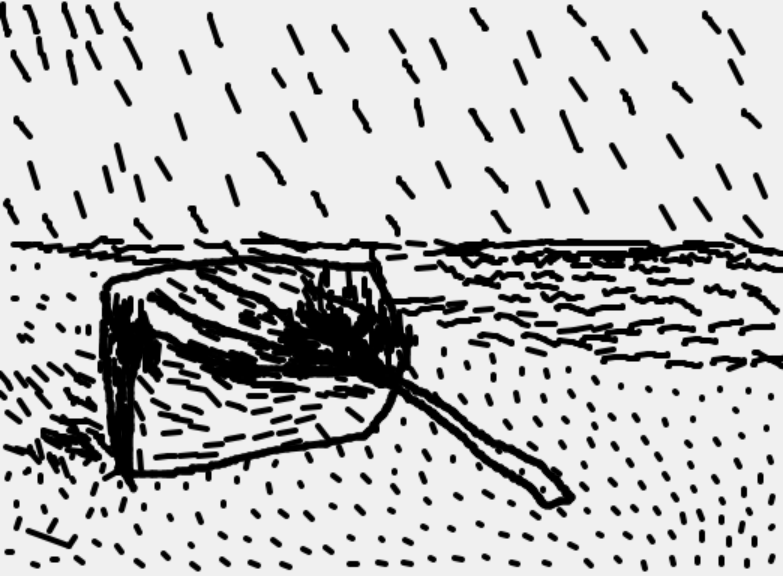}} &
	\subfloat{\includegraphics[width = 0.15\textwidth]{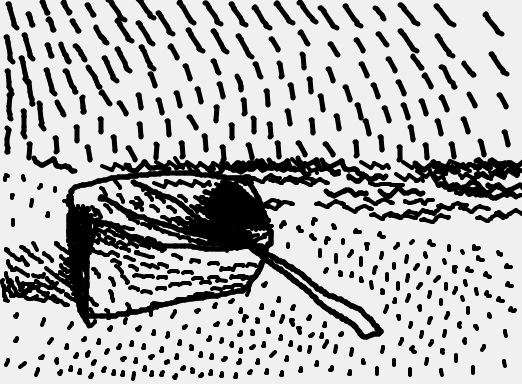}} &
	\subfloat{\includegraphics[width = 0.15\textwidth]{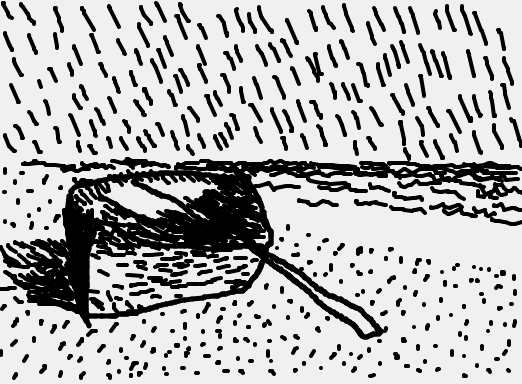}} \\
	\subfloat{\includegraphics[width = 0.15\textwidth]{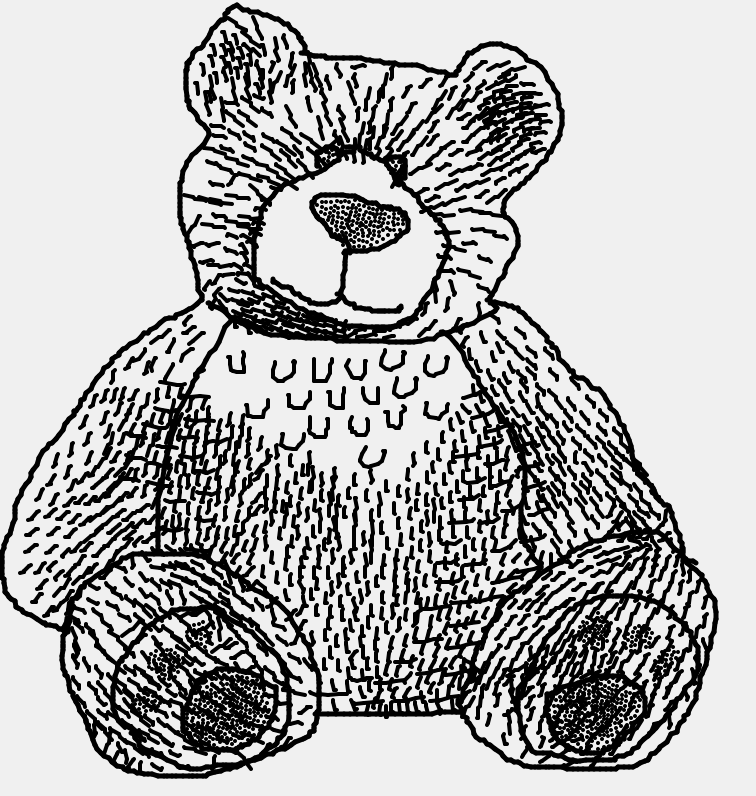}} &
	\subfloat{\includegraphics[width = 0.15\textwidth]{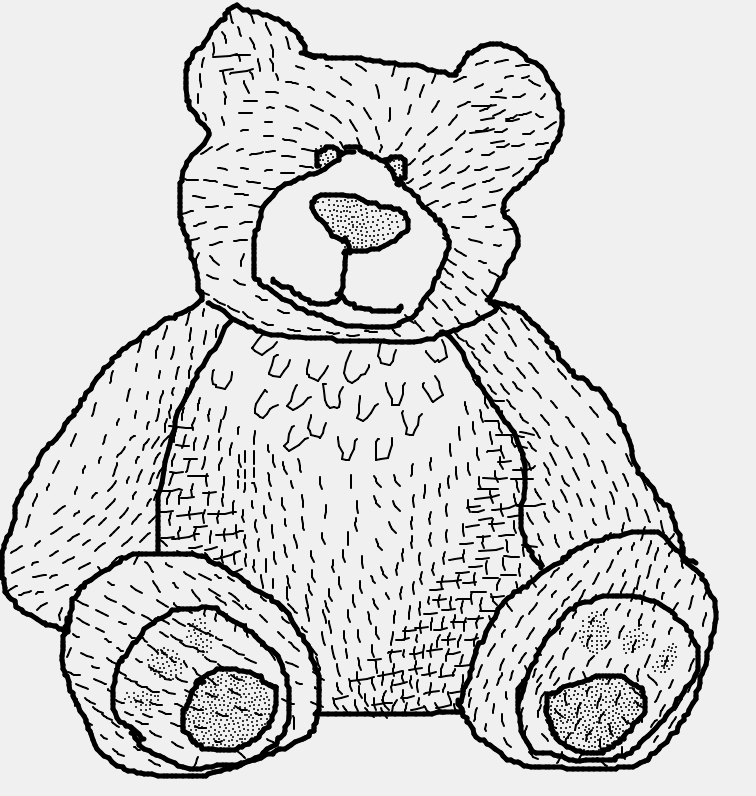}} &
	\subfloat{\includegraphics[width = 0.15\textwidth]{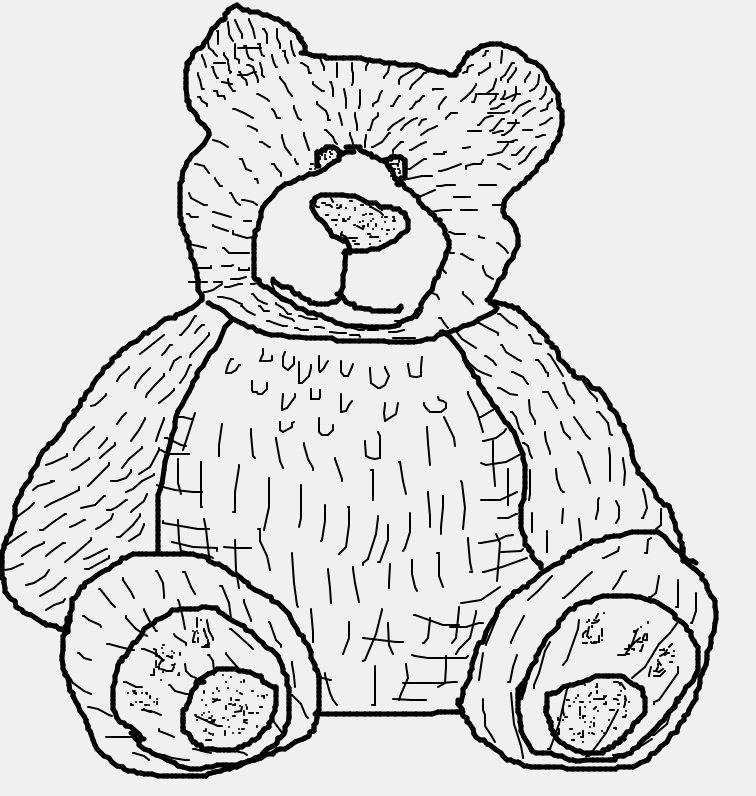}} &
	\subfloat{\includegraphics[width = 0.15\textwidth]{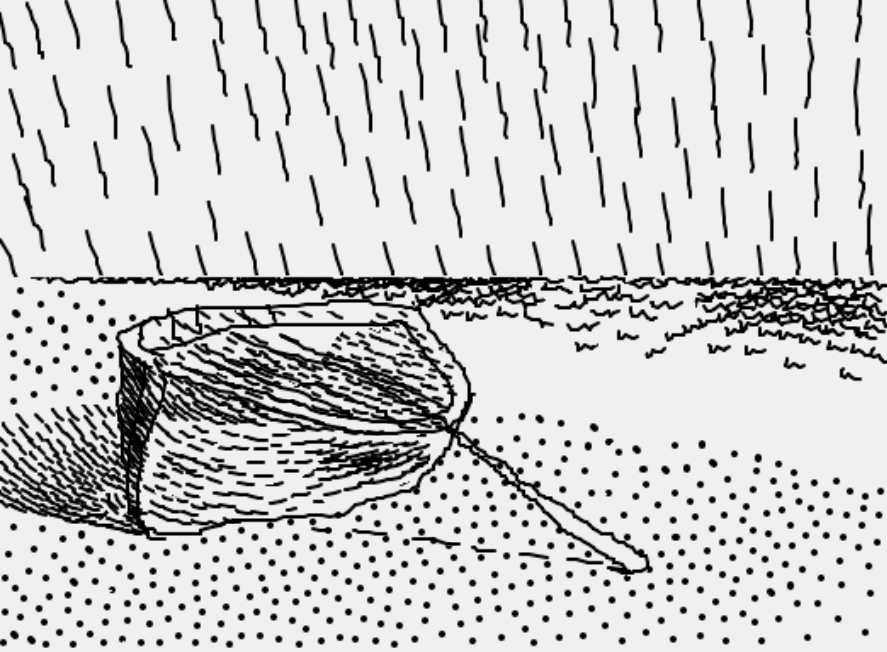}} &
	\subfloat{\includegraphics[width = 0.15\textwidth]{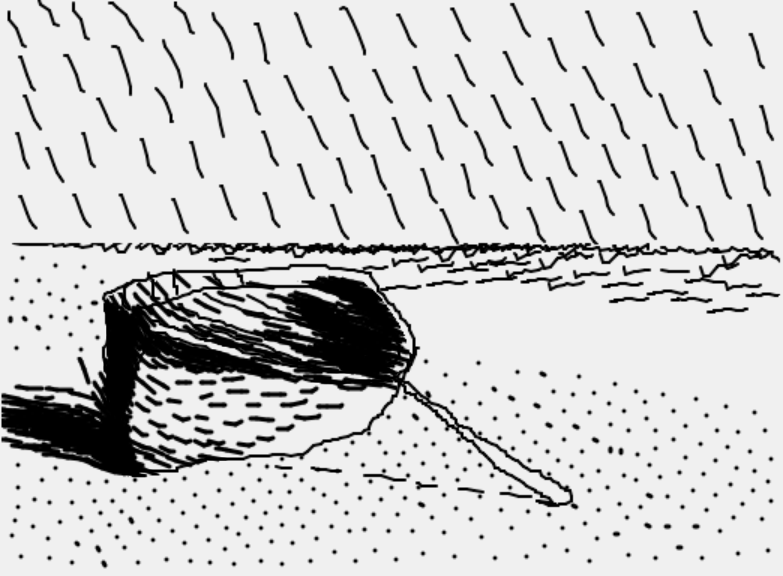}} &
	\subfloat{\includegraphics[width = 0.15\textwidth]{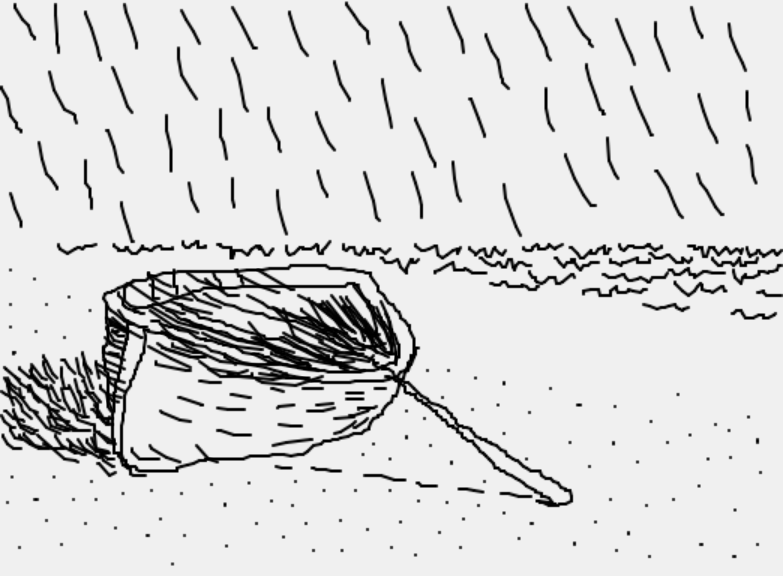}} \\
	\subfloat{\includegraphics[width = 0.15\textwidth]{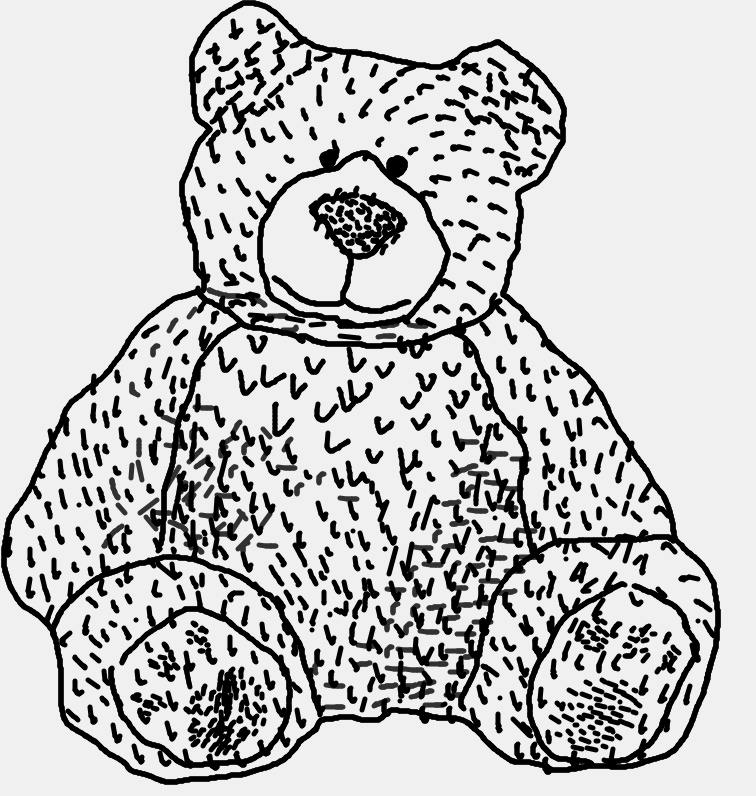}} &
	\subfloat{\includegraphics[width = 0.15\textwidth]{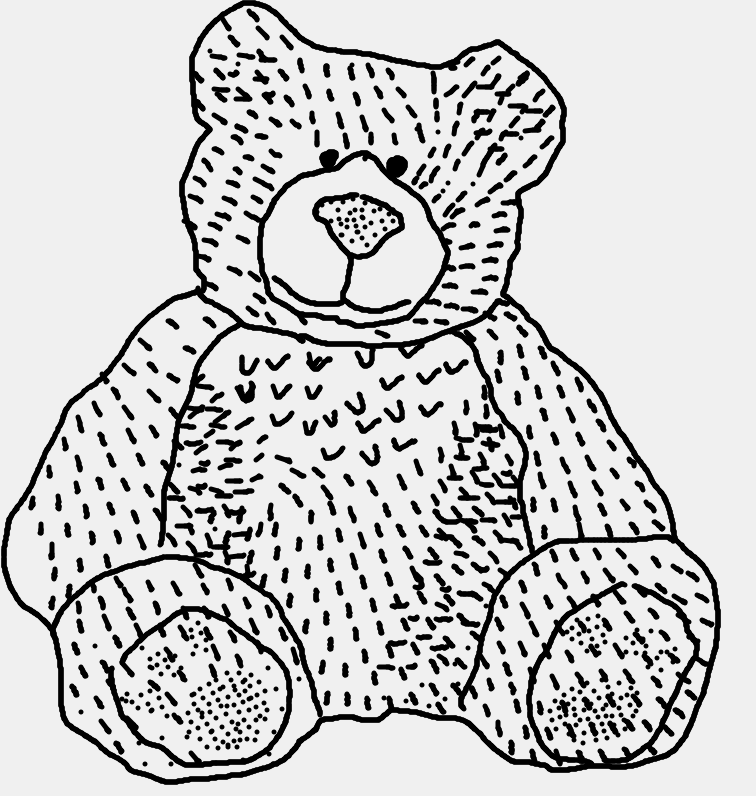}} &
	\subfloat{\includegraphics[width = 0.15\textwidth]{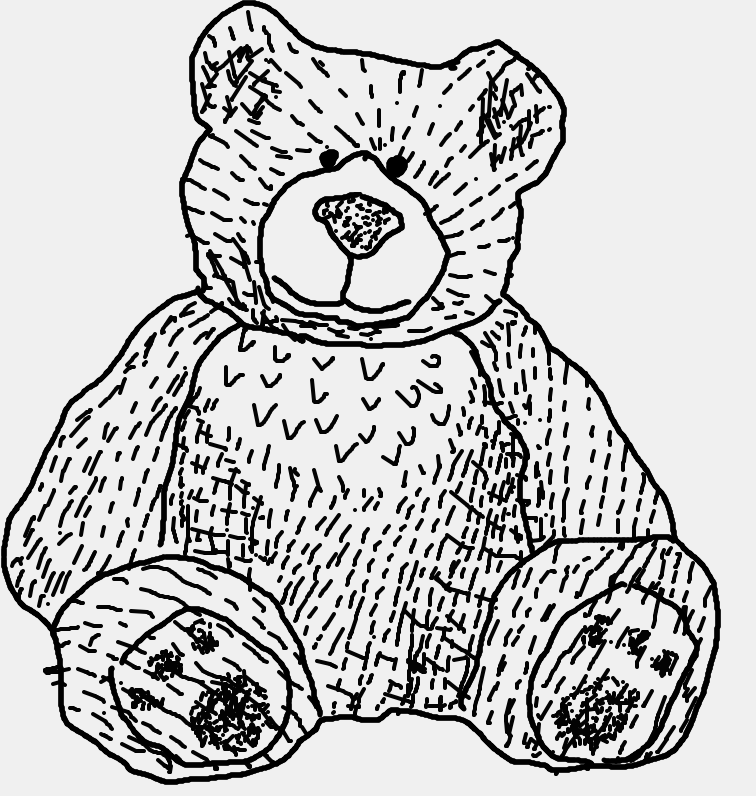}} &
	\subfloat{\includegraphics[width = 0.15\textwidth]{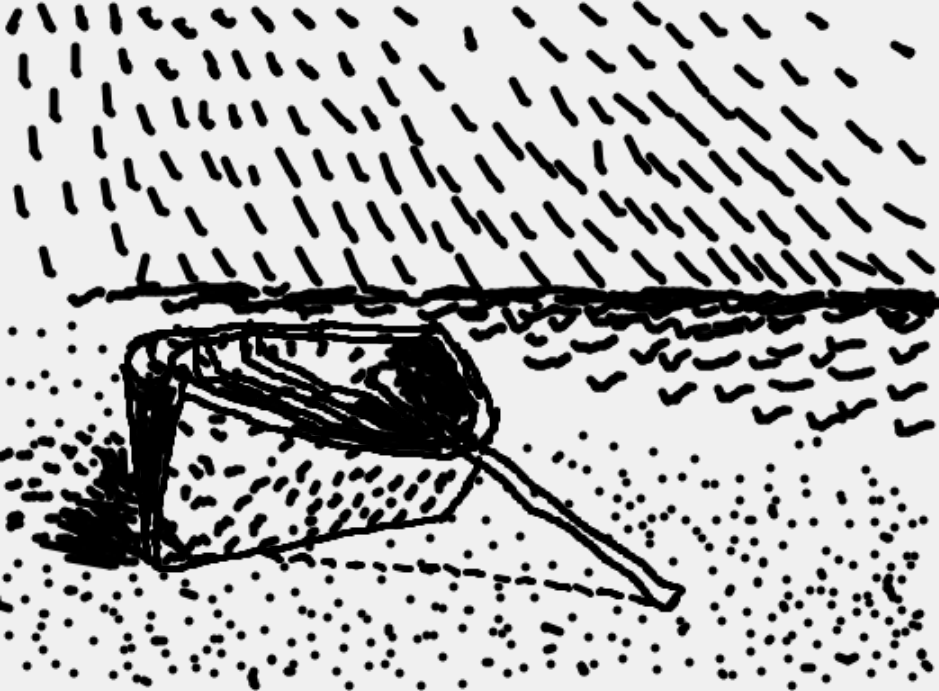}} &
	\subfloat{\includegraphics[width = 0.15\textwidth]{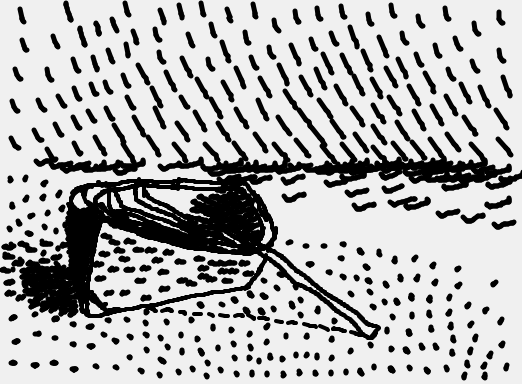}} &
	\subfloat{\includegraphics[width = 0.15\textwidth]{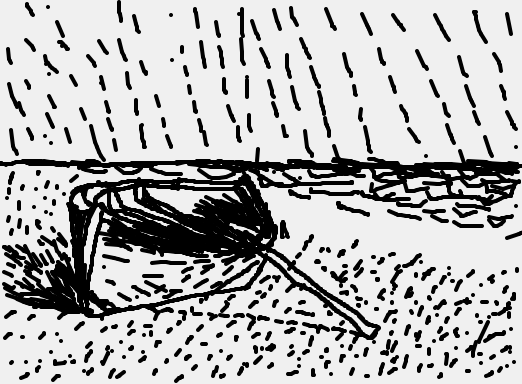}} \\
	\subfloat{\includegraphics[width = 0.15\textwidth]{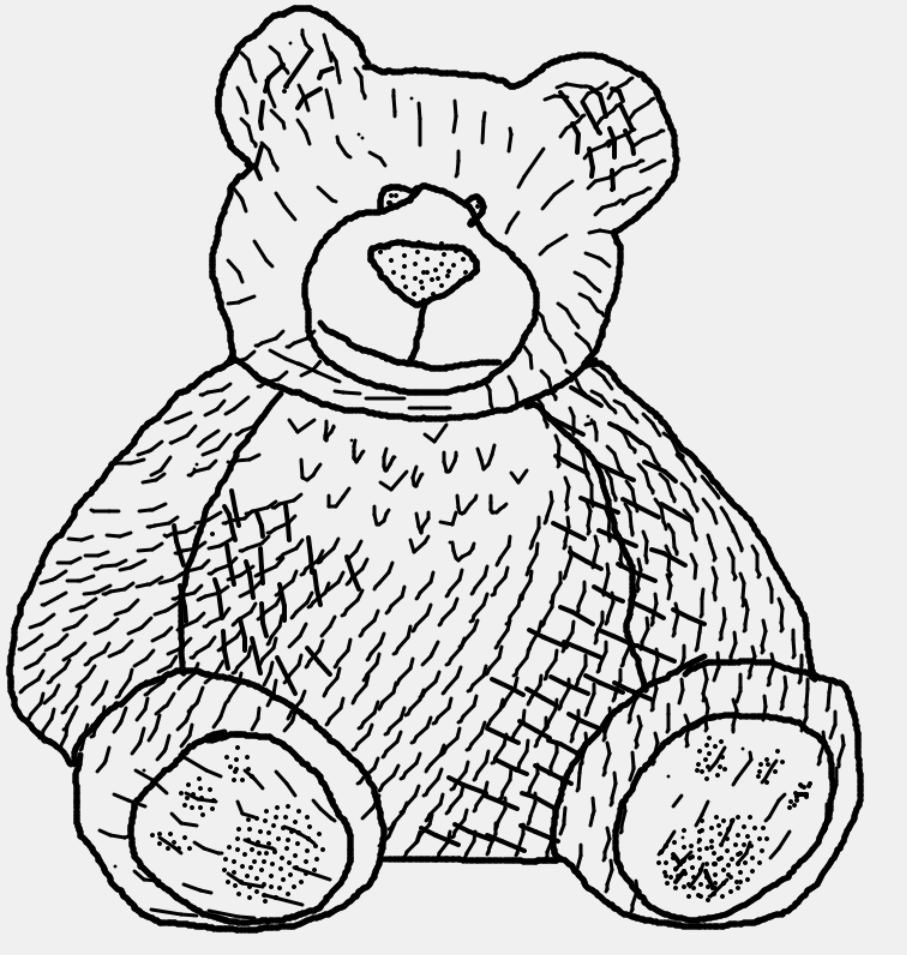}} &
	\subfloat{\includegraphics[width = 0.15\textwidth]{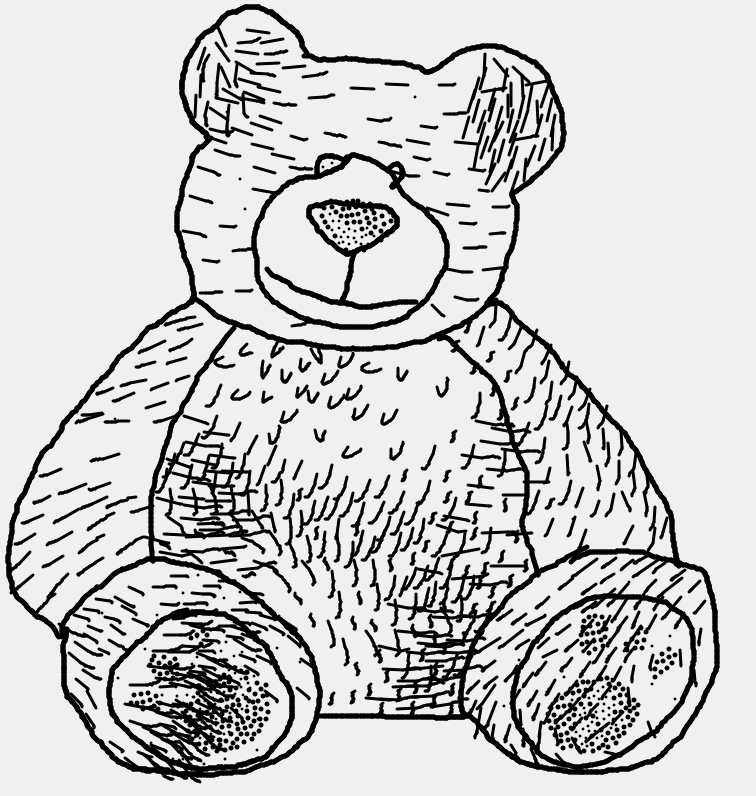}} &
	\subfloat{\includegraphics[width = 0.15\textwidth]{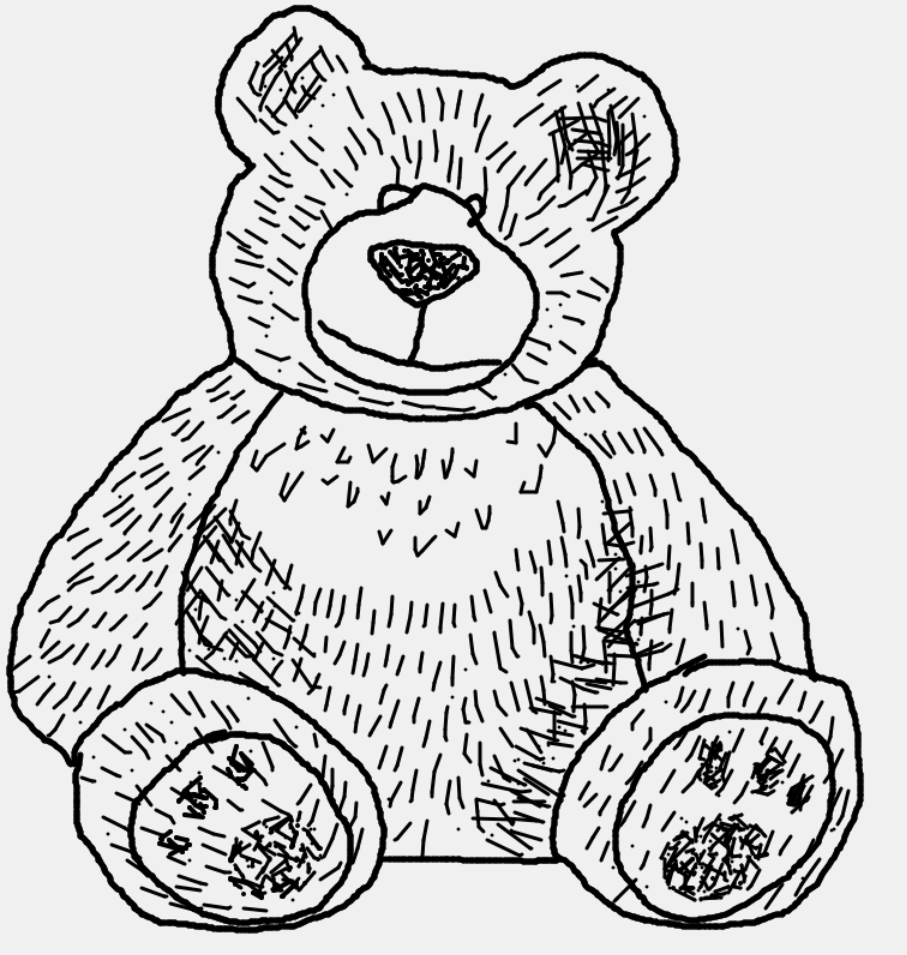}} &
	\subfloat{\includegraphics[width = 0.15\textwidth]{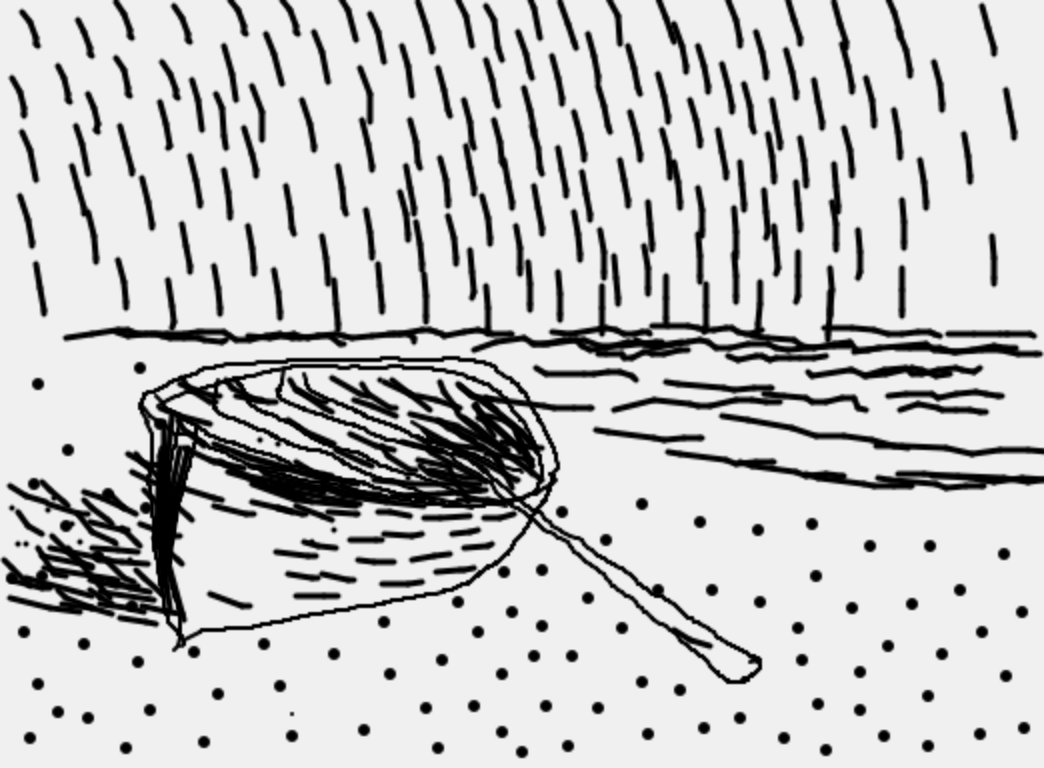}} &
	\subfloat{\includegraphics[width = 0.15\textwidth]{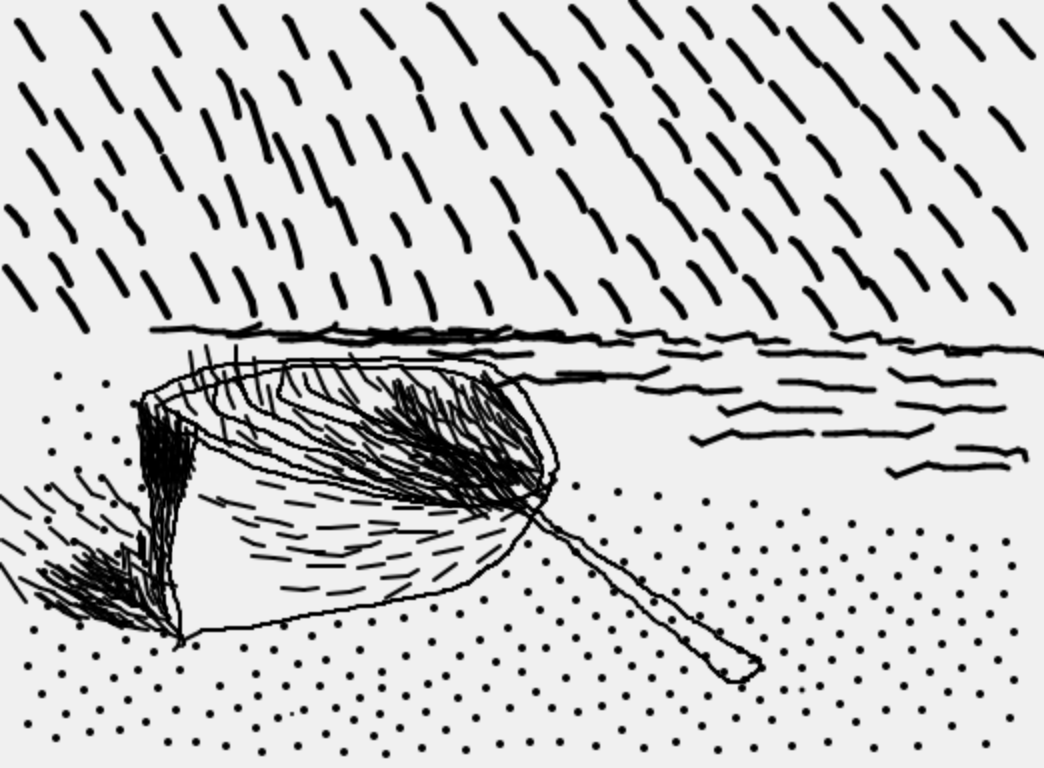}} &
	\subfloat{\includegraphics[width = 0.15\textwidth]{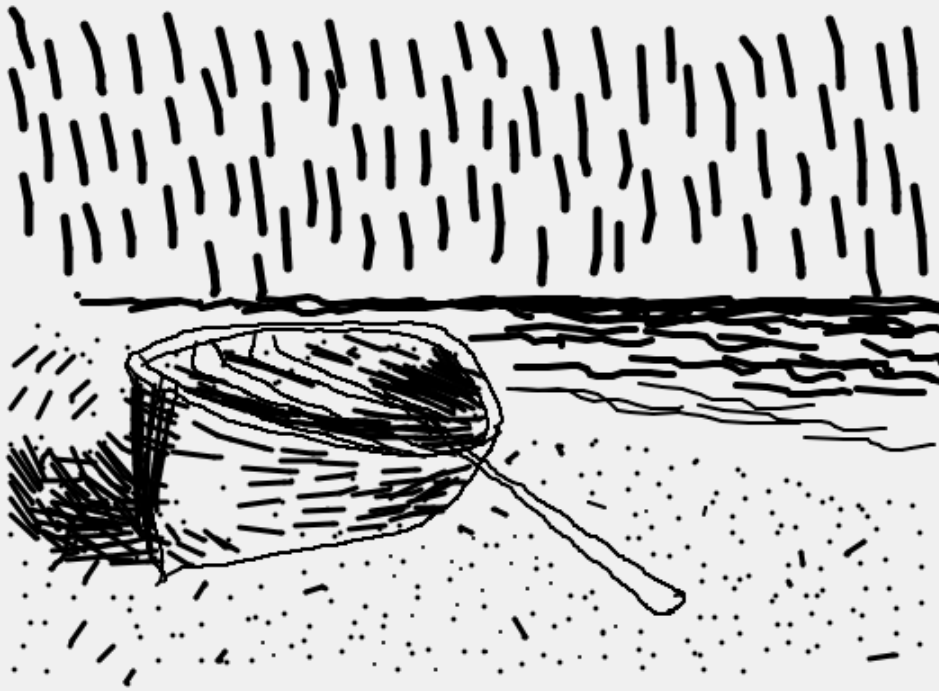}} 
\end{tabular}
\end{figure*}